\documentclass[11pt]{amsart}

 \usepackage{hyperref}
\usepackage{tikz}
\usetikzlibrary{arrows, decorations.markings, decorations.pathmorphing, decorations.pathreplacing, shapes.arrows, patterns, calc}
\usepackage{url}

\tikzset{
  ->-/.style={decoration={markings, mark=at position 0.5 with {\arrow{to}}},
              postaction={decorate}},
}

\tikzset{
  -<-/.style={decoration={markings, mark=at position 0.5 with {\arrow{to reversed}}},
              postaction={decorate}},
}

\tikzset{
  dbl->-/.style={
double, 
double equal sign distance,
shorten >= 1pt,
shorten <= 1pt,
 decoration={markings, mark=at position 0.5 with {\arrow{implies}}},
              postaction={decorate}},
}

\tikzset{
  dbl-<-/.style={
double, 
double equal sign distance,
shorten >= 1pt,
shorten <= 1pt,
 decoration={markings, mark=at position 0.5 with {\arrowreversed{implies}}},
              postaction={decorate}},
}

\usepackage{amsmath,amsthm}
\pdfmapfile{+mathpple.map}
\usepackage{mathrsfs}
\usepackage{amscd,amssymb, amsfonts, verbatim,subfigure, enumerate}
\usepackage[mathcal]{eucal}
\usepackage[super]{nth}

\usepackage{mathpazo}

\usepackage{color,slashed}

\renewcommand{\c}{\mf{c}}

\newcommand{\dbar}{\br{\partial}}

\newcommand{\CP}{\mathbb{CP}}
\newcommand{\wbar}{\br{w}}

\newcommand{\zbar}{\br{z}}

\newcommand{\PV}{\op{PV}}

\newcommand{\dpa}[1]{\frac{\partial}{\partial #1}}

\newcommand{\eps}{\epsilon}
\newcommand{\g}{\mathfrak{g}}

\newcommand{\what}{\widehat}

\newcommand{\til}{\widetilde}
\newcommand{\mscr}{\mathscr}

\newcommand{\br}{\overline}

\newcommand{\iso}{\cong}
\newcommand{\C}{\mathbb C}

\newcommand{\norm}[1]{\left\| #1 \right\|}
\newcommand{\Oo}{\mscr O}
\newcommand{\Z}{\mathbb Z}

\newcommand{\op}{\operatorname}
\newcommand{\mbf}{\mathbf}
\newcommand{\mbb}{\mathbb}
\newcommand{\mf}{\mathfrak}
\newcommand{\mc}{\mathcal}

\newcommand{\ip}[1]{\left\langle #1 \right\rangle}
\newcommand{\abs}[1]{\left| #1 \right|}

\newcommand{\R}{\mbb R}
\renewcommand{\d}{\mathrm{d}}

\DeclareMathOperator{\Sym}{Sym}

\newtheoremstyle{thm}
  {7pt}
  {7pt}
  {\itshape}
  {}
  {\bf}
  {.}
  {5pt}
  {\thmnumber{#2 }\thmname{#1}\thmnote{ (#3)}}

\newtheoremstyle{def}
  {7pt}
  {10pt}
  {\itshape}
  {}
  {\bf}
  {.}
  {5pt}
  {\thmnumber{#2} \thmname{#1}\thmnote{ (#3)}}

\newtheoremstyle{rem}
  {4pt}
  {10pt}
  {}
  {}
  {\itshape}
  {:}
  {3pt}
  {}

\newtheoremstyle{texttheorem}
  {8pt}
  {8pt}
  {\itshape}
  {}
  {\bf}
  {. \hspace{5pt}}
  {3pt}
  {}

\theoremstyle{thm}

\newtheorem*{theorem*}{Theorem}
\newtheorem*{lemma*}{Lemma}
\newtheorem*{corollary*}{Corollary}
\newtheorem*{proposition*}{Proposition}
\newtheorem*{definition*}{Definition}

\newtheorem*{conjecture}{Conjecture}

\newtheorem{theorem}{Theorem}[subsection]
\newtheorem{thm-def}{Theorem/Definition}[theorem]

\newtheorem*{question*}{Question}

\numberwithin{equation}{subsection}

\theoremstyle{def}

\theoremstyle{rem}

\newtheorem*{remark}{Remark}

\usepackage{stmaryrd}
\date{}

\title{Twisted Supergravity and Koszul Duality: \newline A case study in AdS$_3$}
\author{Kevin Costello}
\author{Natalie M. Paquette}
\thanks{CALT-TH 2019-050}

\address{Perimeter Institue for Theoretical Physics}
\address{Walter Burke Institute for Theoretical Physics, California Institute of Technology}
\email{kcostello@perimeterinstitute.ca}
\email{nataliep@caltech.edu}

\begin{document}

\begin{abstract}
In this note, we study a simplified variant of the familiar holographic duality between supergravity on AdS$_3\times S^3\times T^4$ and the SCFT (on the moduli space of) the symmetric orbifold theory $Sym^N(T^4)$ as $N \rightarrow \infty$. This variant arises conjecturally from a twist proposed by the first author and Si Li. We recover a number of results concerning protected subsectors of the original duality working directly in the twisted bulk theory. Moreover, we identify the symmetry algebra arising in the $N\rightarrow \infty$ limit of the twisted gravitational theory. We emphasize the role of \emph{Koszul duality}---a ubiquitous mathematical notion to which we provide a friendly introduction---in field theory and string theory. After illustrating the appearance of Koszul duality in the ``toy'' example of holomorphic Chern-Simons theory, we describe how (a deformation of) Koszul duality relates bulk and boundary operators in our twisted setup, and explain how one can compute algebra OPEs diagrammatically using this notion. \\
Further details, results, and computations will appear in a companion paper.
\end{abstract}

\maketitle
\tableofcontents
\section{Introduction \& Conclusions}
High energy theory and mathematics engage in a rich exchange of ideas from which the AdS/CFT correspondence has so far largely abstained. Much of this cross-talk is mediated by the procedure of (topological) twisting, or restricting the physical observables to the cohomology of a nilpotent supercharge. This maps the physical system under consideration to a theory whose observables often coincide with a protected subset of the observables in the original system. Further, observables in the twisted theory often admit precise mathematical reformulations; equalities among these observables that are predicted from field or string theory dualities then produce numerous conjectural mathematical  equivalences. Running this program for the AdS/CFT correspondence has been stymied by the absence of a general version of twisting---traditionally an off-shell procedure via the path integral---for a gravitational theory \footnote{See, however, \cite{BonettiRastelli} for recent progress identifying the holographic dual of a certain chiral algebra; \cite{BPS} for a study of localization for the holographically renormalized action in gauged supergravity; \cite{DDG14, dWMV18, JM19} for recent progress defining localization in the supergravity path integral; and \cite{LT19} for another proposal of a twisted dual pair, with the twisted gravitational side defined using modified boundary conditions.}, even though a twist of a supersymmetric CFT is expected to have a holographic, gravitational counterpart.

Recent conjectures codified in \cite{CostelloLi} suggest that in natural top-down examples of the AdS/CFT correspondence arising from string theory, twisted supergravity coincides with the closed string field theory associated to familiar topological string theories. One natural choice of twisting supercharge produces Kodaira-Spencer theory and its B-model string theory parent. Further, in spite of the non-renormalizability of both the twisted theory and its parent, twisted supergravity can also be uniquely quantized in perturbation theory \cite{CostelloLi}.  Kodaira-Spencer theory governs a \emph{subclass of metric deformations} present in a full gravitational theory:  deformations of complex structures of the underlying geometry (plus contributions from superpartners). Taking this in conjunction with the fact that twisting on the CFT side produces mathematically well-defined objects like vertex algebras suggests that twisted holography (and hence aspects of the full holographic duality) may be amenable to rigorous mathematical proof. We also anticipate that this correspondence can engender new mathematical results at the interface between complex geometry and chiral algebras\footnote{Further connections with integrability, cohomological Hall algebras, and enumerative invariants are also sure to follow, and we hope to discuss some aspects of these connections in the context of $K3$ holography in future work.}. 

Such developments were recently explored in the context of twisted 4d $\mathcal{N}=4$ super-Yang-Mills and twisted IIB supergravity on AdS$_5 \times S^5$ in \cite{CostelloGaiotto}. OPEs and Witten diagrams were computed, respectively, in the twisted CFT and gravitational theories to leading order in  large $N$. An isomorphism between the $N \rightarrow \infty$ global symmetry algebras of the two theories was established\footnote{This symmetry algebra is an algebra of global symmetries on the CFT side, and is dual to an algebra of gauge symmetries in the gravitational theory. We will refer to it in this paper as the ``global symmetry algebra'', with this understood.}. The global symmetry algebra turns out to be powerful enough to fix almost all of the planar Witten diagrams/correlation functions among the twisted single-trace operators. Since the twisted theory is sensitive to loop effects one might well ask, given the success of working in the twisted theory at large-$N$ how effectively one can compute $1/N$ corrections. One might also wonder the following: since twisting 4d $\mc{N}=4$ produces an effectively two-dimensional theory (a gauged $\beta\gamma$ system), and since its holographic counterpart is also effectively lower-dimensional \footnote{One localizes from Euclidean AdS$_5 \times S^5$ to Euclidean AdS$_3 \times S^3$.}, what lessons can we hope to draw about the physical duality from these lower-dimensional systems?

To ameliorate the second concern, in this work we choose to study a simple top-down model of the AdS/CFT correspondence where the twisting does not reduce the effective dimensionality of the theory: supergravity on AdS$_3 \times S^3 \times T^4$ and the $D1-D5$ system on $T^4$. In other words, both the physical and twisted theories can be viewed as instances of an AdS$_3$/CFT$_2$ correspondence. Of course, gravity in AdS$_3$ is particularly simple (indeed, it is topological), which is a price we pay for this conceptual gain. The physical duality has also been well-studied, enabling us to draw on and compare to existing results in the literature. In our case, the six-dimensional twisted supergravity theory is given by Kodaira-Spencer theory\footnote{Kodaira-Spencer theory needs to be modified a little to work on the supermanifolds we consider. The modification has the feature that the solutions to the equations of motion describe complex supermanifolds of dimension $(3 \mid 4)$ which are fibred over $\C^{0 \mid 4}$ in ordinary Calabi-Yau $3$-folds.}  on a super-manifold we call the \emph{super-conifold}: this is the subvariety  $X^0 \subset \C^{4 \mid 4}$ subject to the equation
\begin{equation} 
	\delta^{\alpha \beta} x_\alpha x_\beta = F^{ab}\eta_a \eta_b \label{eqn:sc} 
\end{equation}
where $x_\alpha$ are bosonic, $\eta_a$ are fermionic, and the tensor $F^{ab}$ is non-degenerate. The singular locus $x_\alpha = 0$ is not present in our geometry.  

We stress the following remarkable fact, which we derive later in the paper: \emph{the backreaction always corrects the flat space computations in the twisted theory by only a finite number of terms!} \\
This truncation is a dramatic qualitative difference from the other instances of twisted holography studied so far. 

We compute the space of single-particle states for Kodaira-Spencer theory on the super-conifold, adopting the standard holographic techniques to our context.  The result matches \footnote{except for a small discrepancy with a handful of states of very small quantum numbers} with the supersymmetric states of the physical supergravity theory as computed by de Boer \cite{deBoer1} (see also \cite{DMW02}).  Hence, the states also match with the corresponding states in the chiral algebra of the $\Sym^N T^4$ in the large $N$ limit. 

Standard holographic techniques will make the states of Kodaira-Spencer theory into a chiral algebra, which we expect to match the $N \to \infty$ limit of the $\Sym^N T^4$ chiral algebra. One of our main results, following \cite{CostelloGaiotto}, is the computation of the global symmetry algebra of this chiral algebra.  We find that it is an infinite-dimensional super-Lie algera containing $\mf{psl}(1,1\mid 2)$ as the lowest-lying elements. 

An explicit description of the whole algebra will be given in section \ref{sec:states}.    We note now that inside this algebra is the infinite-dimensional super Lie algebra $\op{Vect}_0(X^0/\C^{0\mid 4})$ of holomorphic vector fields on the super-conifold $X^0$, which point along the fibres of the projection $X^0 \to \C^{0 \mid 4}$ and preserve a natural volume form on these fibres.  This leads to the following concrete conjecture:
\begin{conjecture}
	The Lie algebra $\op{Vect}_0(X^0/\C^{0\mid 4})$ acts on the large $N$ limit of the $\Sym^N T^4$ chiral algebra by symmetries which preserve the vacuum at $0$ and $\infty$.  
\end{conjecture}
In \cite{W92}, Witten found that the strikingly similar Lie algebra of volume-preserving vector fields on a $3$-dimensional cone arises as the ghost number $1$ states in the $c=1$ bosonic string. See \cite{RW07} for a further intriguing relation between the AdS$_3 \times S^3$ string and non-critical strings.  It would be very interesting to see whether our global symmetry algebra arises as part of the $Q$-cohomology of states of the AdS$_3 \times S^3 \times T^4$ string.  One can further speculate that the ground ring of the AdS$_3 \times S^3 \times T^4$ string is related to the algebra of functions on the super-conifold \eqref{eqn:sc}.  

Let us now introduce the other motivation of this paper: to explain and emphasize the role played in these twisted theories by a mathematical notion known as Koszul duality. We also hope that this will be a useful conceptual framework to keep in mind when studying physical dualities. Speaking loosely, Koszul duality is an operation that takes an algebra $\mc{A}$---the most standard incarnation assumes $\mc{A}$ is a differential graded (dg), associative algebra, though we will presently be interested in generalizations---and produces a new \emph{Koszul dual} algebra $\mc{A}^{!}$ such that $\mc{A}^{!!} = \mc{A}$. Among other properties, the categories of representations of $\mc{A}$ and $\mc{A}^{!}$ are equivalent\footnote{More precisely, there are equivalences between certain subcategories of the derived categories of modules. The subcategories require extra conditions to define and we will not need such technicalities in this paper.}. We will defer the construction of the dual algebra to the body of the paper but for now we notice the following. Let us say a preliminary speculation is: given an operator algebra in a (twisted) CFT $\mc{A}$, its Koszul dual algebra $\mc{A}^{!}$ is the algebra of local operators in the gravitational theory. We view these algebras as graded cochain complexes, where the ghost number is the grading and the BRST operator provides the differential of the complex. If $\mc{A}$ is concentrated in ghost number 0, then $\mc{A}^{!}$ will be supported in positive ghost number. In this way, we evade contradictions with the fact that there are no well-defined local operators in gravity (as was recently illustrated in \cite{C17}). Nonetheless, understanding the structures at ``unphysical''  ghost numbers turns out to be very powerful. Indeed, the derivation of Koszul dual algebras proceeds physically by considering two coupled systems (e.g., an open-closed string theory) and imposing anomaly cancellation on the couplings order by order in perturbation theory. 

The role of Koszul duality in holography was anticipated in \cite{CostelloLi} and was subsequently elucidated in some special systems in \cite{C17, IMZ18}. These models were cooked up with branes and anti-branes such that there was no net flux and therefore no gravitational backreaction from RR-fluxes. Of course, the latter is crucial in top-down models of AdS/CFT. In this paper, we will explain how gravitational backreaction produces a deformation of Koszul duality \footnote{Though we will not make the connection precise in this work, our deformation is expected to be related to the notion of \emph{curved} Koszul duality; see e.g. \cite{HM12}.}.  One of our main results is the following: \\

{\centering \emph{We compute the Koszul dual algebra of our twisted supergravity theory and obtain  the states in a twist of $Sym^N(T^4)$ as $N \rightarrow \infty$.} } \\

More precisely, the twisted gravitational theory maps to a subsector of states in the twist of the Fock space $\oplus_{N=1}^{\infty}Sym^N(T^4)$. We will comment on this further in section \ref{sec:deformed}. On the CFT side, the twisted states are accessible in $\bar{Q}_+$-cohomology familiar from the half-twist, though we emphasize that our analysis in this work is performed on the gravity side. 
 
This paper is meant to be viewed as an invitation to a more complete forthcoming study of the holographic correspondence between the dual pairs of the twisted AdS$_3 \times S^3 \times T^4$/$\Sym^N(T^4)$ theories \cite{CP2}. In the present work, we aim to describe and emphasize the aspects of the study that are unfamiliar to most physicists, namely twisted supergravity and Koszul duality, while concomitantly presenting some new results. We defer the detailed description of the twisted CFT, the matching of observables between the two sides, additional loop-level computations, and the version of the duality with $T^4$ replaced by $K3$ to \cite{CP2}.

The rest of the paper is organized as follows. In the remainder of this section, we will outline some of the many directions for future progress. In Section \ref{sec:twisted} we briefly review the setup of twisted holography discussed in \cite{CostelloLi} (and implemented recently in \cite{CostelloGaiotto}), and apply this procedure to AdS$_3 \times S^3 \times T^4$. We review the Kodaira-Spencer theory that results, and describe its KK-reduction on $T^4$. In Section \ref{sec:states} we enumerate the states in this twisted gravitational theory and identify the Lie algebra formed from the single-particle modes. In Section \ref{sec:koszul} we turn to a discussion of Koszul duality. We provide its mathematical definition, as well as its physical realization, for associative algebras. Motivated by these two realizations, we propose an extension of Koszul duality for chiral algebras from the physical point of view, and illustrate the computation of OPEs for the Koszul dual of an algebra using Feynman diagrams. Finally, in Section \ref{sec:deformed}, we apply this framework to our AdS$_3$ example and show how gravitational backreaction deforms the simple Koszul duality in a concrete, calculable way. We proceed to compute several OPEs of our gravitational mode algebra perturbatively using Koszul duality. An appendix contains some further mathematical facts about Koszul duality and its appearance when studying boundary conditions in TQFTs. 

\subsection{Outlook \& future directions}

We complete this introduction with an incomplete list of interesting extensions for the future.
\begin{enumerate}
	\item Though we study some simple one-loop diagrams in this paper, it would be interesting to push the computation of the OPEs to higher loop order. For the diagrams of relevance to the twisted theory, the Koszul duality approach to extracting OPE coefficients appears more straightforward than traditional Witten diagrams. Particularly exciting would be the possibility of using this mathematical technology to obtain or prove all-orders results for the OPEs. In the model we study, this is not implausible, as only a small number of diagrams can give non-zero contributions: see section \ref{sec:deformed}.  
	\item In this paper, we propose an extension of Koszul duality to chiral algebras (or vertex algebras). It should be fruitful to formalize this notion mathematically. In turn, one may be able to prove aspects of holography using homological algebraic methods once this extension is rigorously established.
\item Previous work has suggested that topological string theories should be connected to large-N dualities \cite{BOV04} and holography \cite{VafaBerkovits}. Among other results, there are connections between computations in the topological string to F-terms of the physical string theory in the hybrid formalism \cite{BOV04}. It would be very gratifying to derive our spacetime approach to twisted holography from the worldsheet perspective discussed in these works. The hybrid formalism has also been used to great effect in studying AdS$_3 \times S^3$ holography \cite{BVW99, EGG18}.   As mentioned above, it would be particularly interesting if the Lie algebra of ghost number $1$ states on the world-sheet could be related to our global symmetry algebra.
\item Further recent progress in deriving the full AdS$_3$/CFT$_2$ duality at the symmetric orbifold point in moduli space has been made in \cite{EGG19} and references therein. The bulk theory enjoys an exact worldsheet string theory description. As mentioned in the previous point, it would be very enlightening to perform a twisted analogue of their computations using the topological string worldsheet and compare the results to the spacetime analysis of this work.
\item It should also be interesting to study AdS$_3$ giant gravitons and other nonlocal excitations using the twisted framework. 
\item Black holes in AdS$_3$ are well known to be quotients of the global Euclidean AdS space (see e.g. \cite{MM01}). Quotients of (Euclidean) AdS$_3 \times S^3 \simeq SL(2, \mathbb{C})$ have counterparts in complex geometry, which should be easily generalized to superspace. We plan to investigate such configurations in the twisted framework in future work.
\item Observables in the B-model topological string provide natural twisted counterparts to standard holographic observables. It would be interesting to understand the ramifications of aspects of the higher-genus B-model, like the holomorphic anomaly, for twisted holography.
\item Similarly, it would be interesting to clarify the holographic interpretation of the A-model mirror of the system discussed here.
\item In many examples, twisting a supersymmetric theory produces a (non-SUSY) lower-dimensional theory (as seen in, say, the twist of a 4d $\mathcal{N}=2$ SCFT producing a non-supersymmetric 2d chiral algebra). It would be interesting to embed other lower-dimensional toy models of holography in the twisted framework, in order to apply lessons gleaned from these toy models to more interesting higher dimensional physical systems. Embedding matrix models and non-critical strings in topological string has already been explored \cite{DV, ADKMV} with fascinating applications to integrable hierarchies. We are currently revisiting 2d Yang-Mills theory at large-$N$ and its closed string dual from this point of view.
\item Perturbative quantum field theories can be studied mathematically using factorization algebras \cite{CG}. It would be interesting to understand the appropriate extension of Koszul duality for  factorization algebras/operator algebras in greater generality (i.e. sans SUSY or twists). Although these algebras are to be understood in Euclidean quantum field theories, it would also be very interesting to understand to what extent this machinery (or at least some intuition) can be applied to Lorentzian operator algebras or von Neumann algebras, which have enjoyed a recent holographic heyday. 
\end{enumerate}

\section{Acknowledgements}
N.P. wishes to thank J. Hilburn, T. Dimofte, and P. Yoo for enjoyable conversations about the many faces of Koszul duality, and M. Cheng, S. Kachru, C. Keller, G. Moore and especially N. Benjamin for earlier collaborations and many enlightening discussions on AdS$_3$. The work of N.P is supported by a Sherman Fairchild Postdoctoral Fellowship. This material is based upon work supported by the U.S. Department of Energy, Office of Science, Office of High Energy Physics, under Award Number DE-SC0011632. N.P. is grateful to the Perimeter Institute, UC Davis, and the University of Amsterdam for hospitality during various stages of this work.

K.C. would like to thank Davide Gaiotto and Si Li for many illuminating collaborations and conversations about twisted supergravity, holography and Koszul duality over the years. The research of K.C. is supported by the Krembil Foundation and the NSERC Discovery program.   Research at Perimeter Institute is supported by the Government of Canada through the Department of Innovation, Science and Economic Development and by the Province of Ontario through the Ministry of Research and Innovation.

\section{Twisted supergravity for AdS$_3 \times S^3 \times T^4$}\label{sec:twisted}
Our goal is to analyze twisted holography in the sense of \cite{CostelloLi, CostelloGaiotto} for the $\Sym^N T^4$ CFT and its gravity dual. To do this, we need to apply a supersymmetric localization not just to the CFT, but to the dual gravitational theory as well. A more extensive introduction to this procedure will appear in \cite{CP2}, as well as the corresponding analysis for the case when $T^4$ is replaced by $K3$.

Much of the logic of obtaining the twisted supergravity theory replicates that of the twisting for AdS$_5\times S^5$ described in \cite{CostelloGaiotto}, so we will be brief here and emphasize the important qualitative differences that arise in our example. Conjecturally, the twisted space-time gravitational theory is \cite{VafaBerkovits, CostelloLi} \emph{Kodaira-Spencer gravity}. On the other hand, the dual operators in the twisted SCFT form a chiral algebra in $\bar{Q}_+$-cohomology, familiar from physical and mathematical studies of the half-twisted $\sigma$-model and, respectively, the chiral de Rham complex \cite{Witten, Kapustin, MSV}.

Our procedure for obtaining the twisted dual pair goes as follows (and parallels Maldacena's original derivation of the AdS/CFT correspondence):
\begin{enumerate}
\item Place the B-model topological string on a Calabi-Yau 5-fold\footnote{The absence of higher-genus anomalies in BCOV theory on a 5-fold was argued for in \cite{CL12, CL3}, by coupling to the open-string sector.} $X$. For us $X=T^4\times {\mbb C^3}$. Recall that the spacetime field theory of the B-model is nothing but Kodaira-Spencer theory \cite{BCOV} and hence the latter is our twisted proxy for the supergravity limit of the string theory.
\item Add B-branes to this configuration. We consider $N_1$ topological D1-branes wrapping ${\mbb C} \subset {\mbb C^3}$ (times a collection of points in $T^4$) and $N_5$ topological D5-branes wrapping the same complex plane ${\mbb C}$, times $T^4$. Denote $N := N_1 N_5$.
\item Compute the backreaction in ${\mbb C^3}$ due to the fields sourced by the B-branes. The decoupling/near-horizon limits required in the standard derivation of the holographic duality are rendered trivial in the twisted theory.
\end{enumerate}

After briefly introducing the notion of twisted supergravity and reviewing Kodaira-Spencer theory, we will report on the backreacted geometry that results from this recipe\footnote{Of course, it would ultimately be preferable, but unfortunately much harder, to perform the twist directly in the AdS background.}.

\subsection{An abridged account of twisted SUGRA}
Twisting supersymmetric field theories is by now a matter of standard practice. However, the fact that supersymmetry is gauged in supergravity means that the usual twisting procedure does not carry through in a straightforward way. Here, we give a flavor of the procedure to twist supergravity introduced by  \cite{CostelloLi}. A longer exposition will also appear in \cite{CP2}. 
The upshot of the discussion is simply: \emph{Twisted supergravity is nothing but ordinary supergravity in an unusual background in which a bosonic ghost acquires a nontrivial vacuum expectation value.}

To figure out how to proceed, we will analyze the coupled system where type IIB on $T^4 \times \C^3$ is coupled to the open string sector, whose IR limit is the $\Sym^N T^4$ $\sigma$-model. One way the coupling between type IIB supergravity and the $\sigma$-model is manifested is that the BRST current of the $\sigma$-model depends on the fields of the bulk supergravity theory.  We are interested in the dependence of the $\sigma$-model BRST current on the ghosts of the supergravity theory.

Because we are gauging the local supersymmetries of the supergravity theory, there are both fermionic ghosts -- for diffeomorphisms -- and bosonic ghosts, for local supersymmetry transformations.  We will denote the bosonic ghosts by ${\mf c}^{\alpha}$, where $\alpha$ runs over a basis of the $16$ dimensional space of fermionic elements in the $(2,2)$ supersymmetry algebra in six dimensions.
Let us expand the BRST charge of the $D1$-$D5$ system as a function of the bulk supergravity fields and ghosts.  There is  a term
\begin{equation} 
Q_{BRST}({\mf c}, g, \dots ) = Q_{BRST}^0 + {\mf c}^{\alpha} Q_{\alpha} + \dots  
 \end{equation}
 where $Q_{BRST}^0$ is the original BRST operator, and $Q_{\alpha}$ are the currents corresponding to the supersymmetry transformations. In this expression, $\alpha$ now runs over the fermionic elements of the $1/2$-BPS subalgebra of the $(2,2)$ algebra which preserves the $D1-D5$ system. 
This expression makes it clear what we need to do to implement supersymmetric localization.  We should work in a supergravity background where the bosonic ghost ${\mf c}^{\alpha}$ has a VEV.  If we do this, then in the $D1-D5$ system, we have added some combination of the super-charges $Q_{\alpha}$ to the BRST charge $Q_{BRST}^0$, which is exactly what we need to do for supersymmetric localization. To obtain a holomorphic twist of the CFT as in a half-twisted $\sigma$-model one can choose $Q_{\alpha} = \bar{Q}_+$, one of the anti-holomorphic supercharges.

Giving some components of the bosonic ghost a VEV will also modify the supergravity action. The most convenient way to proceed is to use the BV formalism, which has been used to great effect in quantizing string field theory (among other contexts). This is a powerful way to perform BRST quantization that works even when the gauge algebra closes only on-shell and operates on an enlarged complex of fields at all ghost numbers, i.e. including not only fields and ghosts but also antifields and antighosts. We will explain this in more detail in \cite{CP2}. It turns out that the ghost VEV modifies the supergravity action by inducing, among other things, mass terms involving the fields and antifields. Integrating out the massive fields yields a gravitational theory with fewer degrees of freedom, that we call \emph{twisted supergravity}.  As advertised, the physical interpretation of twisted supergravity is perhaps more transparent than that of twisted field theory: it is simply ordinary supergravity in an unusual vacuum. As further argued in \cite{CostelloLi}, twisting supergravity in this sense, with respect to an appropriate supercharge, results in the Kodaira-Spencer theory.

Before we describe the field content and action, let us comment on the status of the conjecture that twisting supergravity produces the Kodaira-Spencer theory. Because supergravity is complicated, and the supersymmetry algebra only closes on-shell, it is rather difficult to calculate the fields and action of twisted supergravity.  Therefore, no \emph{explicit} verification of the conjecture of \cite{CostelloLi} has been performed.  However, extensive evidence for this conjecture was presented in \cite{CostelloLi, CostelloGaiotto}, so that one can view the conjecture as verified beyond reasonable doubt. We will assume this conjecture throughout our analysis. 

\subsection{Kodaira-Spencer theory}
In this note we will focus largely on the gravitational side of the correspondence, deferring details of the construction of the boundary chiral algebra and the holographic matching to \cite{CP2}. To this end, let us first describe the field content of Kodaira-Spencer gravity on a Calabi-Yau 5-fold\footnote{The supercharge which is relevant for our analysis is invariant under $SU(5) \subset \op{Spin}(10)$, and so can be defined on a Calabi-Yau $5$-fold $X$.} $X$. Subsequently, we will discuss the addition of branes and the backreaction. 

Kodaira-Spencer is known to be the closed string field theory of the B-model topological string \cite{BCOV}. The B-model governs complex structure deformations of the underlying geometry, and the Kodaira-Spencer theory has been shown to be the appropriate target space theory. The Kodaira-Spencer field content is described in terms of certain tensors called polyvector fields.  If $TX$ is the holomorphic tangent bundle, which is a complex vector bundle of rank $5$, the polyvector fields are
\begin{equation} 
	\PV^{i,j}(X) = \Omega^{0,j}(X, \wedge^i TX). 
\end{equation}
As usual, $\Omega^{(p, q)}(X, {\mc F})$ denotes the space of $(p,q)$-forms on $X$ valued in the bundle ${\mc F}$. A fundamental such field in Kodaira-Spencer theory is the Beltrami differential $\mu \in \Omega^{(0, 1)}(X, TX)$, which parameterizes deformations of complex structures $\bar{\partial} \rightarrow \bar{\partial} + \mu^i \partial_i$ (i.e. $\mu$ is our twisted proxy for a metric tensor field) and is a marginal operator in the B-model. It will play a fundamental role below.
In local coordinates $z_1,\dots,z_5$, a polyvector field can be expressed as 
\begin{equation} 
	\mu^{\br{i}_1 \dots \br{i}_k}_{j_1 \dots j_l}	\d \zbar_{\br{i}_1} \dots \d \zbar_{\br{i}_k} \partial_{z_{j_k}} \dots \partial_{z_{j_l}} 
\end{equation}
where  $\mu^{\br{i}_1 \dots \br{i}_k}_{j_1 \dots j_l}$ is anti-symmetric in both sets of indices.	 

It is convenient to introduce superfields, so that we have fermionic variables $\theta^i$, $\br{\theta}_{\br{i}}$. We are renaming $\d \zbar_{\br{i}}$ as $\br{\theta}_{\br{i}}$ and $\partial_{z_i}$ as $\theta^i$ to avoid confusion; we should bear in mind that $\theta^i$ transforms as a holomorphic vector, and $\br{\theta}_{\br{i}}$ as an anti-holomorphic covector.

Then, a polyvector field is a function $\mu(z_i,\zbar_{\br{i}},\theta^i, \br{\theta}_{\br{i}})$ of the space-time coordinates and these new fermionic variables.   We obtain the coefficients $\mu^{\br{i}_1 \dots \br{i}_k}_{j_1 \dots j_l}$ by expanding in series in $\theta^i$, $\br{\theta}_{\br{i}}$.  

We can identify $\PV^{l,k}(X)$ with $\Omega^{5-l,k}(X)$, by sending
\begin{equation} 
	\mu^{\br{i}_1 \dots \br{i}_k}_{j_1 \dots j_l} \mapsto \eps_{j_1 \dots j_l r_1 \dots r_{5-l}} \mu^{r_1 \dots r_{5-l} \br{i}_1 \dots \br{i}_l}.	
\end{equation}
If we view a differential form as a function of fermionic variables $\theta_i$, $\br{\theta}_{\br{i}}$ then the map from polyvector fields to differential forms sends
\begin{equation}\label{eq:FFT} 
	\mu(z,\zbar,\theta^i, \br{\theta}_{\br{i}}) \mapsto \mu(z,\zbar, \partial_{\theta_i}, \br{\theta}_{\br{i}}) \theta_1 \dots \theta_5. 
\end{equation}
This is a kind of fermionic Fourier transform.

The operators $\partial, \dbar$ on $\Omega^{\ast,\ast}(X)$ transfer to operators $\PV^{l,k}(X)$ with the same name. These operators are given by
\begin{equation}
	\begin{split}
	\dbar = \sum \theta_{\br{i}} \partial_{\zbar_{\br{i}}} \\
	\partial = \sum \partial_{\theta^i} \partial_{z_i}.
	\end{split}
\end{equation} 

The field content of Kodaira-Spencer theory is a superfield
	\begin{equation} 
		\mu(z,\zbar,\theta^i, \br{\theta}_{\br{i}}) \in \oplus_{i,j}  \PV^{i,j}(X) 
	\end{equation}
satisfying the constraint equations
	\begin{equation}\label{eq:constraint}
		\begin{split}
			\partial \mu &= 0 \\
			\partial_{\theta^1} \dots \partial_{\theta^5} \mu &= 0,
		\end{split}
	\end{equation}
	(cf. also the discussion above equation \ref{eq:nonprop} to motivate the second constraint). In particular, if the Beltrami differential is annihilated by $\bar{\partial}$, which is the equation of motion in  Kodaira-Spencer theory in the linear approximation, then the complex structure deformation is integrable. This recovers the familiar geometric fact that $\bar{\partial}$-closed Beltrami differentials in the kernel of $\partial$ correspond to deformations of $X$ as a complex manifold with a holomorphic volume form. 
	The Lagrangian of the theory is
	\begin{equation} 
		\tfrac{1}{2} \int \mu \dbar \partial^{-1} \mu \d^5 z \d^5 \zbar \d^5 \theta \d^5 \br{\theta} + \frac{1}{6} \int \mu^3 \d^5 z \d^5 \zbar \d^5 \theta \d^5 \br{\theta}.   
	\end{equation}
In \cite{CP2} we will show that one can obtain this peculiar kinetic term by starting from a term in the full supergravity BV-action functional involving the metric and the antifield to the gravitino, endowing the superghost with a VEV, and integrating out the resulting massive fields. In spite of this kinetic term, the theory does have a well-defined propagator in perturbation theory \cite{BCOV}; see also \cite{CL12}. We will also defer a proposal for how certain components of physical IIB supergravity fields match to Kodaira-Spencer fields to  \cite{CP2}. For now, we will simply note to orient the reader that, to linear order, there is a conjectural correspondence between a polyvector field and certain components of the physical K{\"a}hler metric
	\begin{equation}
	g^{\br{i} \br{j}} \mapsto \delta^{k \br{i}}\mu_k^{\br{j}}.
	\end{equation}
	
\subsubsection{Ramond-Ramond fields in Kodaira-Spencer theory}
For the purposes of this paper, it is not necessary to understand the exact supergravity origin of every field of Kodaira-Spencer theory. We will, however, need to know the fields sourced by branes in Kodaira-Spencer theory.  The field sourced by a $D_{2k-1}$-brane is a polyvector field of type $(k, 5-k-1)$.  This tells us that such polyvector fields arise from the field strength of the Ramond-Ramond $(9-2k)$-form. 

If the brane wraps the complex submanifold $\C^k \subset \C^5$, with coordinates $z_1,\dots, z_k$, then the field it sources is is a polyvector field $\mu_{i_1 \dots i_k}^{\br{j}_1 \dots \br{j}_{5-k-1}}$ such that the differential form
\begin{equation} 
	F = \eps^{i_1 \dots i_k j_1 \dots j_{5-k}} \mu_{i_1 \dots i_k}^{\br{j}_1 \dots \br{j}_{5-k-1}} \d z_{j_1} \dots \d z_{j_{5-k}} \d \zbar_{\br{j}_1} \dots \d \zbar_{\br{j}_{5-k-1}} 
\end{equation}
satisfies
\begin{equation} 
	\dbar F = \delta_{D_{2k-1}}  
\end{equation}
where $\delta_{D_{2k-1}}$ is the $\delta$-function on the submanifold supporting the brane. It is often convenient to impose the gauge condition
\begin{equation} 
	\dbar^\ast F = 0. 
\end{equation}
with $\dbar^{\ast}$ the codifferential of $\dbar$, so that the gauge condition implies a choice of metric on our manifold. These two constraints fix $F$ uniquely, as long as $F$ tends to zero at $\infty$.
\subsection{Reducing Kodaira-Spencer theory on $T^4$}
Now we have all the necessary ingredients in hand to perform the KK reduction of Kodaira-Spencer theory on $T^4$ and compute the backreaction. In this section we will analyze the result of the dimensional reduction of Kodaira-Spencer theory to $6$ dimensions on $T^4$. When we do this, we will only retain those polyvector fields that are harmonic along the complex surface. This is the twisted version of retaining only the massless field content after compactifying on $T^4$.

A polyvector field on $T^4 \times \C^3$ can be written as a tensor product of one on $T^4$ with one on $\C^3$.  Polyvector fields on $T^4$ are the same as differential forms, because the holomorphic symplectic form on $T^4$ identifies the tangent and cotangent bundles. Therefore, the harmonic polyvector fields are given simply by the cohomology of $T^4$.  Furthermore, polyvector fields on $T^4$ which are harmonic are automatically in the kernel of the operator $\partial$, by standard Hodge theory arguments.   

We conclude that fields of the dimensionally reduced theory are just polyvector fields on $\C^3$ which carry an extra index taking values in the cohomology of $T^4$. It is convenient to express this extra index in terms of a superfield.  The cohomology ring of $T^4$ is generated by four anti-commuting variables $\eta_a$. (This is just the standard fact that all cohomology classes of a torus are cup products of 1-dimensional classes, which follows from viewing the torus as a product of circles.) Therefore, we can view the fields of the dimensionally reduced theory as being polyvector fields on $\C^3$, which also depend on these four fermionic variables $\eta_a$.  

The polyvector fields on $\C^3$ must satisfy the constraint 
\begin{equation} 
	\partial_{z_{i_1}} \mu_{i_1 \dots i_l; a }^{\br{j}_1 \dots \br{j}_k} = 0. 
\end{equation}
As before, let us write the polyvector fields in terms of a superspace by introducing odd variables $\theta^i$, $\br{\theta}_{\br{j}}$.  Then we can write the field content as a collection of superfields
	\begin{equation} 
		\mu(z,\zbar,\theta^i, \br{\theta}_{\br{i}},\eta_a) \in \oplus_{i,j}  \PV^{i,j}(\C^3) \otimes \C[\eta_a]. 
	\end{equation}

The superfield satisfies the equation
	\begin{equation}	
			\partial \mu = 0		
	\end{equation}
where, in the superspace formulation,
\begin{align} 
	\dbar &:= \br{\theta}_{\br{j}} \partial_{\zbar_{\br{j}}} \\
	\partial &:= \partial_{\theta^i} \partial_{z_i}.  
\end{align}
	The Lagrangian is
	\begin{equation} 
		\tfrac{1}{2} \int_{\C^{3 \mid 10}}  \mu \dbar \partial^{-1} \mu \d^3 z \d^3 \zbar \d^3 \theta \d^3 \br{\theta} \d^4 \eta + \frac{1}{6} \int_{\C^{3 \mid 10}} \mu^3 \d^3 z \d^3 \zbar \d^3 \theta \d^3 \br{\theta} \d^4 \eta.   
	\end{equation}
	
Before we turn to the computation of the backreaction, we will simplify the field content somewhat, following \cite{CostelloGaiotto}.  We note that the coefficient of $\theta^1 \theta^2 \theta^3$ does not appear in the kinetic term in the action.  This field does not propagate, so we can (and will) impose the additional constraint
\begin{equation}\label{eq:nonprop} 
	\partial_{\theta^1} \partial_{\theta^2} \partial_{\theta^3} \mu_a(z,\zbar,\theta,\br{\theta}) = 0. 
\end{equation}

Next, let us expand the superfield $\mu$ only in the $\theta^i$ variables:
\begin{equation} 
	\mu = \mu(z,\zbar,\br{\theta},\eta) + \mu_{i}(z,\zbar,\br{\theta},\eta) \theta^i + \dots 
\end{equation}
We note that the constraint $\partial \mu_{ij} = 0$ implies that there is some super-field
\begin{equation} 
	\what{\mu}_{ijk}(z,\zbar,\br{\theta},\eta) = 	\alpha(z,\zbar,\br{\theta},\eta) \eps_{ijk}   
\end{equation}
so that $\partial_{z_i} \what{\mu}_{ijk} = \mu_{jk}$.

It is convenient to rephrase the theory in terms of the field $\alpha(z,\zbar,\br{\theta},\eta)$, which has no holomorphic index. We will also change notation and let $\gamma(z,\zbar,\br{\theta},\eta)$ be the term with no $\theta^i$ dependence in the superfield $\mu(z,\zbar,\theta,\br{\theta},\eta)$.  In terms of these fields, the Lagrangian becomes
\begin{multline} 
	\tfrac{1}{2}\int  \eps^{ijk} \dbar \mu_{i} (\partial^{-1}  \mu)_{jk} \d^3 z \d^3 \zbar \d^3 \br{\theta} \d^4 \eta   + \int  \alpha \dbar \gamma \d^3 z \d^3 \zbar \d^3 \br{\theta} \d^4 \eta 
	\\
	+ \tfrac{1}{6} \int  \eps_{ijk} \mu_{i}\mu_{i} \mu_{c,i} \d^3 z \d^3 \br{z} \d^3 \br{\theta} + \int  \alpha \mu_i \partial_{z_i}  \gamma \d^3 z \d^3 \zbar \d^3 \br{\theta} \d^4 \eta.
\end{multline} 

We note that this Lagrangian has an extra $SL_2(\C)$ symmetry, under which the fields $\alpha$, $\gamma$ form a doublet. In \cite{CostelloGaiotto} it was conjectured that this is $S$-duality. This symmetry should be broken to $SL_2(\Z)$ by asking that the fields $\gamma$, $\alpha$ are functions from the supermanifold $\C^{3 \mid 3+4}$ to $\C^\times$ instead of to $\C$.  This means that locally these fields are defined up to the addition of an integer; thus, they are allowed to have branch-cuts with an integer shift as we cross the branch.

Just as when we twist a field theory, when we twist a supergravity theory the ghost number of the twisted theory  is a mixture of the ghost number and a $U(1)_R$-charge of the original physical theory. To define a consistent ghost number, one can choose any $U(1)_R$ in the physical theory under which the supercharge has weight $1$.  In general, there are many ways to do this.  It is convenient for us to make the following assignments of ghost number.
\begin{enumerate} 
	\item The variables $\eta_a$ are fermionic but have ghost number zero.
	\item The anti-commuting variables $\br{\theta}_i$ have ghost number $1$.
	\item The fields $\alpha$, $\gamma$ have ghost number $-1$, and are fermionic. This means that if we expand each such field in the fermionic variables $\eta_a$, $\br{\theta}_i$, then the coefficient of $\eta^I \br{\eta}^J$ (for some multi-indices $I$, $J$) is a fermionic field if $\abs{I} + \abs{J}$ is even, and is bosonic otherwise; and this field has ghost number $\abs{J} - 1$.  
\end{enumerate}

Note that this differs from the ghost number assignment in \cite{BCOV}.  The choice made in \cite{BCOV} differs from ours by the charge of the Cartan of $SL_2(\C)$, the complexified, perturbative $S$-duality group. Since the choice made in \cite{BCOV} breaks $S$-duality, the ghost number convention here works better for many purposes.   

\subsection{Backreaction, and a truncation}
So far, we have seen that the dimensional reduction of type IIB supergravity on $T^4$, when twisted, is equivalent to Kodaira-Spencer theory on $\C^3$, but where all fields have an extra index living in the cohomology of $T^4$.  At last, we are ready to backreact the $D1$ and $D5$ brane system, following the analysis of \cite{CostelloGaiotto}. We will find that this example boasts some simplifying features that were not present in the higher-dimensional example studied in \cite{CostelloGaiotto}.

We have expressed the fields in terms of the cohomology of $T^4$, instead of the polyvector fields of $T^4$. This can be interpreted physically as performing mirror symmetry (i.e. T-dualities) on the $T^4$ which turns the $B$-model on $T^4$ into the $A$-model.  The $D1$ and $D5$ branes both become $D3$ branes in this duality frame, wrapping $\C \subset \C^3$ and a Lagrangian inside $T^4$.  The charges of these Lagrangian branes are
\begin{align} 
	F_1 &= \sum F_1^{ab} \eta_a \eta_b \\
	F_5 &= \sum F_5^{ab} \eta_a \eta_b  .
\end{align}
The total charge is $F= F_1 + F_5$, and the inner product
\begin{equation} 
	\ip{F,F} = F_1^{ab} F_5^{cd} \eps_{abcd} 
\end{equation}
is $N:=N_1 N_5$. 

In the perturbative analysis of supergravity, the only structure we are using on $T^4$ is the cohomology $H^\ast(T^4,\C)$ with its cup product and Poincar\'e pairing.  The fact that we only need cohomology with complex coefficients appears is an artifact of perturbation theory: non-perturbative effects should rely on integrality constraints for the fields.  

We note that the cohomology ring of $T^4$, with this structure, has an $SL(4,\C) = \op{Spin}(6,\C)$ symmetry. Under this symmetry, $H^1(T^4,\C)$ is the spin representation of $\op{Spin}(6,\C)$ and $H^2(T^4,\C)$ is the vector representation. The charge vector $F$ is an element of the vector representation of non-zero norm, as long as $N_1 N_5 \neq 0$.  This breaks the $\op{Spin}(6,\C)$ symmetry to $\op{Spin}(5,\C)$, which will be a useful symmetry of the backreacted theory.   

Up to a change of basis on the space of fields, the only invariant of the charge $F \in H^2(T^4,\C)$ is the length $\ip{F,F} = N_1 N_5$.   Indeed, two charges with the same length are related by a $\op{Spin}(6,\C)$ transformation. This explains why, from the point of view of perturbative supergravity, all quantities will only depend on $N_1 N_5$. This also enables us to move freely between convenient frames with fixed $N$, such as the frame with a single $D5$-brane and $N$ D1-branes. 

Following \cite{CostelloGaiotto}, the backreaction is given by introducing a supergravity field 
\begin{equation} 
	\mu(\eta_a) \in \PV^{1,1}(\C^3)\otimes \C[\eta_a] \iso \Omega^{2,1}(\C^3)\otimes \C[\eta_a] 
\end{equation}
which satisfies
\begin{equation} 
	\begin{split}
		\dbar \mu &= F^{ab} \eta_a \eta_b \delta_{\C}\\
		\partial \mu &= 0.
	\end{split}
\end{equation}
In addition, we impose the standard gauge constraint that $\dbar^\ast \mu = 0$. 

The field $\mu$ is a Beltrami differential whose equation of motion is now inhomogeneous because of the presence of the brane source. This means the complex structure deformation is integrable away from the support of the brane but not integrable on $\C$. 

If we choose coordinates $z,w_1,w_2$ on $\C^3$, where the branes wrap the $z$-plane, the unique solution to these equations is
\begin{equation} 
	\mu= F^{ab} \eta_a \eta_b  \frac{\eps^{ij} \wbar_i \d \wbar_j   }{ (w_1 \br{w}_1 + w_2 \br{w}_2)^2  } \partial_z. \label{eqn:beltrami} 
\end{equation}
Viewing $\mu$ as a $(2,1)$ form, instead of a Beltrami differential, we have

\begin{equation} 
	\mu = F^{ab} \eta_a \eta_b \frac{\eps^{ij} \wbar_i \d \wbar_j   }{ (w_1 \br{w}_1 + w_2 \br{w}_2)^2  } \d w_1 \d w_2.
\end{equation}

In \cite{CostelloGaiotto}, the backreaction for ordinary Kodaira-Spencer theory, without the index in $H^\ast(T^4)$, was studied. There, it was found that the Beltrami differential in Eqn. \ref{eqn:beltrami} deforms the complex structure from $\C^3 \setminus \C$ to $SL_2(\C)$. The reason is that the holomorphic functions in the deformed complex structure are given by $w_1,w_2$ and
\begin{align} \label{eq:deformation}
	u_1 &= 	w_1 z - N  \frac{\wbar_2}{\norm{w}^2} \\
u_2 &= 	w_2 z + N  \frac{\wbar_1}{\norm{w}^2}.   
\end{align}
and these functions satisfy
\begin{equation} 
	u_2 w_1 - u_1 w_2  = N.  
\end{equation}

What is the analog of this statement in the present set-up, where we have compactified on $T^4$? In that case, everything depends on the fermionic variables $\eta_a \in H^1(T^4)$.   We will therefore be deforming the complex super-manifold $(\C^3 \setminus \C) \times \C^{0\mid 4}$.

The backreaction of the branes gives us a deformation of this space, given by the Beltrami differential in equation \ref{eqn:beltrami}. A function $\Phi(\eta_a, w_i, \br{w}_i, z, \br{z})$ is holomorphic for the deformed complex structure if  
\begin{equation} 
	\d \wbar_i \frac{\partial \Phi}{\partial \wbar_i} +	 F^{ab} \eta_a \eta_b \frac{\eps^{ij} \wbar_i \d \wbar_j   }{ (w_1 \br{w}_1 + w_2 \br{w}_2)^2  }\frac{\partial \Phi}{\partial z} = 0. 
\end{equation}

The following functions are holomorphic for the deformed complex structure:
\begin{align} \label{eq:deformation_k3}
	u_1 &= 	w_1 z - F^{ab} \eta_a \eta_b  \frac{\wbar_2}{\norm{w}^2} \\
	u_2 &= 	w_2 z +  F^{ab} \eta_a \eta_b  \frac{\wbar_1}{\norm{w}^2}.   
\end{align}
These satisfy 
\begin{equation} 
	u_2 w_1 - u_1 w_2 = F^{ab} \eta_a \eta_b. \label{eqn:cone}
\end{equation}
This equation defines a quadratic cone inside $\C^{4 \mid 4}$, associated to a non-degenerate graded antisymmetric quadratic form.  The symmetry group of this quadratic cone is $\op{OSp}(4 \mid 4)$.   

In the backreacted geometry, we exclude the locus where $w_i = 0$.  In order to match with the conical geometry \eqref{eqn:cone}, we must do the same thing.    Note that this locus is not invariant under the group $\op{OSp}(4 \mid 4)$, because the action of $\op{Spin}(4)$ mixes the $u_i$ and the $w_i$.   The orbit containing the locus $w_i\neq 0$ is the subset of the cone where the $u_i, w_j$ are not all zero. 

This larger open subset is the backreaction of the geometry including $z = \infty$.  Indeed, if we include $z = \infty$, $\C^3$ gets completed to the resolved conifold $\Oo(-1) \oplus \Oo(-1)\to \CP^1$.  This is the resolution of the singularity $\eps^{ij} u_i w_j = 0$.  Removing the support of the brane gives the open subset of the singular variety where we remove the tip of the cone.   After including the fermionic variables, the backreaction in the resolved conifold geometry takes us to the geometry given by \eqref{eqn:cone}, with the singular locus where all $u_i,w_j$ are zero removed.   

We will refer to the super-algebraic variety given by \eqref{eqn:cone} as the \emph{super-conifold}, and denote it by $X$.  The open subset where $u_i$, $w_j$ are not all zero will be denoted by $X^0$.

The holomorphic volume form, in the coordinates $z,w_1,w_2$, is unchanged when we deform: it is $\d z \d w_1 \d w_2$. This is because the Beltrami differential is divergence-free. In the coordinates $u_i,w_j$ the holomorphic volume form is
\begin{equation} 
	\Omega =   w_1^{-1}\d u_1 \d w_1 \d w_2. \label{eqn:conifold_volume} 
\end{equation}
We can also write this volume form as
\begin{equation} 
	\Omega = \op{Res} \frac{\d u_1 \d u_2 \d w_1 \d w_2 }{ \eps^{ij}u_i w_j - F^{ab} \eta_a \eta_b  }. 
\end{equation}

It is important to note that this form is only defined along the fibres of the projection map $X^0 \to \C^{0 \mid 4}$, so that it does not involve any $\d \eta_a$.  It is a volume form on these fibres. We will sometimes loosely refer to $\Omega$ as a holomorphic volume form, with it being understood that it is only a volume form in the relative sense.

We pause to note a remarkable feature of this situation: since $F^3 = 0$ in the ring $H^\ast(T^4) = \C[\eta_a] $, \emph{we will only ever find finite-order corrections to the flat space result!}\footnote{We can also see this from the worldsheet point of view. In the topological $A$-model on $T^4$, any worldsheet amplitude with the insertion of more than $3$ copies of the flux $F \in H^2(T^4)$ vanishes.}   The flat space result is $F = 0$, because the complement of $w_i = 0$ in $\C^{3}$ is the same as the complement of $w_i = 0$ in the cone $\eps^{ij} u_i w_j = 0$.  

\subsection{The string coupling and $N$} 
The string coupling $\lambda$ enters the Lagrangian by an overall factor of $\lambda^{-2}$. The Lagrangian depends on the fermionic variables $\eta_a$ via $\d \eta_1 \dots \d \eta_4$. This  scales by $\lambda^2$ under the transformation $\eta \mapsto \lambda^{-1/2} \eta$.   We conclude that on flat space, we can absorb the string coupling constant into a redefinition of the fermionic variables.

In the backreacted geometry this is no longer quite true.  Rescaling the fermionic variables rescales flux $F$ by $\lambda^{-1}$ and hence $N = N_1 N_5$ by $\lambda^{-2}$. Therefore, the theory with parameters $(\lambda, N)$ is equivalent to the theory with parameters $(\lambda N^{-1/2}, 1)$. This is, of course, consistent with the usual large-$N$ identification of the physical string coupling $g_s$ with the CFT central charge $\sim N$ (i.e. $g_s \sim {1 \over \sqrt{N}}$) in the physical duality, so that higher genus corrections in the bulk theory are equivalent to $1/N$ corrections in the CFT.

\section{Supergravity states }\label{sec:states}

Now that we have the full backreacted solution in hand, we would like to study the resulting twisted supergravity states therein.

First, we will enumerate the states in our twisted supergravity theory.  Scattering processes among these states as computed by Witten diagrams, in principle, lead to the determination of the OPE coefficients. This is the more traditional route towards determining the algebra from gravity calculations. In Section \ref{sec:koszul} we will present a different approach to this result, which also conceptually incorporates loop corrections in $1/N$ order-by-order, based on the mathematical notion of Koszul duality.

\subsection{Enumerating supergravity states}
In this section we will show how to enumerate states of twisted supergravity compactifed on $T^4$.  For the physical supergravity theory, the calculation was presented in \cite{deBoer1} and, pleasingly, the states we find from the twisted perspective recover the supersymmetric sector of the single-particle states found in \cite{deBoer1}. 

In \cite{CostelloGaiotto}, Witten diagrams were studied from the holomorphic point of view for Kodaira-Spencer theory on the deformed conifold $SL_2(\C)$. The idea is that we should compactify the deformed conifold $SL_2(\C)$, with coordinates $u_i,w_j$ satisfying $\eps^{ij} u_i w_j = N$, to the projective variety in $\mbb{CP}^4$ with homogeneous coordinates $U_i,W_j, Z$ and relation $\eps_{ij} U^i W^j = N Z^2$.  The boundary of this space is the subvariety of $\mbb{CP}^2$ with homogeneous coordinates $U^i, W^j$ satisfying $\eps_{ij} U^i W^j = 0$. This is a copy of $\mbb{CP}^1 \times \mbb{CP}^1$. We can translate this into more familiar AdS/CFT language.  The deformed conifold $SL_2(\C)$ is AdS$_3 \times S^3$ (for Euclidean AdS).   The boundary of $SL_2(\C)$ is $\mbb{CP}^1 \times \mbb{CP}^1$. One of these $\mbb{CP}^1$'s is the boundary of AdS$_3$; the other $\mbb{CP}^1$ is the base of the Hopf fibration of the $S^3$.  

We give coordinates $z$ to the $\mbb{CP}^1$ which is the boundary of AdS$_3$, and $w$ to the other $\mbb{CP}^1$.  We let $n$ be a coordinate in the normal direction to the boundary of $\mbb{CP}^1 \times \mbb{CP}^1$ in the compactification of $SL_2(\C)$.  This coordinate $n$ has a first-order pole at $w = \infty$ and at $z = \infty$.  In the setting of \cite{CostelloGaiotto}, it was shown that this coordinate system was not holomorphic: rather we had a Beltrami differential
\begin{equation} 
	N n^2  \d \wbar \frac{1}{(1 + \abs{w}^2)^2}   \partial_z \label{eqn:beltrami_N}
 \end{equation}
deforming the complex structure. 
The complement of $n = 0$ in this complex manifold is the deformed conifold. There are holomorphic functions with poles at $n = 0$ given by $w_1 = 1/n, w_2 = w/n$ and
\begin{align} 
u_1 &= z/n -  N n    \frac{\wbar}{(1 + \abs{w}^2)}\\ 
u_2 &= wz/n + N n    \frac{1}{(1 + \abs{w}^2)}.\label{eqn_boundary_coords} 
 \end{align}
We have $u_2 w_1 - u_1w_2 = N$.

In these coordinates, the holomorphic volume form is, up to a factor independent of $N$, 
  \begin{align}
\Omega &= \frac{1}{w_1} \d u_1 \d w_1 \d w_2 \\
	  &= - n^{-3} \d n \d w \d z  + N n^{-1} \d n \d w \d \wbar \frac{1}{(1 + \abs{w}^2)^2}.\label{eqn:volume_form} 
  \end{align}

In our setting, things are almost exactly the same.  The super-conifold $X^0$ has a completion $\br{X^0}$ to a super-projective variety inside $\CP^{4} \times \C^{0 \mid 4}$\footnote{Note that this is \emph{not} the same as $\CP^{4 \mid 4}$.  We choose this particular completion to match better with the approach of \cite{CostelloGaiotto}.}. We give $\CP^{4}$ homogeneous bosonic coordinates $U_i, W_j, Z$. The completion $\br{X^0}$ is defined by 
\begin{equation} 
	\eps^{ij} U_i W_j = F^{ab} \eta_a \eta_b Z^2. 
\end{equation}
The boundary is the locus $Z = 0$, given by the variety in $\CP^3 \times \C^{0 \mid 4}$ defined by $\eps^{ij} U_i W_j = 0$.  This variety is $\CP^1 \times \CP^1 \times \C^{0 \mid 4}$.  
Now we can simply apply the analysis of \cite{CostelloGaiotto} of the complex structure on a neighbourhood of the boundary, where we everywhere replace $N$ by $F^{ab} \eta_a \eta_b$. 
We have bosonic coordinates $w,z,n$ and their complex conjugates, as well as fermionic coordinates $\eta_a$. The coordinate $n$ has a pole at $w = \infty$, $z = \infty$, and the complex structure is deformed by the Beltrami differential
\begin{equation} 
	F^{ab} \eta_a \eta_b  n^2  \d \wbar \frac{1}{(1 + \abs{w}^2)^2}   \partial_z \label{eqn:beltrami_F}.
 \end{equation}
The holomorphic volume form is obtained by simply replacing the appearance of $N$ in \eqref{eqn:volume_form} by $F^{ab} \eta_a \eta_b$.  

To describe supergravity states in the holomorphic language, we should specify the ``vacuum'' boundary conditions for our supergravity fields $\mu,\alpha,\gamma$.  Then, a boundary state is a field configuration which satisfies the equation of motion, and also satisfies the vacuum boundary conditions except at $z = 0$.   Enumerating such boundary states will give us the supergravity index, and scattering such states (using Witten diagrams) will give us the $2$ and $3$-point functions.
 This was done successfully in the case of the chiral algebra associated to 4d $\mc{N}=4$ gauge theory in \cite{CostelloGaiotto}, and we mimic that approach here. The fields $\alpha,\gamma$ are in the Dolbeault complex of our coordinate patch, and also depend on the fermionic variables $\eta_a$. Our vacuum boundary condition is that $\alpha,\gamma$ are divisible by $n$.

The field $\mu$ can be viewed as a $(2,\ast)$ form on the space with coordinates $n,z,w$, which also depends on the fermionic variables $\eta_a$.  The boundary condition chosen in \cite{CostelloGaiotto} for $\mu$ is that it is a $(2,\ast)$ form with logarithmic singularities along the boundary $n = 0$. This means it can be expressed as a wedge product of $\d \log n$, $\d w, \d z$, $\d \br{n}$, $\d \wbar$, $\d \zbar$ with a coefficient which is regular at $n = 0$.  
The boundary states were classified (up to gauge equivalence) in \cite{CostelloGaiotto}, and the same argument applies here.  Let us recall the argument for the fields $\alpha,\gamma$. The vacuum boundary condition states that $\alpha$ (or $\gamma$) is divisible by $n$.  We can modify this, just at the point $z = 0$ in the boundary, by taking as an ansatz  
\begin{equation} 
\alpha = n^{-k}w^l \delta^{(r)}_{z = 0} \label{configuration_initial}. 
 \end{equation}
In this expression, $k \ge 0$, and $l \le k$ to ensure that there are no poles at $w = \infty$. Also $\delta^{(r)}_{z = 0}$ indicates the $r$th $z$-derivatives of the delta function. The R-symmetry group $SU(2)_R$ acts by change of coordinates on the $ w $ plane. This expression transforms in the $ SU (2)_R$ representation of dimension $ k +1 $. 

This ansatz, however, does not satisfy the equations of motion, because of the Beltrami differential \eqref{eqn:beltrami_F}.  One has to apply corrections which, in the setting of \cite{CostelloGaiotto} depend on $N$, and here depend on $F$.  

For the field $\mu$, the possible gravitational states were also classified in \cite{CostelloGaiotto}.  It was found that, up to gauge equivalence, every solution to the equation of motion which satisfies the boundary condition except at $z = 0$ is given by expressions of the form
\begin{equation}
	\begin{split}
		& w^l \d \log n \d z  n^{-k}   \delta_{z =0} + \dots \\
		& w^l \d \log n \d w  n^{-k}    \delta_{z = 0} - \frac{1}{k}  \d w \d z n^{-k} \partial_z \delta_{z = 0 } + \dots 
	\end{split} \label{eqn_beltrami_state} 
	\end{equation}
	In the first expression, $l \le k$ and $k > 0$. The field configuration transforms in a representation of spin $k/2$ of $SU(2)_R$. In the second expression, $l \le k-2$ and $k \ge 2$. The field transforms in a respresentation of spin $k/2  - 1$ of $SU(2)_R$.  The first field is of conformal weight $k/2$, and the second is of conformal weight $k/2+1$. In each case, we only find a supergravity state for $k > 0$, as otherwise the form has logarithmic poles even at $z = 0$ and so does not violate the boundary condition. 

	In each equation, the ellipses indicate terms that we need to add to ensure the equations of motion hold, because of the Beltrami differential \eqref{eqn:beltrami_N}, \eqref{eqn:beltrami_F}.  These extra terms were not determined explicitly in \cite{CostelloGaiotto} although they were shown to exist by a cohomological argument. Again, we will present the full solutions below.

Now that we understand how to modify the boundary conditions, let us enumerate all the single-particle gravitational states. We will write down only those states that are Virasoro primaries and $SU(2)_R$ heighest weights.  Each state will come with a multiplicity of $H^\ast(T^4)$, possibly with a parity shift, denoted $\Pi$. 

\begin{tabular}{c|c|c|c}\label{tbl:states}
	State & Multiplicity & $(J_0^3,L_0)$ & Range \\
	$\alpha \sim \delta_{z=0} n^{1-k}+ \dots $ &  $\Pi H^\ast(T^4)$ & $(\frac{ k-1}{2}$, $ \tfrac{k+1}{2})$  & $k \ge 1$ \\
	$\gamma \sim \delta_{z=0} n^{1-k}+ \dots $ &  $\Pi H^\ast(T^4)$ & $(\frac{ k-1}{2}$, $ \tfrac{k+1}{2})$  & $k \ge 1$ \\	
	$\mu \sim    n^{-k}   \d \log n \d z \delta_{z =0} + \dots  $ & $H^\ast(T^4)$ & $(\tfrac{k}{2}, \tfrac{k}{2})$ & $k \ge 1$ \\
	$ \mu \sim   n^{-k}    \d \log n \d w \delta_{z = 0} + \dots $ & $H^\ast(T^4)$ & $ (\tfrac{k-2}{2}, \tfrac{k+2}{2}) $& $k \ge 2 $ 
\end{tabular}.

The counting of states is very similar, except for the cohomological contribution of the internal manifold, to that appearing in \cite{CostelloGaiotto}. 

These states also can be easily seen to rearrange themselves to fill out short multiplets with respect to $\mf{psu}(1, 1|2)$, the global subalgebra of the $\mc{N}=4$ superconformal algebra, after acting on each of these four towers with the appropriate $SL(2, \R)$ generators and $SU(2)_R$ generators. As usual, the bosonic subalgebra is (the chiral half of) the isometry group of Euclidean AdS$_3 \times S^3$. This of course recovers a key observation of \cite{deBoer1} (see also \cite{DMW02}). The highest weight states of the multiplet are, on the CFT side, well known to be the purely left-moving, or holomorphic, chiral primary states with conformal weight equal to $SU(2)_R$ spin.

Let us denote by $(\mbf{\tfrac{m}{2}})_S$ the short representation of $\mf{psu}(1,1 | 2)$ whose heighest weight vector has $(J_0^3, L_0)$ eigenvalues $(m/2,m/2)$.  (Our notation differs by the factor of $\tfrac{1}{2}$ from that of \cite{deBoer1, deBoer2}).  The highest weight vector of the short representation $(\mbf{\tfrac{k}{2}})_S$ is represented by the state
\begin{equation} 
	\mu \sim n^{-k} \d \log n \d z \delta_{z = 0} + \dots 
\end{equation}
for $k > 0$.   In the table \ref{tbl:states}, if we fix the value of $k$ in each row, we will find those states in the short representation $(\mbf{\tfrac{k}{2}})_S$ which are heighest weight vectors under the bosonic subalgebra $\mf{su}(2)_R \oplus \mf{sl}(2)$.  

To sum up, we find that the single-particle states are
\begin{equation} 
	\oplus_{m \ge 1} (\mbf{\tfrac{m}{2}})_S \otimes H^\ast(T^4). 
\end{equation}
If we were to perform this analysis in the case of $K3$, we would find in the same way that the single-particle states are
\begin{equation} 
	\oplus_{m \ge 1} (\mbf{\tfrac{m}{2}})_S \otimes H^\ast(K3). 
\end{equation}

Let us compare this to the results of \cite{deBoer1, DMW02} for the counts of supergravity states for the physical theory.  Denote by $(\mbf{\tfrac{m}{2}}, \mbf{\tfrac{n}{2}})_S$ the representation of the left and right moving copies of $\mf{psu}(1,1\mid 2)$ which is the tensor product of $(\mbf{\tfrac{m}{2}})_S$ on the left and $(\mbf{\tfrac{n}{2}})_S$ on the right.    

Then, \cite{deBoer1} found that the supergravity states are\footnote{We re-index slightly and correct some small typos in \cite{deBoer1}.}
\begin{equation}
	\bigoplus_{m \geq 0}\bigoplus_{i, j}H^{i, j} \otimes (\mbf{\tfrac{m+i}{2}}, \mbf{\tfrac{m+j}{2} } )_S
\end{equation}
in terms of the cohomology groups $H^{i,j}$ of either $K3$ or $T^4$.

When we perform supersymmetric localization, we only retain the highest weight vector in the right moving sector, giving 
\begin{equation}
	\bigoplus_{m \geq 0}\bigoplus_{i, j}H^{i, j} \otimes (\mbf{\tfrac{m+i}{2}})_S = \bigoplus_{i,j} \bigoplus_{m \ge i} H^{i,j} \otimes (\mbf{\tfrac{m}{2}})_S 
\end{equation} 
Let us compare this to what we find, which is simply
\begin{equation} 
	\bigoplus_{m \ge 1} \bigoplus_{i,j} H^{i,j} \otimes (\mbf{\tfrac{m}{2}})_S. 
\end{equation}
Clearly, our answer matches those of  \cite{deBoer1, DMW02} in the range when the heighest weight of the representation is at least two. There are some small low-lying discrepancies. First, the analysis of \cite{deBoer1} yields an extra $H^{0,i} \otimes (\mbf{0})_S$.  The representation $(\mbf{0})_S$ is one-dimensional, and so must correspond to an operator killed by $L_{-1}$.   These are topological operators, which have non-singular OPE with all other operators.  In the case of $K3$, there are two extra bosonic topological operators, and  in the case of $T^4$, two bosonic and two fermionic operators. These extra topological operators are both charged under an extra ``degree'' quantum number of \cite{deBoer1, deBoer2} conjugate to the variable $p$, which tracks the multiparticle number in the Fock space of supergravity states (dually, tracks the symmetric orbifold number of the CFT, as in $p^n \op{Sym}^n$).  In the analysis of \cite{deBoer2} in the $K3$ case, the contribution of these two operators is removed by multiplying the index by $(1-p)^2$.  The fact that our answer does not contain these operators in the first place is perhaps a feature of our approach.

The other discrepancy concerns the groups $H^{2,j} \otimes (\mbf{\tfrac{1}{2}})_S$.  These are present in our analysis but not in that of \cite{deBoer2, DMW02}.  In the case of $K3$, there are two bosonic representations, and in the case of $T^4$, two bosonic and two fermionic.  While these operators are identified with the cohomology classes in $H^{2,j}(X)$ in the standard analysis, it is not strictly necessary that they be interepreted in this way. Since, in the computation of the index, only the dimensions of the spaces $H^{2,j}(X)$ plays any role, we can just say that our analysis yields an extra two bosonic copies of the short representation $(\mbf{\tfrac{1}{2}})_S$ and (in the case of $T^4$) two extra fermionic representations.  In the case of $T^4$, it is possible to remove the extra two bosonic representations while preserving the $\op{Spin}(5)$ symmetry, if we identify the bosonic representation in $H^2(T^4)$ with the class $F \in H^2(T^4)$, and that in $H^4(T^4)$ with the only non-zero element.    We can proceed similarly in the case of $K3$, preserving $SO(21)$ symmetry.

We note that, in the case of $T^4$, the extra \emph{fermionic} short representations can not be removed from our theory without breaking the $\op{Spin}(5)$ symmetry.  The group $H^3(T^4)$ transforms in an irreducible spin representation of $\op{Spin}(5)$, so we can not remove two elements in $H^3(T^4)$ without breaking the $\op{Spin}(5)$ symmetry.  It is interesting to note, however, that $\op{dim} H^{2, 1}(T^4) = \op{dim} H^{1, 0}(T^4)$ and the two states in the latter space correspond to non-propagating degrees of freedom in the bulk supergravity \cite{deBoer1, DMW02}. We leave a better understanding of this discrepancy to future work.

\subsubsection{Indices of short representations}
Let us also spell out the appearance of short $SU(1,1|2)$ representations using indices. We implicity choose a $U(1)_R \subset SU(2)_R$ quantum number. The $U(1)$ generator satisfies $j_0 = 2J_0^3$, where $J_0^3$ is the Cartan generator of $SU(2)$. In these conventions, the contributions from these four towers of states (two bosonic and two fermionic)  comprise short $SU(1, 1|2)$ multiplets, which arise by acting on a chiral primary by the generators of $\mathfrak{psu(1,1|2)}$. In particular, the character of these short representations are (for spin $j >1 \in \mathbb{Z}_{\geq 0}$)

\begin{equation}
\chi_{j}(q, y)=\text{Tr}((-1)^F q^{L_0}y^{2 J^3_0}) = {q^{j/2} \over (1-q)(y - y^{-1})}\left((y^{j+1} - y^{-j-1}) - 2 q^{1/2}(y^j - y^{-j}) + q (y^{j-1} - y^{1-j}) ) \right)
\end{equation} 
and \begin{equation}
\chi_1(q, y)= \text{Tr}((-1)^F q^{L_0}y^{2 J^3_0}) = {q^{1/2} \over (1-q)(y - y^{-1})}\left((y^{2} - y^{-2}) - 2 q^{1/2}(y - y^{-1}) ) \right)
\end{equation} for $j=1$; $\chi_0(q, y)= 1$ for spin 0. These short $SU(1, 1|2)$ characters encompass precisely the states enumerated above.  
The factors of  $(y^j - y^{-j})/(y - y^{-1})$ and $1/(1-q)$ account for, respectively, the $SU(2)_R$ representation carried by each single-particle state and repeated applications of the $SL_2$ generator $L_{-1}$.  Each short  representation then occurs with multiplicity dictated by the Hodge numbers of $T^4$ (with similar results holding for $K3$). For example, the short multiplet $(\mbf{1})_S$ with highest weights $(J_0^3=1, h=1)$\footnote{Notice that our gravitational states also capture the fact that the representation at $J_0^3=1/2$ is further truncated.} also contains two fermionic states with quantum numbers $(1/2, 3/2)$ and a bosonic state $(0, 2)$. In terms of the states in the table above these come from, respectively:  $\mu \sim  n^{-2} \d z \delta_{z=0} + \ldots$;  $\alpha, \gamma = n^{-1} \delta_{z = 0} + \dots$; and $\mu \sim n^{-2} \d \log n \d w\delta_{z = 0} + \ldots$. Acting on these with our $SU(2)_R, SL_2$ modes gives us the rest of the short representation. The generating function for our single-particle states is then given in terms of these characters as
\begin{equation}
\sum_{n,m,l}c_{sugra}(m,n,l)p^m q^n y^l = \sum_{m \geq 0}\sum_{i,j}h^{i,j}\chi_{m+i}(q, y)p^{m+1}
\end{equation} and the $c_{sugra}$ appear as exponents in the formula for the full multiparticle Fock space assembled from these single-particle contributions \cite{deBoer2, DMVV}:
\begin{equation}
\prod_{m>0,n,l}{1 \over (1 - p^m q^n y^l)^{c_{sugra}(n,m,l)}}.
\end{equation}

At this stage, we have simply observed that the single-particle states we have just enumerated can only organize themselves into (short) representations of $SU(1,1|2)$; in section \ref{sec:deformed} we will see that the states must organize with respect to this algebra more explicitly. Of course, the complete large-$N$ elliptic genus of the $T^4$ theory (but not the $K3$ theory), which is the graded trace in a multi-particle Fock space assembled from these single-particle contributions, vanishes due to the presence of fermionic zero modes\footnote{One can still construct, however, a non-vanishing index if one includes fugacities from the $Spin(5)$ symmetry which acts on the $T^4$ cohomology. There is an $SU(4)$ rotating the classes in $H^1(T^4)$, but to preserve our charge vector of length $N$ in $H^2(T^4)$ we must break the symmetry to $USp(4) = Spin(5)$. From a CFT perspective, an $SO(5)$ is well-known to act on the $Sym^{N>1}(T^4)$ chiral ring, which produces an interesting index \cite{BT}.}.

\subsection{Closed forms for the supergravity states}

In order to study the symmetry algebra of the supergravity states we have just enumerated, we need to first write down explicit formulae for these states. It turns out that once we accomplish this, we will already be able to derive a powerful result about the vanishing of many two-point functions in this theory, which we will use in the sequel. Some aspects of the model we are considering here are easier than those considered in \cite{CostelloGaiotto}:  the main simplification is that the formula presenting the gravitational state contains less than $3$ terms since $F^3 = 0$.

We will defer the derivation of these states to \cite{CP2} and just quote the results below. One can check that these field configurations on the backreacted geometry both solve the equations of motion and respect the boundary conditions everywhere except at $z=0$ on the boundary. For convenience, we also define 
\begin{equation} 
	\what{z} := z -  F^{ab} \eta_a \eta_b  n^2 \frac{\wbar}{1 + \abs{w}^2}    
\end{equation}
which is holomorphic in the deformed complex structure (unlike $z$). Notice that an expansion of any function of $\hat{z}$ will terminate due to the presence of the fermionic coordinates $\eta_a$.
First, let us write down the fields corresponding to the fermionic states $\alpha, \gamma$:
\begin{equation}
	\mc{E}_{k,m} := \begin{cases}
		n^{-k} \delta^{(m-1)}_{\what{z} = 0} & \text { if } k \ge 2 \\
		n^{-1} \delta^{(m-1)}_{\what{z} = 0} + F^{ab} F^{cd} \eta_a \eta_b \eta_c \eta_d m(m+1) z^{-m-2} (n')^3 \delta_{w = \infty}  & \text{ if } k = 1.
	\end{cases}
\end{equation} A supergravity state is obtained by setting $\alpha$ or $\gamma$ to be equal to $\mc{E}_{k, m}$. What we have written is the highest weight vector of an $SU(2)_R$ representation of spin $k/2$ with conformal dimension $m + k/2$. The second term in the $k=1$ case is incorporated precisely to deal with a pole at $w=\infty$ that localizes the field away from $z=0$ (that is, at $w = \infty$), which can be seen by expanding $\hat{z}$ in terms of the $z, w, n$ coordinates and studying the poles of the resulting terms.

It is also helpful to write explicit expressions for the states which are not highest weight states for $ SU (2)_R$. These are obtained by applying the lowering operator $J_{-1} = \partial_{\bar{w}} - w^2 \partial_w$ of $SU(2)_R$. The $SU(2)_R$ action is simply that on $\mbb{CP}^1_w$ by M\"obius transformations. The coordinate $z$ is invariant, while $n$ transforms as $(\d w)^{1/2}$.  In the coordinates $\what{z}$, $n$, $w$ one can calculate that the lowering operator satisfies 
\begin{equation}
	\begin{split}
		J_{-1}\what{z} &= -F^{ab} \eta_a \eta_b  n^2 \\
		J_{-1} n &= -w n \\
		J_{-1} w &= -w^2. 
	\end{split}
\end{equation}
We find that the rest of the states in the $SU(2)_R$ representation with highest weight $\mc{E}_{k,m}$ are given by 
\begin{equation}
    J_{-1}^l \mc{E}_{k,m} = \dbar \left( \left(w n \dpa{n} -F^{ab} \eta_a \eta_b  n^2 \dpa{\what{z}}\right)^l n^{-k} \right). 
\end{equation}
Note that, because $\mc{E}_{k,m}$ is localized at $z = 0$ for $k \ge 2$, the same holds for $J_{-1}^l \mc{E}_{k,m}$. 

Similarly, the supergravity state corresponding to the Beltrami differential $\mu$ (corresponding to highest weight vectors of $SU(2)_R$) are:
\begin{equation}\label{eq:beltrami}
\mc{E}'_{k,m}:=
	\begin{cases}	
	\dbar \left(  n^{-k} \d \log n \d \what{z}  \what{z}^{-m}  \right) &  \text{ for } \ k \geq 1  \\
	\dbar \left( n^{-k-2} \d \log n \d w \what{z}^{-m}  - \frac{m}{k+2}  \d w \d \what{z} n^{-k-2} \what{z}^{-m-1} \right)&  \text{ for }k \geq 2.
	\end{cases}
\end{equation}
In each expression, the field is written as $\dbar$ of a closed $(2,0)$ form, since $\mu$ can be viewed as a $(2, 1)$ form.

As in the case for the fields $\alpha,\gamma$, one can expand $\what{z}^{-m}$ in terms of $z$ and $n^2$. By an argument similar to that given earlier, we see that most of the time each field is localized at $z = 0$.  However, in this case, we find that for the first expression, it is localized at $z = 0$ for $k \ge 3$, and in the second case the localization occurs for $k \ge 2$ as before \footnote{The difference between this case and the previous case is that $\d \log n$ has a first-order pole at $w = \infty$. Indeed, writing $n' = n/w$ as before,
\begin{equation} 
	\d \log n = \d \log n' + \d \log w 
\end{equation}
and $\d \log w$ has a first-order pole at $w = \infty$. In the second expression, $\d \log n$ is appears in the wedge product $\d \log n \d w$, which has a second-order pole at $w = \infty$ only coming from the $\d w$ factor.}.

This implies the fact: \emph{all  holographic two-point functions for the single-particle states in $SU(2)_R$ representations of spin $\ge 1$ vanish}. \\
The supergravity solution corresponding to any state in such a representation is localized at $z = 0$. To compute the two-point function, we would normally evaluate the supergravity solution corresponding to a state placed at $z = 0$ at some other point, say $z =1$.  Because the state is localized at $z = 0$, this automatically gives us zero.

This statement holds purely at infinite $N$, so there is no contradiction with the fact that at finite $N$ the two-point functions are non-trivial.

\section{The gravitational global symmetry algebra}
\label{sec:global_symmetry}

In this section we will describe a natural subalgebra of the algebra of modes of the gravitational theory.  This subalgebra has the special feature that it preserves the vacuum at both $0$ and $\infty$, and hence preserves all correlation functions.  We refer to it as the \emph{global symmetry algebra}. The study of the global symmetry algebra played a key role in the analysis of \cite{CostelloGaiotto}.   

We will find that the global symmetry algebra has a beautiful geometric interpretation: it can be described in terms of symmetries of the super-conifold.   Among other features, this algebra very strongly constrains tree-level two and three-point functions.  For this study we may readily borrow the technology which was used in \cite{CostelloGaiotto} to identify the symmetry algebra acting in the planar limit of twisted 4d $\mc{N}=4$ super Yang-Mills, and that of its gravitational dual. 

We will conclude this section by identifying some simple subalgebras of the global symmetry algebra, which are related to familiar higher spin algebras.

Let us first define the \emph{gravitational mode algebra}.  A single-particle element of the gravitational mode algebra is given by a solution to the equations of motion of the gravitational theory which satisfies the boundary conditions everywhere except on the \emph{circle} $\abs{z}= 1$ on the boundary. The definition is the obvious gravitational counterpart of the definition of the mode algebra in an ordinary chiral or vertex algebra, where one integrates modes of local operators around a circular contour. In the limit we are considering, where the gravitational theory is treated classically, the single-particle elements of the mode algebra form an infinite-dimensional Lie algebra.  The Lie bracket is defined in a similar way to the OPE coefficients, and the central extension is defined in a similar way to the two-point function.

Given any single-particle state $\mc{O}$ of the gravitational theory, which is a Virasoro primary, we can form the corresponding modes $\oint_{\abs{z} = 1} z^n \mc{O}(z) \d z $ by averaging the state over the circle $\abs{z} = 1$. All single-particle gravitational modes arise uniquely in this way, hence we can enumerate them using the analysis  of single-particle states above.  

For each short representation $(\mbf{\tfrac{m}{2}})_S$ of $\mf{psu}(1,1\mid 2)$, there are four collections of Virasoro primaries.  Denote by $(r)_k$ the finite-dimensional representation of $SU(2)_R$ on which $J_0^3$ acts on the highest weight vector with weight $r$, and which is   acted on by $L_0$ with charge $k$.  Then the Virasoro primaries in the short representation $(\mbf{\tfrac{m}{2}})_S$ are 
\begin{equation} 
	(\tfrac{m}{2})_{\tfrac{m}{2}}  \oplus 2 \Pi (\tfrac{m-1}{2})_{\tfrac{m+1}{2}}   \oplus (\tfrac{m-2}{2})_{\tfrac{m+2}{2}}  
\end{equation}
where $\Pi$ indicates fermionic states.  Since we know the list of single particle states \eqref{tbl:states} in terms of short representations, we immediately read off the list of single-particle modes.  

Given an operator $\mc{O}$ of charge $k$ under $L_0$, not all the modes $\oint z^m \mc{O}(z)$ will preserve the vacuum at zero and infinity. Only those modes 
\begin{equation} 
	\oint z^m \mc{O}(z) \d z  \ \ \ \ \ \ \ \ \ \ \ \ (0 \le m \le 2k-2)
\end{equation}
preserve the vacua at $0$ and $\infty$.  Clearly $m$ must be non-negative in this expression, otherwise we would generate a non-zero state when we apply this mode to the vacuum at zero. Since $\mc{O}$ transforms as a section of $K^{-k}$, where $K$ is the canonical bundle, the transformation $z \mapsto z^{-1}$ picks up a factor of $z^{2k}$ from $\mc{O}(z)$ and $z^{-2}$ from $\d z$.  The vacuum at infinity is preserved if $z^{2k-2-m}$ does not have a pole at $z = 0$, leading to the upper bound on $m$.

For example, if $\mc{O}$ has spin $1$, only the zero-mode $\oint \mc{O}(z) \d z$ is a global symmetry. If $\mc{O} = T$ is the stress-energy tensor, then the global $SL_2$ charges $\oint z^m T (z) \d z$, $ \ 0 \le m \le 2$, are the global symmetries.  In general, an operator of charge $k$ under $L_0$ will give rise to a collection of global symmetries which transform under the $SL_2$ global conformal transformations as the representation of spin $k-1$.  

Applying this to our situation, we find that each short representation $(\mbf{\tfrac{m}{2}})_S$ of the superconformal algebra gives rise to two fermionic and two bosonic colletions of global symmetries.  The two bosonic global symmetries transform in representations of spin $(\tfrac{m}{2}, \tfrac{m}{2} - 1)$ and $(\tfrac{m}{2} -1, \tfrac{m}{2})$ under $SU(2)_R \times SL_2$, and the fermionic representations both transform in the representations of spin $(\tfrac{m-1}{2}, \tfrac{m-1}{2})$.   For instance, if $m=2$, the two bosonic and two fermionic representations arrange into the adjoint representation of $\mf{psu}(1,1\mid 2)$. 

Now, let us present our theorem on the global symmetry algebra, which parallels one of the results of \cite{CostelloGaiotto}.  As before, let $X^0$ be the super-conifold, with the singular point removed.  This has a fibration $X^0 \to \C^{0 \mid 4}$.  Let $\op{Vect}(X^0 / \C^{0 \mid 4})$ be the Lie algebra of holomorphic vector fields which point along the fibres of this fibration. (In coordinates, this means they do not involve any $\partial_{\eta_a}$).  Let $\op{Vect}_0(X^0/\C^{0 \mid 4})$ be the subalgebra which are divergence free on every fibre (recalling that there is a holomorphic volume form on the fibres of this map).

Let $\Oo(X^0)$ be the super-vector space of holomorphic functions on $X^0$. This is the same as the vector space of holomorphic functions the whole manifold $X$, including the conical point, by Hartog's theorem. In turn, this is the algebra generated by bosonic variables $u_i, w_j$ and fermionic variables $\eta_a$, subject to the relation $\eps^{ij} u_i w_j = F^{ab} \eta_a \eta_b$.

\begin{theorem} 
\label{thm:main}
	The Lie algebra of global symmetries of the gravitational chiral algebra is, as a vector space,
	\begin{equation} 
		\op{Vect}_0(X^0/\C^{0 \mid 4}) \oplus \left( \C^2 \otimes \Pi \Oo(X^0) \right) 
	\end{equation}
	where $\Pi$ indicates parity shift.  An element of this vector space will be denoted $(V,f_1,f_2)$.  The Lie bracket is as follows:
	\begin{equation} 
		\begin{split}
			[V,V'] &= \text{ usual commutator of vector fields } \\
			[V,f_i] &= V( f_i) \\
			[f_1,g_1] = [f_2,g_2] &= 0.
		\end{split}	
	\end{equation}
	The commutator of $f_1, g_2$ is non-trivial and defined as follows: $[f_1,g_2]$ is a vector field characterized by
	\begin{equation} 
		[f_1,g_2] \vee \Omega = \partial f_1 \wedge \partial g_2 
	\end{equation}
	where $\Omega$ is the volume form on the fibres of the map $X^0 \to \C^{0 \mid 4}$. 
\end{theorem}
The most interesting part of this result is that the global symmetry algebra includes the infinite-dimensional Lie algebra $\op{Vect}_0(X^0/ \C^{0 \mid 4})$.  

The proof of this theorem is not hard.  Let us explain concretely how the subalgebra $\op{Vect}_0(X^0/\C^{0 \mid 4})$ lives inside the global symmetry algebra.  This Lie algebra is the Lie algebra of gauge transformations of the Beltrami differential field of our Kodaira-Spencer theory on $X^0$, which preserve the vacuum solution to the equations of motion.  Recall that such gauge transformations must be divergence free, and do not involve any $\partial_{\eta_a}$. Preserving the vacuum solution to the equations of motion means that they are holomorphic.

Fix any $V \in \op{Vect}_0(X^0/\C^{0\mid 4})$.
In the coordinates $(n,w,z)$ defined near the boundary, define a gravitational field-configuration by
\begin{equation} 
	V \delta_{\abs{z} = 1}^{(0,1)}. \label{eqn:field_config} 
\end{equation}
Here, we take the $(0,1)$-form component of $\delta_{\abs{z} = 1}$.  This is a $\dbar$-closed $(0,1)$ form, even though $z$ is not a holomorphic function because of the Beltrami differential \eqref{eqn:beltrami_F}.  This means that this (mildly singular) field configuration satisfies the equations of motion. This field configuration satisfies the boundary conditions except at $\abs{z} = 1$, because it vanishes on the boundary except at $\abs{z} = 1$.  Therefore it defines an element of the gravitational mode algebra.

To check that this mode preserves the vacuum at $0$, we note that, in the absence of any state at $z = 0$, we can trivialize the field configuration \eqref{eqn:field_config} by the gauge transformation
\begin{equation} 
	V \delta_{\abs{z} \le 1}. 
\end{equation}
A similar argument applies to show that this configuration preserves the vacuum at $z =\infty$.   

To understand the commutator of these states in the mode algebra, we should study the two-particle state
\begin{equation} 
	\left( V \delta_{\abs{z} = 1}^{(0,1)} \right)  \left( V' \delta_{\abs{z} = 1+\eps}^{(0,1)} \right)  - \left( V \delta_{\abs{z} = 1}^{(0,1)} \right)  \left( V' \delta_{\abs{z} = 1-\eps}^{(0,1)} \right) , 
\end{equation}
where $V,V' \in \op{Vect}_0(X^0/\C^{0\mid 4})$. 

In the absence of any other states, the gauge transformation 
\begin{equation} 
	V' \delta_{1-\eps \le \abs{z} \le 1 + \eps} 
\end{equation}
gives a gauge equivalence 
\begin{equation} 
	V' \delta_{\abs{z} = 1 - \eps} \sim V' \delta_{\abs{z} = 1 + \eps}, 
\end{equation}
moving the circle on which we are studying modes from $1-\eps$ to $1 + \eps$.  In the presence of the state $V \delta_{\abs{z} = 1}$, however, we get an extra term
\begin{equation} 
	[V',V] \delta_{\abs{z} = 1}. 
\end{equation}
This leads to the gauge equivalence
\begin{equation} 
	\left( V \delta_{\abs{z} = 1}^{(0,1)} \right)  \left( V' \delta_{\abs{z} = 1+\eps}^{(0,1)} \right)  - \left( V \delta_{\abs{z} = 1}^{(0,1)} \right)  \left( V' \delta_{\abs{z} = 1-\eps}^{(0,1)} \right)  \sim [V',V'] \delta_{\abs{z} = 1}, 
\end{equation}
giving the desired commutator in the mode algebra.

A similar argument shows that the modes in $\Pi \Oo(X^0)\otimes \C^2$ satisfy the desired commutator. 

Now let us show that the algebra we have just described has the same quantum numbers (under $SU(2)_R$ and $SL_2$) as those we obtained from the global symmetries built as modes of single-particle states.  The computation of the quantum numbers does not depend on the flux $F$, so we can set it to zero.   We let $Z^0$ denote the bosonic conifold $\eps^{ij} u_i w_j = 0$.  As a representation of $SU(2)_R \times SL_2$, we have 
\begin{equation}
	\begin{split}
		\op{Vect}_0(X^0/\C^{0 \mid 4}) &= \op{Vect}_0 (Z) \otimes H^\ast(T^4) \\
		\Oo(X^0) &= \Oo(Z)\otimes H^\ast(T^4)
	\end{split}
\end{equation}
	(here $\op{Vect}_0(Z)$ indicates divergence-free holomorphic vector fields).

	As we have seen, a single-particle state in the short representation $(\mbf{\tfrac{m}{2}})_S$ contributes global symmetries in the representations
	\begin{equation} 
		(\tfrac{m-2}{2}, \tfrac{m}{2}) \oplus (\tfrac{m}{2}, \tfrac{m-2}{2}) \oplus  \Pi (\tfrac{m-1}{2}, \tfrac{m-1}{2})^{\oplus 2}  
	\end{equation}
	where we use the notation $(\frac{m}{2}, \tfrac{n}{2})$ for the representation of $SU(2)_R\times SL_2$ of these spins.  For example, the short representation $(\mbf{\tfrac{2}{2}})_S$ contains both the stress-energy tensor $T$,  the $SU(2)_R$ current $J$, and the two super-currents of spin $3/2$. The stress-energy tensor contributes global symmetries in the adjoint of $SL_2$, hence in the representation $(0,1)$; and the $SU(2)_R$ current contributes global symmetries in the adjoint of $SU(2)_R$.   The supercurrents give global symmetries the spin $1/2$ representation of each of $SU(2)_R$ and $SL_2$.

To show that the quantum numbers of the global symmetry algebra match with what we found by analyzing modes of single particle states, we need to show that
	\begin{equation} 
		\begin{split}
			\op{Vect}_0 (Z) &= \oplus_{m \ge 1} (\tfrac{m-2}{2}, \tfrac{m}{2}) \oplus (\tfrac{m}{2}, \tfrac{m-2}{2}) \\
			\Oo(Z)  &= \oplus_{m \ge 1} (\tfrac{m-1}{2}, \tfrac{m-1}{2}). 
		\end{split}
	\end{equation}
	The first statement was verified in the appendix to \cite{W92} (it is not hard to check directly).  The second statement is an easy exercise (one can use, for instance, the fact that holomorphic functions on $Z$ are the same as those on the resolved conifold, where it is almost obvious).


	\subsection{$\mf{psu}(1,1| 2)$ inside the global symmetry algebra}
We have been describing the supergravity states as representations of $\mf{psu}(1,1|2)$, although so far we have not explained how the fermionic part of this algebra acts (the action of the bosonic part is obvious). In this section, we will show that the lowest-lying modes of the global symmetry algebra give a copy of $\mf{psu}(1,1\mid 2)$.  	

The bosonic $\mf{sl}(2) \oplus \mf{sl}(2)$ consists of those $4 \times 4$ complex matrices which preserve the quadratic form $\eps^{ij} u_i w_j$. Every such matrix gives rise to a vector field on $\C^{4 \mid 4} $, acting trivially on the fermionic variables, which evidently preserves the super-conifold. 

To understand the fermionic part, recall that an element of the global symmetry algebra can be written as a triple $(V,f_1,f_2)$ where $V$ is a vector field on $X^0$ with certain properties, and $f_1,f_2$ are functions on $X^0$ (with reversed parity).     The fermionic elements of $\mf{psu}(1,1 | 2)$ are given by such triples with $V = 0$: we will write such an element as $(f_1,f_2)$ with $V = 0$ understood.  

The fermionic part of the global symmetry algebra is given by specifying the eight elements
	\begin{equation} 
		(u_i,0) \ \ (0, u_i) \ \ (w_i,0) \ \ (0,w_i)		
	\end{equation}
	We need to check that these fermionic symmetries do indeed commute according to the algebra $\mf{psu}(1,1\mid 2)$. 

	It is convenient to collect the four coordinates $u_i$, $w_j$ into a four-index vector $x_\alpha$, transforming in the vector representation of $\mf{so}(4)= \mf{sl}_2 \oplus \mf{sl}_2$.  The variety $X^0$, in these coordinates, is cut out by the equation $\sum x_{\alpha}^2 = F^{ab} \eta_a \eta_b$. The volume form on $X^0$ is
\begin{equation}	
			\Omega = \op{Res} \frac{ \prod \d x_\alpha  }{ \sum x_{\alpha}^2 - F^{ab} \eta_a \eta_b} 	
\end{equation}
where the residue is taken along the hypersurface $X^0$ in $\C^{4 \mid 4}$.  This is only a volume form on the fibres of the map $X^0 \to \C^{0 \mid 4}$.  
	
The commutator $[(f_1,f_2), (g_1,g_2)  ]$ is a vector field determined by
	\begin{equation} 
		[(f_1,f_2),(g_1,g_2)] \vee \Omega = \partial f_1 \wedge \partial g_2 + \partial g_1 \wedge \partial f_2. \label{eqn:function_bracket}
	\end{equation}
Implementing this definition, we will find that 
	\begin{equation} 
		[ (f,0), (0,g)] =  \eps^{\alpha \beta \gamma \delta} (\partial_{\alpha} f) (\partial_{\beta} g)  x_\gamma \partial_{\delta} \label{eqn:function_bracket2}. 
\end{equation}
(after an overall rescaling of $f$, which can be absorbed into a change of basis). 

	To see this, note that the vector field on the right hand side of \eqref{eqn:function_bracket2} always preserves the quadratic form $\sum x_\alpha^2$, and so defines a vector field on $X^0$.  Since the formation of residues commutes (up to a sign) with contraction with a vector field, to check that the vector field in \eqref{eqn:function_bracket2} satisfies \eqref{eqn:function_bracket},  we need to show that
	\begin{equation} 
		 \op{Res} \frac{ (\partial_\alpha f) (\partial_\beta g) x_\gamma \d x_\alpha \d x_\beta \d x_\gamma } { \sum x_\mu^2 - \sum F^{ab} \eta_a \eta_b }  = C (\partial_{\alpha} f)(\partial_{\beta} g) \d x_\alpha \d x_\beta.  
	\end{equation}
for some overall non-zero constant $C$ (the residue is taken at the hypersurface given by the super-conifold). This can be verified by explicit integration.

Applying this to $f = x_\alpha$, $g = x_\beta$, we find that
\begin{equation} 
	[(x_\alpha,0) , (0,x_\beta)] = \eps^{\alpha \beta \gamma \delta} x_{\gamma} \partial_{x_{\delta}}  
\end{equation}
Interpreting the right hand side as an element of $\mf{so}(4) = \mf{sl}(2) \oplus \mf{sl}(2)$, this is the commutation relation for $\mf{psl}(2 \mid 2)$.

The global symmetry algebra we have just computed bears many similarities with that obtained in \cite{CostelloGaiotto}. In that case the bosonic part of the Lie algebra was $\op{Vect}_0(SL(2, \C))$, the divergence-free vector fields on the deformed conifold $SL(2, \C)$ which satisfies the defining equation $u_2 w_1 - u_1 w_2 = N$. The Lie algebra $\op{Vect}_0(SL(2, \C))$ then captures the symmetries of this geometry as a Calabi-Yau manifold (i.e. gauge symmetries of vacuum field configurations). In our case, the defining equation of the backreacted geometry, whose symmetries we must characterize, is instead $u_2 w_1 - u_1 w_2 = F^{a b}\eta_a \eta_b$, where the backreaction is governed by fermionic fields and, due to the truncation discussed earlier, can only give finite corrections to the flat space algebra. As in \cite{CostelloGaiotto} the algebra contains the algebra of holomorphic divergence-free vector fields over the complement of the boundary-condition-violating locus.

\subsection{The global symmetry algebra and higher spin algebras}

The global symmetry algebra is the symmetries of a three-dimensional space.  As such, it is closely related to higher-spin algebras, which are symmetries of two-dimensional spaces. In this section we analyze the relationship.   Our results are partially negative: we find that there is a natural homomorphism from our global symmetry algebra to a higher spin algebra, but that this homomorphism is not surjective. For some original references on $\mc{W}_{\infty}$ in the physics literature see, e.g., \cite{B89, PRS90, W92, BS93} whose presentations we largely follow.  The analysis of \cite{W92} is particularly relevant for our story. 

The higher spin algebras that we will study are very closely related to the $\mc{W}_{\infty}$ algebra.  The $\mc{W}_{\infty}$ vertex algebra is generated by operators of spins $1,2,\dots$.  There are a few minor variants of the definition. We will be interested in the version of $\mc{W}_{\infty}$ characterized by the fact that its mode algebra is a central extension of the Lie algebra of area-preserving (i.e.\ divergence-free) vector fields on the cylinder.  (In some contexts, it is more natural to study the closely related Lie algebra of differential operators on the circle; that is not relevant for the present work).  Since we work in a holomorphic context, we can replace divergence-free vector fields on a cylinder by divergence-free holomorphic vector fields on the holomorphic cylinder $T^\ast \C^\times$. 

There is no way that the full $\mc{W}_{\infty}$ algebra can possibly live inside our global symmetry algebra. The best we can hope for is that the global symmetry algebra of the $\mc{W}_{\infty}$ algebra will live inside our global symmetry algebra.  Because $\mc{W}_{\infty}$ is generated by fields of all positive integer spin, the global symmetry subalgebra will be given by fields which transform under the representations of integer spin $(k)$ of the global $SL_2$ symmetry, for $k \ge 0$.   We will see shortly that this subalgebra is a higher spin algebra. 

What is this subalgebra geometrically? Let us choose coordinates $X,Y$ on $\C \times \C^\times$, so that the symplectic form is $\d X \wedge Y^{-1} \d Y$.  The global $SL_2$ symmetry is represented by the vector fields associated to the Hamiltonian functions $X/Y$, $X$, $XY$ (and the Poisson bracket is $\{X,Y\}= Y$).  The Hamiltonian functions which are in the global symmetry algebra which transform in the representation $(k)$ of the global $SL_2$ are easily seen to be given by
\begin{equation} 
	X^k Y^{-k}, X^k Y^{1-k}, \dots, X^k Y^k.  
\end{equation}
Note that 
\begin{equation} 
	\begin{split}
		\{X Y^{-1}, X^k Y^{-k}\} &= 0 \\
		\{XY, X^k Y^k \} &= 0
	\end{split}
\end{equation}
so that these Hamiltonians are highest and lowest weight vectors for the $SL_2(\R)$ global symmetry.

The Hamiltonians on $\C \times \C^\times$ which are in the global symmetry algebra can be characterized by the fact that they extend without poles to an algebraic symplectic surface birational to $\C\times \C^\times$. Indeed, the Hamiltonians in the global symmetry algebra are closed under taking the product, and as a commutative algebra it is generated by
\begin{align*} 
	A&:=X \\
	 B&:= XY \\
	   C&:=XY^{-1} 
\end{align*}
subject to the relation 
\begin{equation} 
	A^2 = BC.  \label{eqn:conic_winfty} 
\end{equation}
This relation defines an affine algebraic surface in $\C^3$. We remove the singular locus where all coordinates vanish. The locus where $A \neq 0$ is isomorphic to the subset of $\C\times \C^\times$ where $X \neq 0$.  The symplectic form $Y^{-1}\d X\wedge \d Y$ extends to this affine algebraic surface.  On the locus $B \neq 0$, it can be written as $B^{-1}\d A\wedge \d B$.  On the locus $C\neq 0$, it can be written as $-C^{-1}\d A \wedge \d C$.  The fact that
\begin{equation} 
	2 \d A = B\d C + C \d B 
\end{equation}
implies that these two expressions for the symplectic form agree on their common domain of definition.

The Poisson bracket between the elements $A,B,C$ is the Lie bracket on $\mf{sl}_2$. From this we see that the Poisson algebra generated by the $A,B,C$, subject to the relation \eqref{eqn:conic_winfty}, is the quotient of the Poisson algebra of functions on the Lie algebra $\mf{sl}_2$ by the quadratic Casimir.

Recall that the higher spin algebra $HS[\lambda]$ is defined to be the quotient of $U(\mf{sl}_2)$ by the relation that the quadratic Casimir is set to $\lambda$.  The Poisson algebra we are studying is the Poisson version of this construction, at $\lambda = 0$.  We denote this Poisson higher-spin algebra by $hs[\lambda]$, and we note that it deforms in a natural way to $HS[\lambda]$.

To sum up, we find that global symmetry algebra of the $\mc{W}_{\infty}$ algebra is the Lie algebra of Hamiltonian functions on the affine symplectic surface defined by \eqref{eqn:conic_winfty}, which we view as the Poisson higher spin algebra $hs[\lambda]$.  

Let us now explain a variant of this picture.  Let us introduce fermionic variables $\eta_a$, as before, and deform equation \eqref{eqn:conic_winfty} to the equation
\begin{equation} 
 A^2  - F^2/4 = BC\label{eqn:conic_winfty_deformed}
\end{equation}
where $F= F^{ab}\eta_a \eta_b$ is the flux, as before.  If we again remove the singular locus, the equation \eqref{eqn:conic_winfty_deformed} defines a super-algebraic variety of dimension $2 \mid 4$, which is fibred over $\C^{0 \mid 4}$. This super-variety is equipped with a two-form along the leaves of the fibration, defined as above by $B^{-1} \d A \wedge \d B$ in an appropriate patch.  

It is perhaps more natural to invert the $2$-form to give us a Poisson tensor, given by the expression $B \partial_A \wedge \partial_B$. Thus, we have the Poisson brackets
\begin{equation} 
	\begin{split}
		\{A,B\} &= B \\
		\{A,C\} &=- C \\
		\{B, C\}&= - 2 A. 
	\end{split}
\end{equation}
The last equation is derived by applying $\{B,-\}$ to both sides of the  relation \eqref{eqn:conic_winfty_deformed}.  As before, the Poisson brackets between the generators $A,B,C$ form a copy of the Lie algebra $\mf{sl}_2$.  This Poisson algebra is then a version of the higher-spin algebra. It is a deformation of the tensor product $hs[\lambda = 0] \otimes H^\ast(T^4)$ obtained by setting the quadratic Casimir to $F$ rather than to zero.

This construction generalizes: for any commutative algebra $R$ and element $F \in R$, we can form a higher-spin algebra $hs^R[F]$ obtained as the quotient of $R \otimes \Sym^\ast \mf{sl}_2$ by the relation $A^2 - BC = F$.  In the case $R = H^\ast(X)$, for $X = T^4$ or $K3$, and $F \in H^2(X)$, we denote this by $hs^X[F]$.  Of course, we will find that $hs^{T^4}[F/4]$ is closely related to the global symmetry algebra we defined above.

It is convenient in all these definitions to remove from the higher-spin Lie algebra the central elements which are just in the algebra $R$ (i.e.\ which are in $\C[\eta_a]$ and do not depend on the $A,B,C$ coordinates).  Then, if by $Y^0$ we denote the super-variety given by equation \eqref{eqn:conic_winfty_deformed}, the Lie algebra $hs^{T^4}[F]$ is the Lie algebra of vector fields on $Y^0$, which point along the fibres of the fibration $Y^0 \to \C^{0 \mid 4}$, and which preserve the fibrewise volume-form. 

\subsubsection{Higher spin algebras and the global symmetry algebra} 
How does this relate to the global symmetry algebra we discussed earlier? As above, let $X^0$ be the super-conifold, with the singular locus removed.  Let us give $X^0$ an action of the group $\C^\times$, given by the Cartan of $SU(2)_R$.  Then, the quotient of $X^0$ by $\C^\times$ is the variety given by \eqref{eqn:conic_winfty_deformed}.  To see this, we note that the $\C^\times$ action gives the coordinates $w_1,u_1$ weight $1$ and $w_2,u_2$ weight $-1$. The coordinates $\eta_a$ are given weight $0$.  We have
\begin{equation} 
	\begin{split}
		A-F/2 &= u_2 w_1  \\
		A + F/2 &= u_1 w_2\\
		B &= w_1 w_2 \\
		C &= u_1 u_2.
	\end{split}
\end{equation}
Relation \eqref{eqn:conic_winfty_deformed} is immediate. 

Similarly, the holomorphic $2$-form on the variety defined by \eqref{eqn:conic_winfty_deformed} is determined from the holomorphic $3$-form on $X^0$. If we denote by $Y$ the super-variety given by \eqref{eqn:conic_winfty_deformed}, and $Y^0$ the complement of the singular locus where $A,B,C$ are all zero, there is a $\C^\times$ fibration
\begin{equation} 
	X^0 \to Y^0. 
\end{equation}
We can take the residue of the holomorphic $3$-form 
\begin{equation} 
	\Omega = \frac{1}{w_1} \d w_1 \d w_2 \d u_1 
\end{equation}
along the fibres of this fibration, we will find the $2$-form on $Y^0$.  (In each case, they are forms defined on the fibres of the map to $\C^{0 \mid 4}$).

It follows immediately from this that there is a homomorphism
\begin{equation} 
	\op{Vect}_0 (X^0/\C^{0 \mid 4})^{\C^\times} \to \op{Vect}_0(Y^0/\C^{0 \mid 4}) = hs^{T^4}[F].  
\end{equation}
Indeed, if we have a vector field on $X^0$ which commutes with the $\C^\times$ action, then it gives one on the quotient $Y^0 = X^0 / \C^\times$.  If the vector field on $X^0$ leaves the functions $\eta_a$ fixed, then it does so on the quotient $Y^0$.  Finally, if the vector field on $X^0$ preserves the volume form on the fibres of the map $X^0 \to \C^{0 \mid 4}$, then it must do so on $Y^0$, as this volume form is obtained as a residue of the one on $X^0$. 

However, this homomorphism is not exactly what we would like. We would like to realize the higher-spin algebra as a subalgebra of $\op{Vect}_0(X^0/\C^{0 \mid 4})$, and not as a quotient.  To do this, we must understand when a vector field on $Y^0$ can be lifted to one on $X^0$.

The threefold $X^0$ forms a $\C^\times$ bundle over $Y^0$. Let us describe it as a line bundle.  Sections of the associated line bundle are generated (as a module over functions on $Y^0$)  by functions on $X^0$ which are of charge $1$ under the $\C^\times$ action.  These are the functions $u_1,w_1$ which satisfy the relations
\begin{equation}
      \begin{split} 
	(A - F/2) u_1 &= C w_1 \\
	(A + F/2) w_1 &= B u_1	 
	\end{split}
\end{equation}
These relations define a locally-free rank one module over the structure sheaf of $Y^0$.  On the locus $C \neq 0$, the module is generated by $u_1$. In this region, $u_1$ defines a coordinate on the fibres of the map $X^0 \to Y^0$. On the locus $B \neq 0$, $w_1$ provides a coordinate. These coordinates are related by 
\begin{equation} 
	u_1 = \frac{C}{A - F/2} w_1.
\end{equation}
Consider the divisor $C = 0$. This has two components (separated only in the fermionic directions) according to whether $A-F/2$ is zero or $A + F/2$ is zero.   On the locus where $C = 0$ and $A+F/2 = 0$, we have the relation $F u_1 = 0$. Thus, $u_1$ fails to be a coordinate on the fibres of the map $X^0 \to Y^0$. 

We can try to lift a vector field $V$ on $Y^0$ to a vector field $\til{V}$ $X^0$ by declaring $\til{V} (u_1) = 0$ (and $\til{V}$ acts in the same way as $V$ on the coordinates $A,B,C$ of the base). This makes sense except on the region where $u_1$ fails to be a coordinate on the fibres.  In this region, $w_1 = \tfrac{B }{ A + F/2 } u_1$ is the appropriate coordinate. Declaring that $\til{V} u_1 = 0$ gives us 
\begin{equation} 
	\til{V} (w_1) = \frac{A + F/2}{B} V  \left( \frac{B}{A + F/2}   \right) w_1. 
\end{equation}
For $\til{V}$ to be a vector field on $X^0$ without any poles, the expression on the right hand side must be a regular function on $Y^0$.  

Let us check whether Poisson bracket with the coordinate functions $A,B,C$ has this property. We have
\begin{equation} 
	\begin{split}
		\frac{A + F/2}{B} \left\{A, \frac{B}{A + F/2} \right\} &= 1\\ 
		\frac{A + F/2}{B} \left\{B, \frac{B}{A + F/2} \right\} &= \frac{B }{A + F/2} \\
		\frac{A + F/2}{B} \left\{C, \frac{B}{A + F/2} \right\} &= - \frac{C}{A + F/2} + \frac{2 A}{B} . 
	\end{split}
\end{equation}
From these expressions, we see that Poisson bracket with $A$ lifts to a regular vector field on $X^0$, but that Poisson bracket with $B$ and $C$ does not, because the right hand side has poles. 

A basis, over the ring $\C[\eta_a]$, of functions on $Y^0$ is given by expressions like $(A+F/2)^k B^l$, $A^k$,  $(A-F/2)^k C^l$.   From the Poisson brackets listed above we see that $(A+F/2)^k B^l$ gives a regular vector field on $X^0$ as long as $k > l$, and that the Poisson bracket with $A^k$ always gives a regular vector field on $X^0$. The Poisson bracket with $(A+F/2)^k C^l$ is never a regular vector field.

In a similar way, the vector fields $(A-F/2)^k C^l$, for $k > l$, give rise to a regular vector field on $X^0$ which leaves the coordinate $w_1$ invariant, but modifies the coordinate $u_1$.  

We have learned that the map
\begin{equation} 
	\op{Vect}_0 (X^0 / \C^{0 \mid 4})\to hs^{T^4}[F]  
\end{equation}
is not surjective, but its image consists of the elements of the form $A^k$, $(A+F/2)^k B^l$for $k > l$, $(A-F/2)^k C^l$ for $k > l$.  Let us call this sub-Lie algebra $hs(T^4)[F]^+$.  The kernel of this map is necessarily given by vector fields which point along the fibre of the map $X^0 \to Y^0$, are divergence free, and are fixed by rotating the fibres.  Such vector fields are all of the form $G(A,B,C,\eta_a) u_1 \partial_{u_1}$, for $G$ any regular function on $Y^0$.  These vector fields all commute with one another. 

This tells us that there is a short exact sequence of Lie algebras 
\begin{equation} 
	\Oo(Y^0) \to  \op{Vect}^0 (X^0 / \C^{0 \mid 4})\to hs^{T^4}[F]^+ . 
\end{equation}

We may also hope to formulate similar relations between our full global symmetry algebra, including the fermionic generators, and suitable $\mc{N}=4$ supersymmetric higher spin algebras. The natural candidate that arises in the study of AdS$_3 \times S^3 \times T^4$ is a certain matrix-extended supersymmetric higher spin algebra described in, e.g., \cite{EGR18, GG15} and references therein. It may be described in terms of the algebra $\mf{shs}[\lambda]$ whose bosonic part may be expressed as $\mf{shs}[\lambda=0]^{bos} \simeq \mf{hs}[0] \oplus \mf{hs}[1/4]$\footnote{Our convention here differs slightly from the existing literature. Our convention is that $\mf{hs}[\lambda]$ and $\mf{shs}[\lambda]$ are essentially defined to be universal enveloping algebras quotiented by the two-sided ideal that sets their respective Casimirs to the value $\lambda$, whereas in the literature the ideal for $\mf{hs}[\lambda]$ is generated by $\mc{C}^{sl_2} -{1 \over 4}(\lambda^2 - 1)\mathbb{1}$ and for $\mf{shs}[\lambda]$ by $\mc{C}^{osp(1|2)} - {1 \over 4}\lambda(\lambda-1)\mathbb{1}$, so that in this notation $\mf{shs}[\lambda]^{bos} \simeq \mf{hs}[\lambda]\oplus \mf{hs}[1-\lambda]$ \cite{GG12}.}. The generators of $\mf{shs}[\lambda]$ are often written as $V^{(s), \pm}_m$ where $s = 1 + n/2, n \geq 0, -s+1 \leq m \leq s-1$ so that there are two generators for each $s$ (except for a subtlety at $s=1$, where there turns out to be a single generator). From these one can construct a matrix extension $\mf{shs}_2[\lambda]$ by appending to the generating basis the generators of $U(2)$; this extension has as its largest finite dimensional subalgebra the wedge subalgebra of the large $\mc{N}=4$ superconformal algebra. The corresponding higher spin extension of the small $\mc{N}=4$ wedge subalgebra can then be obtained by specialization of the parameters to $\lambda=0$.  It has been argued \cite{GG14} that the super $\mc{W}$-algebra based on this higher spin algebra is a subalgebra of the chiral algebra of the symmetric orbifold theory (the full chiral algebra is known to have additional generators). In fact, it was explicitly shown in \cite{GG15} that the \emph{single-particle} generators of the asymptotic symmetry algebra organize into representations of $\mf{shs}_2[0]$.   We plan to comment further on the connections between our global symmetry algebra and these algebras in \cite{CP2}\footnote{We also leave potential interesting connections between the higher spin square \cite{GG15} and Yangian \cite{GLPZ18} algebras studied in the context of AdS$_3$ to future work.}.

\section{Koszul duality: an introduction for physicists}\label{sec:koszul}

We have seen that the effective supergravity theory obtained by compactifying twisted type IIB supergravity on $T^4$ is a version of Kodaira-Spencer theory on $\C^{3|4}$.  The $D1$ and $D5$ brane systems source the flux given by the Beltrami differential
\begin{equation} 
	F^{ab} \eta_a \eta_b \frac{\eps^{ij} \wbar_i \d \wbar_j   }{ (w_1 \br{w}_1 + w_2 \br{w}_2)^2  } \partial_z 
\end{equation}
which deforms the complex structure. As mentioned before, we could in principle calculate the chiral algebra of the large $N$ CFT by studying scattering of particles in the backreacted geometry, using Witten diagrams.  This was the approach taken in \cite{CostelloGaiotto}.  Here, we will introduce a new method, which gives the same answer, based on the idea of \emph{Koszul duality}. Although there are technical challenges to studying Witten diagrams at loop-level in AdS$_3$ directly from the bulk perspective (though see e.g. \cite{GST} for recent progress), we will see that this algebraic perspective which is based on studying \emph{coupled} bulk-defect systems naturally incorporates quantum corrections to the chiral algebra.

\subsection{Koszul duality and universal defects}
Before explaining how Koszul duality applies to the model we are studying, let us make some remarks on the role of Koszul duality in field theory in general.  Let us consider some field theory living on $\R \times \R^n$ with coordinates $(t,x_i)$. We will treat $\R$ as a ``time'' direction, and assume that the theory is topological in this direction, i.e.\ has no Hamiltonian.  We ask that the system is topological in a strong sense: we require that there exists some fermionic symmetry $\what{Q}$, of ghost number $-1$, such that $\{Q, \what{Q}\} = \partial_t$ (where $Q$ is the BRST operator).  This will be satisfied, for instance, if we took a partially topological twist of a system which had enough supersymmetry. 

In particular, and in contrast to the most common incarnation of twisting in the literature, we do \emph{not} assume that the system is Lorentz invariant.  A good model field theory to have in mind is the four-dimensional Chern-Simons theory of \cite{C13}\footnote{For recent applications of this four-dimensional Chern-Simons theory to the study of integrable systems, see \cite{CWY17, CWY18, CY19}}, where the time direction is taken to be one of the two topological directions.

We let $\mc{A}$ be the algebra of local operators of the system.  We do not take BRST cohomology, so we treat $\mc{A}$ as a cochain complex, with differential the BRST operator. The operator product in the $t$-direction gives $\mc{A}$ the structure of an algebra.  If we work in the greatest generality, this algebra structure will be an $A_\infty$ algebra\footnote{See \cite{GO} for a recent discussion of Koszul duality and line defects for $A_{\infty}$ algebras in the context of $\Omega$-deformed M-theory.}.  We will work under the assumption\footnote{This is not in fact a restrictive assumption, as one can always enlarge $\mc{A}$, without changing the cohomology, so that the multiplication is strictly associative.} that $\mc{A}$ is a differential graded algebra, meaning the multiplication is strictly associative and the Leibniz rule
\begin{equation} 
Q (a \cdot b) = (Q a) \cdot b \pm a \cdot (Q b) 
\end{equation} 
holds, where $Q$ is the BRST operator.

One can then ask the question: \emph{what is the most general topological line defect that we can couple to our system along the $t$-direction?}

Most, perhaps all, such line defects are obtained by coupling to some topological quantum mechanical system. If $\mc{B}$ is the algebra of operators of an auxiliary quantum mechanical system, the question becomes: what data on $\mc{B}$ do we need to couple to our field theory? 

For this question to have a good answer, we will need to make some additional assumptions. One sufficiently strong assumption which we will make, and which holds in all of the examples we discuss, is the following. We will assume that the algebra $\mc{A}$ is a small deformation of $\mc{A}^{free}$, the dg-algebra for a free theory.  We assume that the cohomology $H^\ast(\mc{A}^{free})$ is an exterior algebra freely generated by operators $\mf{c}_i$ in ghost number $1$. Under this assumption, the algebra $\mc{A}$ is equivalent to the algebra $H^\ast(\mc{A}^{free}) = \C[\mf{c}_i]$, but with some possibly non-trivial differential and $A_\infty$ structure, which encode the perturbative interactions of the theory.

Then, the answer to our question is the following: there is a new algebra $\mc{A}^!$, built algebraically from the algebra of operators $\mc{A}$, such that \emph{coupling to quantum mechanics with algebra of operators $\mc{B}$ is the same as giving an algebra homomorphism}
$$ \mc{A}^! \to \mc{B}.$$
The algebra $\mc{A}^!$ is the Koszul dual of $\mc{A}$.

We see that the Koszul dual algebra is the algebra of operators of a ``universal defect'', which encodes the most general possible consistent couplings of our given theory to a defect. We will elaborate on this claim with various examples shortly. So far everything we have said is quite general and so we expect Koszul duality to play a role when studying field theories coupled to more general types of defects, and indeed it does; see appendix \ref{app:koszulTFT} for a short review. However, we can also view an open-closed string theory as a bulk theory coupled to a ``defect'' theory given by the branes, which includes some interaction term coupling the two pieces together. If we view the system at low energies as some effective theory with a Lagrangian description, we therefore have $\mc{L} = \mc{L}_{bulk} + \mc{L}_{brane} + \mc{L}_{int}$. 

Even though we ultimately take a decoupling limit in the holographic correspondence that sends $\mc{L}_{int} \rightarrow 0$ (a limit which is automatic when we pass to the twisted theory), one might naively hope from these considerations that the operator algebra on the branes as $N \rightarrow \infty$ becomes the Koszul dual of the algebra of local operators of the bulk gravitational theory--- again, at least in a twisted version of the story where we have a hope of making this precise. 
It will turn out that gravitational backreaction of the branes on the bulk, which is central to the AdS/CFT correspondence as we look at the near-horizon geometry of the branes, will ultimately produce a certain calculable deformation of the Koszul dual algebra, as mentioned in the Introduction. 

\subsection{Koszul duality in Chern-Simons theory}
Let us explain how ordinary (i.e. without backreaction) Koszul duality works in the simplest example:  that of Chern-Simons theory with Lie algebra $\mf{g}$. We will find that in this case the Koszul dual $\mc{A}^!$ to the algebra $\mc{A}$ of local operators is the universal enveloping algebra $U(\g)$.  The universal enveloping algebra is generated by $\rho_a$, where $a$ is a Lie algebra index, with relations
\begin{equation} 
[\rho_a,\rho_b] = f^c_{ab}\rho_c. 
\end{equation}
  We will derive this algebra both by analyzing the algebra of operators of the universal line defect, as above; and by using the mathematical definition of Koszul duality.

\subsubsection{Physical derivation}
Let us first perform the analysis from the physics point of view, by asking what data is needed to couple a topological quantum mechanical system with algebra of operators $\mc{B}$ to Chern-Simons theory.  Since the system we are coupling is topological, all local operators are of dimension $0$. The most general dimensionless coupling we can imagine involves coupling the Chern-Simons gauge field $A_t^a$ to local operators $\rho_a \in \mc{B}$. The action of the coupled system takes the form
\begin{equation} 
\op{Exp}\int_{\R^3} CS(A) + \op{PExp}\left( \int_{\R} A_t^a \rho_a( \phi) \right) + \op{Exp}\int_{\R} \mc{L}(\phi) \label{eqn:cs_coupling} 
\end{equation} 
where we have schematically written the fields of the quantum mechanical system we are introducing as $\phi$, and the Lagrangian as $\mc{L}(\phi)$.  The operators $\rho_a$ are, of course, the currents for the gauge symmetry which we use to couple the quantum mechanical system to Chern-Simons theory. As such, local operators $b \in \mc{B}$ transform under gauge symmetry by $b \mapsto [\rho_a,b]$. 

We must ask when \eqref{eqn:cs_coupling} is gauge invariant.  It is obvious that gauge invariance can only hold if the currents $\rho_a$ satisfy the  gauge algebra $[\rho_a,\rho_b] = f^{c}_{ab} \rho_c$, but let us derive this directly.  Let us perform a gauge transformation 
$$\delta A^a = \d \mf{c}^a + f^a_{bc} \mf{c}^b A^c. $$
The path-ordered exponential in  \eqref{eqn:cs_coupling} changes by 
\begin{equation} 
\sum_{n \ge 1} \sum_{i = 1}^n \int_{t_1 \le \dots \le t_n} A_t^{a_1}(t_1) \rho_{a_1}(t_1) \dots \left(\d \mf{c}^{a_i}(t_i) + f^{a_i}_{bc} \mf{c}^b(t_i) A_t^c(t_i) \right) \dots A_t^{a_n}(t_n).\label{eqn:pexp_variation} 
\end{equation} 
Integrating by parts the terms with $\d \mf{c}$ picks up boundary terms, when the points $t_i$, $t_{i+1}$ come together.  There are two ways that this can happen: either $\d \mf{c}$ is on $t_i$, or $\d \mf{c}$ is on $t_{i+1}$.  When two points collide, the corresponding operators $\rho_{a_i}$, $\rho_{a_{i+1}}$ are multiplied according to the algebra structure on $\mc{B}$.  The two types of boundary contribute with opposite signs, so that the contribution of the terms with a $\d \mf{c}$ is 
\begin{equation} 
\sum_{n \ge 1} \sum_{i = 1}^n \int_{t_1 \le \dots \le t_n} A_t^{a_1}(t_1) \rho_{a_1}(t_1) \dots \left( - \mf{c}^{b} A_t^c \rho_{b} \rho_c + \mf{c}^b A_t^c \rho_c \rho_b    \right)(t_i) \dots A_t^{a_n}(t_n). 
\end{equation} 
Inside the brackets, $\mf{c}^{b} A_t^c \rho_{b} \rho_c$ and  $\mf{c}^{b} A_t^c \rho_{c} \rho_b $ correspond to two types of boundary contribution.  

Introducing again the second term in the gauge variation of the path-ordered exponential \eqref{eqn:pexp_variation} , we  find that the path-ordered exponential is gauge invariant provided
\begin{equation} 
\rho_b \rho_c - \rho_c \rho_b = f^a_{bc} \rho_a. 
\end{equation} 
This means that the operators $\rho_a$ give us a homomorphism from the universal enveloping algebra $U(\g)$ to $\mc{B}$, as desired.

\subsubsection{Algebraic derivation}
Let us now sketch the algebraic derivation of the same result. We will be rather brief here, since we will use the physical picture given above as our working definition of Koszul duality. 

First, we need to understand the algebra of local operators of Chern-Simons theory.  Typically, one would say that Chern-Simons theory has no local operators.  This is because the equations of motion for the Chern-Simons gauge field implies that it is locally gauge trivial.  All gauge-invariant operators must therefore vanish on-shell.

This argument implies there are no local operators of ghost number $0$. There are, however, operators of positive ghost number, generated by the $\mf{c}$ ghost, where $\mf{c} \in \mf{g}^*$ as always. \footnote{Here we encounter a slightly subtle point: if our gauge group is compact, and we work non-perturbatively, then the correct thing to do is to drop the $\mf{c}$-ghost itself, and only include its derivatives. Gauge invariance for constant gauge transformations is imposed by simply taking operators invariant under the compact group. Here, we treat Chern-Simons theory as a theory that only knows about the Lie algebra and not the any real form of the group, so that it is natural to include the $\mf{c}$-ghost.}  This operator is not BRST closed, but satisfies the familiar equation
\begin{equation} 
 Q \mf{c}^a =\tfrac{1}{2} f^a_{bc} \mf{c}^b \mf{c}^c . \label{eqn:BRST}
\end{equation} 
In particular, if we take the Lie algebra $\mf{g}$ to be Abelian, so that our Chern-Simons theory is free, the algebra of local operators is simply the exterior algebra $\wedge^\ast \mf{g}^\ast$ on the linear dual of the Lie algebra $\mf{g}$. 

We need to understand what the Koszul dual of this algebra is, and what happens when we re-introduce the BRST operator \eqref{eqn:BRST}. First, let us recall some basic algebraic facts about Koszul duality, which are standard in the mathematics literature but which we do not prove here (see, for instance, \cite{LV} for a nice mathematical introduction). See appendix \ref{app:koszulTFT} for some key definitions in what follows, as well as a review of another physical definition of Koszul duality using topological field theory. 

Suppose that $\mc{A} = \C[\theta^i]$ is the algebra generated by fermionic variables $\theta^i$. Then, using the definition \ref{eq:koszuldef}, it is easy to prove that the Koszul dual $\mc{A}^!$ is generated by bosonic, commuting variables $x_i$.  

Deformations of $\mc{A}$ lead to deformations of $\mc{A}^!$.  Suppose we deform $\mc{A}$ by turning on a differential
\begin{equation} 
\d = \tfrac{1}{2} f^{i}_{jk} \theta^j \theta^k \partial_{\theta_i}. 
\end{equation}
This differential is nilpotent, $\d^2 = 0$, if and only if $f^{i}_{jk}$ are the structure constants of a Lie algebra.  Then, the corresponding deformation of $\mc{A}^{!}$ is given by asking that the variables $x_i$ no longer commute, but satisfy 
\begin{equation} 
[x_i, x_j] = f^{k}_{ij} x_k. 
\end{equation}

Applied to Chern-Simons theory, these results tell us that the Koszul dual of the algebra of operators of Abelian Chern-Simons theory, which is $\wedge^\ast \mf{g}^\ast$, is the symmetric algebra $\Sym^\ast \mf{g}$ on $\mf{g}$, generated by bosonic commuting variables $\rho_a$.  When we turn on the Chern-Simons interaction term, we turn on a BRST operator \eqref{eqn:BRST}, which on the Koszul dual side make the variables $\rho_a$ non-commutative, as desired.

\subsubsection{Koszul duality for four-dimensional Chern-Simons theory}
Let us briefly mention a more algebraically interesting example of the same phenomenon.  In \cite{CWY17}, the authors studied line defects that can be coupled to four-dimensional Chern-Simons theory.  (We refer to \cite{C13, CWY17} for the definition and basic properties of this theory).  The theory is topological in one plane, with coordinates $x,y$, and holomorphic in another plane, with coordinates $z,\zbar$. The line defects are placed on a straight line, say $y = 0$, in the topological plane.  

In \cite{CWY17}, the authors considered coupling the four-dimensional gauge field $\mc{A}$ to a quantum mechanical system living on the line defect, by a coupling of the form
\begin{equation} 
	\op{PExp} \int t_{a,k} \partial_z^k A^a	 
\end{equation}
where $t_{a,k} \in \mc{B}$ are operators. When the gauge theory is treated classically, gauge invariance implies that the $t_{a,k}$ satisfy the relation \begin{equation} 
	[t_{a,k}, t_{b,l}] = f^c_{ab} t_{c,k+ l}. 
\end{equation}
However, there is a two-loop anomaly to gauge invariance, given by the diagram
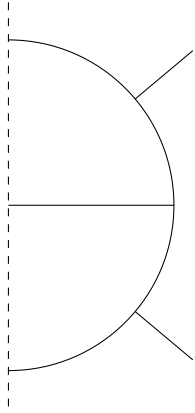
\begin{center}
\begin{tikzpicture}

\draw[dashed](0,2.7) to (0,-2.7);
\draw (0,-2.2) arc (-90:90:2.2);
\draw(2.2,0) to (0,0); 
\draw (40:2.2) to (40:3.2);
\draw (-40:2.2) to (-40:3.2);

\end{tikzpicture}
\end{center}
where the dashed lines indicate the defect and the solid lines are bulk gluons.  This two-loop anomaly can be cancelled by asking that the commutation relation for the operators $t_{a,k}$ is modified, so that they satisfy the relations of the Yangian algebra $Y(\g)$.  This relation is rather complicated, but relates the commutator of $[t_{a,1}, t_{b,1}]$ with a certain cubic expression in the $t_{c,0}$.    

The calculation of \cite{CWY17} tells us that the Koszul dual of the algebra of local operators of four-dimensional Chern-Simons theory is the Yangian. This result was obtained by other methods in \cite{C13, IMZ18}. A similar algebraic relation was derived in \cite{C16, C17} for a $5$-dimensional version of Chern-Simons. 

\subsubsection{Connection between physical and algebraic Koszul duality}
We have explained that the algebraic and physical interpretations of Koszul duality give the same answer.  One can ask in what generality this holds, and why. In the section we will show how to connect the two approaches.

To understand the connection, let us analyze algebraically the data we need to couple the bulk system to the defect system. If we just take the uncoupled product of the two systems, the algebra of local operators on the line is the tensor product $\mc{A} \otimes \mc{B}$ of the bulk and defect algebras of local operators.  To first order, a deformation of the coupled system should be given by the integral over the $t$-line of some local operator $\mc{O}^{(0)}$. If we work to all orders, the deformation takes the form 
\begin{equation} 
	\op{Tr}	\op{PExp} \int_{S^1} \mc{O}^{(0)}.
\end{equation}
(Here we take the path ordered exponential on a circle and the trace in some representation of $\mc{B}$, just like we do for a Wilson loop).

We can ask when this deformation is BRST closed. To first order in $\mc{O}^{(0)}$,  we need that $\mc{O}^{(0)}$ is BRST closed up to total derivative:
\begin{equation} \label{eq:descent}
	Q \mc{O}^{(0)} = \partial_t \mc{O}^{(1)} 
\end{equation}
where $\mc{O}^{(1)}$ is BRST closed.  If we work to all orders in $\mc{O}^{(1)}$ this equation gets replaced by the condition that
\begin{equation} 
	Q \mc{O}^{(0)} = \partial_t \mc{O}^{(1)} + [\mc{O}^{(1)}, \mc{O}^{(0)}] .\label{eqn:gauge_trans} 
\end{equation}
Here the commutator is taken in the tensor product of the algebras of bulk and defect local operators, $\mc{A} \otimes \mc{B}$.  We should compare this equation with the variation of a connection under gauge transformation, where $\mc{O}^{(1)}$ plays the role of the ghost.

Equation \eqref{eqn:gauge_trans} is sufficient to guarantee that the operator built as the path-ordered exponential of $\mc{O}^{(0)}$ is gauge invariant.  To see this, we use an argument almost identical to the familiar argument that shows gauge invariance of a Wilson line.  

Applying the BRST operator to the path ordered exponential, we find   
\begin{equation} 
	\sum_{n \ge 1} \sum_{i = 1}^n \int_{t_1 \le \dots \le t_n} \mc{O}^{(0)} (t_1) \dots \left(\partial_t \mc{O}^{(1)} + [\mc{O}^{(1)}, \mc{O}^{(0)} ] \right)(t_i) \dots \mc{O}^{(0)}(t_n) \label{eqn:pexp_brst_variation} 
\end{equation} 
Integrating by parts the terms with $\partial_t \mc{O}^{(1)}$ picks up boundary terms, when the points $t_i$, $t_{i+1}$ come together.  There are two ways that this can happen: either $\partial_t \mc{O}^{(1)}$  is on $t_i$ or on $t_{i+1}$.  When two points collide, the corresponding operators are multiplied according to the algebra structure on $\mc{A} \otimes \mc{B}$.  The two types of boundary contribute with opposite signs, so that the contribution of the terms with a $\partial_t \mc{O}^{(1)}$  is 
\begin{equation} 
	\sum_{n \ge 1} \sum_{i = 1}^n \int_{t_1 \le \dots \le t_n}\mc{O}^{(0)} (t_1) \dots \left( [\mc{O}^{(0)}, \mc{O}^{(1)}]    \right)(t_i) \dots \mc{O}^{(0)}(t_n). 
\end{equation} 
The second term in the BRST variation \eqref{eqn:pexp_brst_variation} cancels this, as desired. 

Now, let us assume that both our bulk and defect theories are equipped with fermionic symmetries $\what{Q}$ of ghost number $-1$ such that 
\begin{equation} 
	[Q,	\what{Q} ] = \partial_t. 
\end{equation}
(This happens whenever our system arises as the supersymmetric localization of a supersymmetric system. In that case $\what{Q}$ is one of the supercharges of the original system).   Then, it is reasonable to express
\begin{equation} 
	\mc{O}^{(0)} = \what{Q} \mc{O}^{(1)}. 
\end{equation}
Equation \eqref{eqn:gauge_trans}, which guarantees BRST invariance of the path-ordered exponential, holds as long as $\mc{O}^{(1)}$ satisfies the Maurer-Cartan equation
\begin{equation} 
	Q \mc{O}^{(1)} + \tfrac{1}{2} [\mc{O}^{(1)}, \mc{O}^{(1)} ] = 0. 
\end{equation}

We have sketched that BRST invariant couplings to a quantum mechanical system with algebra of operators $\mc{B}$ are the same as solutions to the Maurer-Cartan equation in $\mc{A} \otimes \mc{B}$.  We let $\op{MC}(\mc{A} \otimes \mc{B})$ denote the set \footnote{Normally one considers solutions to the Maurer-Cartan equation up to a notion of equivalence, which physically corresponds to a field redefinition.  Here we will assume, for simplicity, that we take a model for $\mc{A}$ which is generated by operators of ghost number $1$ and higher, and $\mc{B}$ is entirely in ghost number $0$.} of solutions to the Maurer-Cartan equation.  

In the mathematics literature it is known that there is a bijection\footnote{The precise statement also involves augmentations, which we discuss in the next paragraph.} 
\begin{equation} 
	\op{MC}(\mc{A} \otimes \mc{B}) \to \op{Hom}(\mc{A}^!, \mc{B})  
\end{equation}
where $\mc{A}^!$ is the Koszul dual of $\mc{A}$. 

In other words, the solutions to the Maurer-Cartan equation, which parameterize the space of possible deformations of the uncoupled line defect into an anomaly-free (at least perturbatively) coupled defect, are in one-to-one correspondence with homomorphisms $ \op{Hom}(\mc{A}^!, \mc{B})$ from the Koszul dual algebra. 

We should comment briefly on the role of augmentations, which play an important role in the mathematics definition.  An augmentation of $\mc{A}$ is a maximal two-sided ideal $m \subset \mc{A}$, preserved by the differential. This is equivalent to a homomorphism of dg algebras $\mc{A} \to \C$.    A careful mathematical treatment would define the Koszul dual of an algebra $\mc{A}$ as depending on the augmentation.  The Maurer-Cartan functor we should use should really by $\op{MC}(m \otimes B)$, where $m$ is the maximal ideal.  In all the examples of interest to us, we can model $\mc{A}$ by a differential graded algebra sitting entirely in positive cohomological degree, with only the identity operator in ghost number $0$.   If we assume that $\mc{A}$ is of this form, and that $\mc{B}$ is entirely in ghost number $0$, then there is a unique augmentation of $\mc{A}$, and the formulation in terms of the Maurer-Cartan set $\op{MC}(\mc{A} \otimes \mc{B})$ is correct. 

Physically, the augmentation (see appendix \ref{app:koszulTFT}) of $\mc{A}$ corresponds to a choice of vacuum and the augmentation of $\mc{A}^{!}$  corresponds to coupling to the trivial defect. This provides the connection between the physical and mathematical pictures of Koszul duality.

Notice also that equation \ref{eq:descent} is of the type that appears in topological descent: one trades a (say) local 0-form operator of ghost number 1, $\mc{O}^{(1)}$, for a one-form operator of ghost number 0 that can then be integrated over a non-trivial one-cycle: in our case, the support of the line defect. Anticipating the appearance of Koszul duality in our holographic system, then if our gravity theory has local operators supported only in nonzero ghost number then this descent procedure allows one to take an operator in the product of the uncoupled theories bulk/defect theories of ghost number 1 (since the bulk operator has ghost number 1 and the defect operator has ghost number 0) and solve for a 1-form of ghost number 0, valued in operators of the coupled system, on the line. (In the BV formalism, solving the Maurer-Cartan equation is equivalent to the 1-form operator we have obtained by descent satisfying the quantum master equation. Hence, it is a consistent deformation of the bulk/defect system). In turn, this is equivalent to a choice of homomorphism from the Koszul dual algebra of the bulk to the defect algebra.

\subsection{Koszul duality for chiral algebras}

In the example we are interested in, we wish to perform Koszul duality for chiral algebras (a.k.a.\ vertex algebras\footnote{Our vertex algebras may not have a stress-energy tensor.}), not just associative algebras.  Unfortunately a general mathematical framework for this is not available\footnote{The mathematically inclined reader should be cautioned that the chiral Koszul duality discussed in \cite{FG12} is \emph{not} what we are considering here.  In that paper, Francis and Gaitsgory analyze a kind of meta-Koszul duality, which interchanges two different ways of describing the concept of chiral algebra.  The Koszul duality we study, in contrast, is an operation which turns one chiral algebra into a new chiral algebra.}.    

We can, however, simply \emph{define} Koszul duality for chiral algebras mimicking the physical definition given above.  Thus, let us consider any field theory (which will not be Lorentz invariant, but will be translation invariant) on $\C \times \R^n$. We assume that the theory is holomorphic along $\C$. The theories we consider will be some mixture of holomorphic and topological in the remaining $\R^n$ directions.  We let $\mc{A}$ be the algebra of local operators of the theory. We view this as a differential graded vertex algebra, with differential the BRST operator and vertex algebra structure coming from the OPE along $\C$.  

We then define the Koszul dual $\mc{A}^!$ to be the universal vertex algebra that can be coupled to our theory as the algebra of operators of a defect wrapping $\C$.  In other words, $\mc{A}^!$ is characterized by the universal property that a homomorphism of vertex algebras from $\mc{A}^!$ to the algebra of operators $\mc{B}$ of some chiral theory is the same as a way of coupling the chiral theory as a defect in our original theory on $\C \times \R^n$.

We expect that this definition is reasonable only under certain additional hypothesis: we ask that $\mc{A}$ is a deformation of the VOA $\mc{A}^{free}$ of a free theory, and that the cohomology of $\mc{A}^{free}$ is generated by a collection of operators $\mf{c}_i$ of ghost number $1$ and trivial OPE.   When we deform to the interacting theory, the operators $\mf{c}_i$ will acquire a non-trivial BRST operator and OPEs. It should be straightforward to relax some of these assumptions (for instance, in a theory with higher form gauge symmetries, one needs additional generators at ghost number greater than 1, the ghosts-of-ghosts), but we will not need to do so here.

\begin{remark}
	It is very plausible that coupling a chiral algebra $\mc{B}$ to the bulk theory is given by a Maurer-Cartan element in a dg Lie algebra built from the vertex algebra $\mc{A} \otimes \mc{B}$.  The relevant Lie algebra is the quotient of $\mc{A} \otimes \mc{B}$ by those operators which are total derivatives, with Lie bracket giving by the coefficient of $z^{-1}$ in the OPE.  Indeed, Si Li \cite{L16} proved a version of this statement in a special case. 

	This suggests a purely algebraic definition of Koszul duality for chiral algebras: the Koszul dual $\mc{A}^!$ is the chiral algebra with the universal property that a map $\mc{A}^! \to \mc{B}$ is a Maurer-Cartan element in the dg Lie algebra built from $\mc{A} \otimes \mc{B}$, in the sense of the previous paragraph.  We will not pursue this here. 
\end{remark}

In order to demonstrate how to work with our definition of the Koszul dual algebra as the universal defect algebra,  let us illustrate how it works when the bulk theory we consider is holomorphic Chern-Simons\footnote{Holomorphic Chern-Simons suffers from a one-loop rectangle anomaly, and so is not defined at the quantum level without coupling to some version of Kodaira-Spencer theory \cite{CL3}.  We will perform some tree level calculations that happen to be insensitive to this anomaly.    }   on $\C \times \C^2$, with gauge algebra some simple Lie algebra $\mf{g}$. This is a $3$-complex dimensional holomorphic theory, and so has similar features to the effective twisted supergravity theory that is holographically dual to the $\Sym^N T^4$ CFT.

The fundamental field of holomorphic Chern-Simons theory on $\C^3$ is a $(0,1)$ gauge field
\begin{equation} 
	A = A_{\zbar} \d \zbar + A_{\wbar_1} \d \wbar_1 + A_{\wbar_2} \d \wbar_2,
\end{equation}
where we use holomorphic coordinates $z,w_1,w_2$.  The Lagrangian is
\begin{equation} 
	\int \d z \d w_1 \d w_2 CS(A). 
\end{equation}
The equations of motion of holomorphic Chern-Simons theory are that $F^{0,2}(A) = 0$. Locally, every field satisfying the equations of motion is gauge trivial. The trivial field configuration is preserved by a very large group of gauge transformations, namely the holomorphic maps to the group $G$.

Because of this, all the local operators in holomorphic Chern-Simons are generated by the ghosts $\c^a$ and their holomorphic derivatives. The derivatives $\partial_{w_1}^k \partial_{w_2}^l \c^a$ are Virasoro primaries.  At the classical level, the operators all have trivial OPE and a BRST operator
\begin{equation} 
	Q  \partial_{w_1}^k \partial_{w_2}^l   \c^a = \tfrac{1}{2} f^a_{bc} \partial_{w_1}^k \partial_{w_2}^l \left(  \c^b \c^c \right). 
\end{equation}
By analogy with the associative algebra case, one would hope that the Koszul dual of the algebra of operators of holomorphic Chern-Simons theory is a chiral analog of a universal enveloping algebra.  This is what we will find: the Koszul dual is generated by currents $J_a[k,l]$, dual to $\partial_{w_1}^k \partial_{w_2}^l \c^a$, which have the OPE
\begin{equation} 
	J_a[k,l] (0) J_b[r,s](z) \sim \frac{1}{z} f^c_{ab} J_c[k+r,s+l](0). 
\end{equation}

\subsubsection{Koszul duality for free holomorphic Chern-Simons theory} 
Now let us turn to calculating this Koszul dual. Suppose we couple to a chiral theory living on the plane $w_1 = w_2 = 0$, with algebra of operators $\mc{B}$.  Our goal is to write down the most general way to couple this theory to holomorphic Chern-Simons theory.  

Let us start with the case that the Lie algebra $\mf{g}$ is Abelian.  Then, the most obvious coupling we can write down involves coupling the gauge field $A^a_{\zbar}$ to a current $J_a \in \mc{B}$ (where $a$ is a Lie algebra index).  The coupling
\begin{equation} 
	\int_{z} A^a_{\zbar} J_a \label{eqn:hcs_coupling_initial} 
\end{equation}
is BRST invariant.  To see this, we assume that $\mc{B}$ is concentrated in ghost number $0$, and so has no BSRT operator.  Then, the BRST variation of $A^a_{\zbar}$ is by $\partial_{\zbar} \mf{c}^a$, where $\mf{c}$ is the ghost.  Integration by parts, together with the fact that $\partial_{\zbar} J_a = 0$, tells us that \eqref{eqn:hcs_coupling_initial} is BRST invariant.

 A more general class of couplings is obtained by  coupling the derivatives $\partial_{w_1}^{k_1} \partial_{w_2}^{k_2} A^a_{\zbar}$ to a sequence of currents $J_a[k_1,k_2] \in \mc{B}$, so that the term coupling the two systems looks like
\begin{equation} \label{eqn:hcs_coupling_general} 
	\sum_{k_1,k_2 \ge 0} \int \tfrac{1}{k_1! k_2!} \partial_{w_1}^{k_1} \partial_{w_2}^{k_2} A^a_{\zbar} J_a[k_1,k_2]. 
\end{equation}
This expression is still BRST invariant, because the BRST variation of $ \partial_{w_1}^{k_1} \partial_{w_2}^{k_2} A^a_{\zbar}$ is a $\partial_{\zbar} \partial_{w_1}^{k_1} \partial_{w_2}^{k_2} \mf{c}^a$.  

It will be helpful to understand how the operator $J_a[k_1,k_2]$ transforms under the positive part of the Virasoro algebra.   This operator is sourced by coupling to holomorphic Chern-Simons theory with background gauge field 
\begin{equation} 
	A_{\zbar} = \mbf{t}_a \delta_{z = 0}w_1^k w_2^l  
\end{equation}
where $\mbf{t}_a$ is a Lie algebra element.  

Holomorphic Chern-Simons theory is preserved by the Lie algebra of divergence-free vector fields on $\C^3$.  Holomorphic vector fields on $\C$ are represented by the vector fields 
\begin{equation} 
	L_{-n} = -z^{n+1} \partial_z + \tfrac{1}{2}(n+1) \left( w_1 \partial_{w_1} + w_2 \partial_{w_2}  \right)  
\end{equation}
(which is divergence free). It is elementary to check that
\begin{equation}
	L_{-n} \left(w_1^{k_1} w_2^{k_2} \delta_{z = 0} \right)= \begin{cases}
		0 & \text{ if } n > 0 \\
		\left(\frac{1}{2} (k_1 + k_2) + 1 \right)\left(w_1^{k_1} w_2^{k_2} \delta_{z = 0} \right) & \text{ if } n = 0. 
	\end{cases}
\end{equation}
From this we see that $J_a[k_1,k_2]$ are Virasoro primaries of dimension $\tfrac{1}{2}(k_1 + k_2) + 1$.

It turns out that \eqref{eqn:hcs_coupling_general} represents the most general possible coupling. To understand this, note that the equations of motion imply $F_{\wbar_1 \wbar_2}(A)= 0$.  We can use this equation together with a gauge transformation to set $A_{\wbar_j}= 0$. Once we have done this, we can only couple to polynomials in $A_{\zbar}$ and its derivatives.  Derivatives with respect to $\wbar_i$ vanish, using the equation of motion $F_{\wbar_i \zbar}= 0$.  The Lagrangian must transform as a $(1,1)$-form on the $z$-plane, which forces us to have only a single copy of $A_{\zbar}$ and no $\zbar$-derivatives. Any derivatives of $A_{\zbar}$ by $z$ can, by integration by parts, be absorbed into the operator in $\mc{B}$ to which we are coupling.

To sum up, we have found that coupling to \emph{free} holomorphic Chern-Simons theory amounts to specifying an infinite number of Virasoro primaries  $J_a[k_1,k_2] \in \mc{B}$, of dimension $1 + (k_1 + k_2)/2$.  

For Abelian holomorphic Chern-Simons theory, these currents have non-singular OPEs with each other. To see this, let us exponentiate the coupling between the defect and bulk theories. It is represented by the following expression, which is analogous to the path-ordered exponential:
\begin{equation} 
	\sum_{ n \ge 0} \tfrac{1}{n!} \int_{z_1,\dots,z_n \in \C} \prod_{i = 1}^n \left( 	\int \tfrac{1}{k^i_1! k^i_2!} \partial_{w_1}^{k^i_1} \partial_{w_2}^{k^i_2} A^{a_i}_{\zbar}(z_i) J_{a_i}[k_1,k_2](z_i) \right). \label{eqn:chiral_pexp} 
\end{equation}
If the operators $J_a[k_1,k_2]$ have non-trivial OPEs with each other,  this expression can have short-distance divergences.  We can perform a path-splitting regularization by asking that the distance between any two $z_i$ is at least $\eps$, where $\eps$ is a small regulator.  

We calculate the OPEs between the operators $J_a[k_1,k_2]$ by asking that \eqref{eqn:chiral_pexp} is BRST invariant. The BRST variation is
\begin{equation} 
	Q A^a_{\zbar} = \partial_{\zbar} \mf{c}^a
\end{equation}
If we insert the BRST variation into equation \eqref{eqn:chiral_pexp}, the term involving $\partial_{\zbar} \mf{c}^a (z_i)$ can be removed by integration by parts, at the price of introducing boundary terms where $\abs{z_i - z_j} = \eps$.  The boundary term is -- suppressing the points $z_k$ for $k \neq i,j$ --
\begin{equation} 
	- \int_{\abs{z_i - z_j} = \eps}  \tfrac{1}{k^i_1! k^i_2!}  \tfrac{1}{k^j_1! k^j_2!}  \partial^{k_1^i}_{w_2} \partial^{k_2^i}_{w_2} \mf{c}^{a_i}(z_i)   J_{a_i}[k_1^i, k_2^i] (z_i)   \partial^{k_1^j}_{w_2} \partial^{k_2^j}_{w_2} A_{\zbar}^{a_j}(z_j) J_{a_j}[k_1^j,k_2^j] (z_j)  
\end{equation}
Asking that this boundary term vanishes amounts to asking that the operators $J_a[k_1,k_2]$ have non-singular OPE with each other.  Indeed, any singularity in the OPE between these operators will yield factors of $(z_1 - z_2)^{-n}$ in the above contour integral. Performing the integral will give us a non-zero expression of the form 
\begin{equation} 
	\partial^{k_1^i}_{w_2} \partial^{k_2^i}_{w_2} \mf{c}^{a_i}(z_i)   \partial^{k_1^j}_{w_2} \partial^{k_2^j}_{w_2} \partial_{z_i}^{n-1} A_{\zbar}^{a_j}(z_i) 
\end{equation}
times some operator in the chiral algebra $\mc{B}$.    We thus see that BRST invariance of the coupling requires that the operators $J_a[k_1,k_2]$ all have non-singular OPE with each other.

\subsubsection{Koszul duality for interacting holomorphic Chern-Simons theory}
Now let us turn on the interaction in holomorphic Chern-Simons. We will see that this modifies the condition that the operators $J_a[k_1,k_2]$ have trivial OPE.  To see this, we note that the BRST variation for the interacting theory is
\begin{equation} 
	Q A^a_{\zbar} = \partial_{\zbar} \mf{c}^a + f^a_{bc} \mf{c}^b A^c_{\zbar}.  
\end{equation}
In the interacting theory, we need the second term in the BRST variation to cancel with the boundary term discussed above.  This cancellation will hold provided that 
\begin{multline} 
	\int_{z} \sum \tfrac{1}{k_1!k_2!} \partial_{w_1}^{k_1} \partial_{w_2}^{k_2} \left(f^{a}_{bc} \mf{c}^b(z) A^c_{\zbar}(z) \right)  J_{a}[k_1,k_2](z) \\ = \int_{z} \oint_{\abs{u} = \eps}\sum  \tfrac{1}{l_1! l_2!}  \tfrac{1}{m_1! m_2!}  \partial^{l_1}_{w_2} \partial^{l_2}_{w_2} \mf{c}^{d}(z+u)   J_{d}[l_1, l_2] (z+u)   \partial^{m_1}_{w_2} \partial^{m_2}_{w_2} A_{\zbar}^{e}(z) J_{e}[m_1,m_2] (z)\label{eqn:ope_equality} 
\end{multline}
where the indices $k_i$, $l_i$, $m_i$ are summed over.

To obtain constraints on the OPEs, we can set 
\begin{equation} 
	\begin{split}
		A_{\zbar} &= \mbf{t}_b w_1^{m_1} w_2^{m_2} \delta_{z = 0}\\
		\mf{c} &= \mbf{t}_a w_1^{l_1} w_2^{l_2} z^r. 
	\end{split}\label{eqn:testfields}
\end{equation}
Inserting these values for $A$, $\mf{c}$ into equation \eqref{eqn:ope_equality} we find 
\begin{equation} 
	f^{a}_{bc} J_a[l_1 + m_1, l_2 + m_2] \delta_{r = 0} = \oint_{\abs{u} = \eps} u^r J_b[l_1,l_2](0) J_c[m_1,m_2](u) \d u.  
\end{equation}
Taking $r > 0$, we find the OPE between $J_b[l_1,l_2]$ and $J_c[m_1,m_2]$ has at most a first-order pole.  Taking $r = 0$, and changing $u$ to $z$, we find that the coefficient of $1/z$ in the OPE is $f^{a}_{bc} J_a[l_1 + m_1, l_2 + m_2]$.  That is, we have the OPE
\begin{equation} 
	J_b[l_1,l_2] (0) J_c[m_1,m_2](z) \sim \frac{1}{z} f^{a}_{bc} J_a[l_1 + m_1, l_2 + m_2] .  
\end{equation}

We have shown that this OPE is \emph{necessary} for the coupling between holomorphic Chern-Simons and the chiral defect to be anomaly free. It is also sufficient, as long as we treat holomorphic Chern-Simons at the classical level. This can be proved by a cohomological argument. The key point is the following.  The anomaly is given by the integral over the plane $w_i = 0$ of some Lagrangian of ghost number $1$ which depends on the fields of holomorphic Chern-Simons, multiplied by local operators in the chiral algebra $\mc{B}$.  Any expression involving the $\wbar_i$ or $\zbar$ derivatives of $A$ and $\mf{c}$ is automatically BRST exact and does not contribute to the BRST cohomology class of the anomaly.  The most general anomaly only involves $w_i$ and $z$ derivatives, and can be detected by inserting fields $A$ and $\mf{c}$ of the form in \eqref{eqn:testfields}. 

\subsection{Feynman diagram interpretation of Koszul duality}
So far, we have explained how to compute the Koszul dual of the algebra of operators of holomorphic Chern-Simons theory at the classical level.  To be able to work at the quantum level, we need to understand how to compute the anomalies in the coupled system using Feynman diagrams.    

\begin{figure}
	\begin{tikzpicture}
		\node[circle, draw] (A) at (0,0) {$J$};
		\node (B) at (2,0)  {$A_{\zbar}$};
		\draw[decorate, decoration={snake}] (A) --(B);
		\draw (A) -- (0,1); 
		\draw (A) -- (0,-1);
	\end{tikzpicture}
	\caption{A schematic depiction of the coupling between the bulk and defect theories.The vertical line indicates the defect plane. \label{fig:defectdiag}} 
\end{figure}
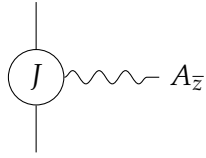

\begin{figure}
	\begin{tikzpicture}
	\begin{scope}
		\node[circle, draw] (J1) at (0,1) {$J$};
		\node[circle, draw] (J2) at (0,-1) {$J$};
		\node (A1) at (2,1)  {$A_{\zbar}$};
		\node (A2) at (2,-1)  {$A_{\zbar}$};
		\draw[decorate, decoration={snake}] (J1) --(A1);
		\draw[decorate, decoration={snake}] (J2) --(A2);
		\draw (0,2) -- (J1) --(J2) -- (0,-2); 
	\end{scope}	
	\begin{scope}[shift={(4,0)}];
		\node[circle, draw] (J) at (0,0) {$J$};
		\node (A1) at (3,1)  {$A_{\zbar}$};
		\node (A2) at (3,-1)  {$A_{\zbar}$};
		\node[circle,draw,fill=black, minimum size = 0.2pt]  (V) at (1.5,0) {};  
		\draw[decorate, decoration={snake}] (J) -- (1.5,0) --  (A1);
		\draw[decorate, decoration={snake}] (1.5,0) --(A2);
		\draw (0,2) -- (J)-- (0,-2);
	\end{scope}
	\end{tikzpicture}
	\caption{Cancellation of the gauge anomaly of these two diagrams leads to the equation for the OPEs of the currents $J[k,l]$ presented above. \label{fig:cancel}}
\end{figure}
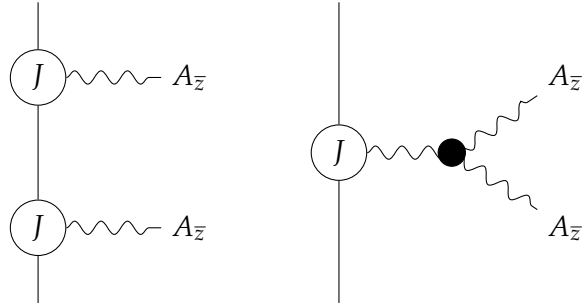

Let us illustrate the idea by recasting the calculation of the Koszul dual of the algebra of operators of holomorphic Chern-Simons in the language of Feynman diagrams.  The coupling between the defect theory and the bulk theory will be illustrated as in figure \ref{fig:defectdiag}.
The equation we derived above for the OPE of the currents $J[k,l]$ is equivalent to the cancellation of the gauge variation of the two Feynman diagrams depicted in figure \ref{fig:cancel}.

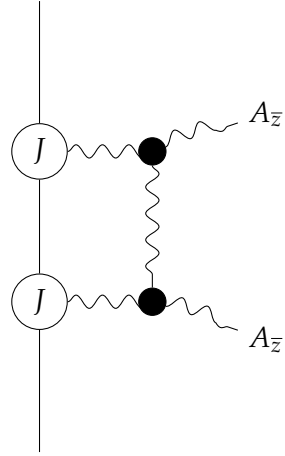
\begin{figure}
	\begin{tikzpicture}	
		\node[circle, draw] (J1) at (0,1) {$J$};
		\node[circle, draw] (J2) at (0,-1) {$J$};
		\node (A1) at (3,1.5)  {$A_{\zbar}$};
		\node (A2) at (3,-1.5)  {$A_{\zbar}$};

		\node[circle,draw,fill=black ]  (V1) at (1.5,1) {}; 
		\node[circle,draw,fill=black]  (V2) at (1.5,-1) {};  
		\draw[decorate, decoration={snake}] (J1) -- (V1) --  (A1);
		\draw[decorate, decoration={snake}] (J2) -- (V2) -- (A2);
		\draw[decorate, decoration={snake}] (V1) -- (V2); 
		\draw (0,3) -- (J1) -- (J2) -- (0,-3);

	\end{tikzpicture}
	\caption{This diagram has a gauge anomaly. \label{fig:anomaly}}
\end{figure}

If we go beyond the tree level, one finds that gauge anomalies can be cancelled only if the currents $J[k,l]$ have very particular quantum-corrected OPEs.  For example, one can show (we will not analyze the explicit integral here) that the diagram in figure \ref{fig:anomaly} has a gauge anomaly.
The gauge anomaly is proportional to
\begin{equation} 
	\hbar \int_{w_1 = w_2 = 0} \eps_{ij} \left( \partial_{w_i} A_{\zbar}^a \right) (\partial_{w_j} \mf{c}^b ) K^{fe}  f^c_{ae} f^d_{bf}   J_c J_d    + \dots 
\end{equation}
where the ellipses indicate terms with more than two derivatives applied to the bulk fields and $K^{fe}$ is the Killing form on $\mf{g}$.   We would like this anomaly to be cancelled by the first Feynman diagram in figure \ref{fig:cancel}.   A necessary condition for this to happen is that the classical OPE of the operators $J[1,0]$ and $J[0,1]$ acquires a quantum correction:
\begin{equation} 
	J_a [1,0] (0)  J_b [0,1] (z) \simeq \frac{1}{z} f^c_{ab} J_c[1,1] + \hbar \frac{1}{z}  K^{fe} f_{ae}^c f_{bf}^d J_c[0,0] J_d[0,0].  
\end{equation}
If we have such a quantum-corrected OPE, then the gauge variation of the expression
\begin{equation} 
	\int_{z_1, z_2} J_a[1,0] (z_1) \partial_{w_1}A_{\zbar}^a(z_1) J_b[0,1] (z_2)  
\end{equation}
gives us, at order $\hbar$, 
\begin{equation} 
	\hbar \int_{w_1 = w_2 = 0} \eps_{ij} \left( \partial_{w_i} A_{\zbar}^a \right) (\partial_{w_j} \mf{c}^b ) K^{fe}  f^c_{ae} f^d_{bf}   J_c J_d    
\end{equation}
which cancels the anomaly from the diagram in figure \ref{fig:anomaly}.

We have sketched this case to indicate how to compute the Koszul dual algebra at the quantum level.  We work order by order in perturbation theory, and at each order study which connected\footnote{Connected in the sense that any two vertices are linked by bulk propagators.} diagrams fail to be gauge invariant.  The only diagrams that can contribute an anomaly are those which have two external lines at which we place gauge fields, and are connected to the defect theory by an arbitrary number of vertices\footnote{The gauge variation of a diagram with more than $2$ external lines would be a quantity which depends quadratically on the gauge field and linearly on the ghost. Every potential gauge anomaly of this nature can be made BRST exact in a rather trivial way.  }.  If we have such an anomalous diagram which is connected to the defect at $k$ vertices, and which occurs at order $l$ in the loop expansion, then the anomaly can be cancelled by asking that at order $\hbar^l$ the OPE of two currents $J[r,s]$ can be expressed in terms of $k$ currents, and possibly their $z$ derivatives.

\subsection{Deformed Koszul duality}
For the purposes of holography, we are interested in a deformation of Koszul duality where one includes the effect of backreaction.  In this section we will illustrate how to do this, continuing with the example of holomorphic Chern-Simons theory.

Holomorphic Chern-Simons theory is not, of course, a gravitational theory. However, it does couple to a gravitational theory: Kodaira-Spencer theory.  We can study holomorphic Chern-Simons theory in the background of the Kodaira-Spencer field sourced by a brane wrapping $\C \subset \C^3$.  This field is
\begin{equation} 
	\mu_{BR} = 	N \frac{\eps^{ij} \wbar_i \d \wbar_j   }{ (w_1 \br{w}_1 + w_2 \br{w}_2)^2  } \partial_z.  
\end{equation}
Here $N$ indicates the number of $D1$ branes we place in the topological string, wrapping the curve $w_i = 0$.  The Beltrami differential $\mu_{BR}$  satisfies
\begin{equation} 
	\dbar \mu_{BR} = N \delta_{w_i, \br{w}_i = 0} \partial_z. 
\end{equation}
This is why it is the field sourced by the brane. This Beltrami differential deforms $\C^3 \setminus \C$ to the manifold $SL_2(\C)$ \cite{CostelloGaiotto}.

This Beltrami differential couples to holomorphic Chern-Simons theory by 
\begin{equation} 
	S_{BR} = \tfrac{1}{2} \int_{\C^3}  A^a \mu_{BR}  A^a \d z \d w_1 \d w_2 =  -\tfrac{1}{2}	N \int_{\C^3} \frac{\eps^{ij} \wbar_i \d \wbar_j   }{ (w_1 \br{w}_1 + w_2 \br{w}_2)^2  }  A^a \partial_z A^a.   
\end{equation}
(We also modify the gauge transformations by deforming the $\dbar$ operator using the Beltrami differential). 

It is important to note that, although $S_{BR}$ defines a gauge-invariant deformation of holomorphic Chern-Simons theory away from $w_i = 0$, there is a gauge anomaly localized on the 
complex line $w_i = 0$. This is because $\dbar \mu_{BR} \neq 0$. Indeed, the gauge variation of $S_{BR}$ is
\begin{equation} 
	\int A^a (\dbar \mu_{BR}) \c^a = \int_{z} A^a_{\zbar} \partial_z \c^a. 
\end{equation}

We incorporate the backreaction into the Koszul duality as follows. Instead of asking for universal chiral algebra which couples to the bulk gravitational theory in an anomaly-free way, but we ask for the universal chiral algebra \emph{whose anomaly cancels that of } $S_{BR}$.  


When we include the backreaction, the Feynman diagrams whose anomalies compute the Koszul dual algebra get a new ingredient. We use a dotted line to indicate the field sourced by the brane, which deforms us away from flat space, as in figure \ref{fig:br}.   This field is sourced by the identity operator on the defect.  
\begin{figure}
	\begin{tikzpicture}
	\begin{scope}
		
		\node (A1) at (3,1) {};
		\node (A2) at (3,-1) {};
		\node[circle,draw,fill=black, minimum size = 0.2pt]  (V) at (1.5,0) {};  
		\draw[dashed] (0,0) -- (1.5,0); 
		\draw[decorate, decoration={snake}] (1.5,0) --  (A1);
		\draw[decorate, decoration={snake}] (1.5,0) -- (A2);
		\draw (0,2) -- (0,-2);
	\end{scope}

		\begin{scope}[shift={(5,0)}]		
			\node (A1) at (3,1.5) {};
			\node (A2) at (3,-1.5){};

		\node[circle,draw,fill=black]  (V1) at (1.5,1) {}; 
		\node[circle,draw,fill=black]  (V2) at (1.5,-1) {};  
		\draw[decorate, decoration={snake}]  (V1) --  (A1);
		\draw[decorate, decoration={snake}] (V1) -- (V2) -- (A2);
		\draw[dashed](0,-1) -- (V2);
		\draw[dashed] (0,1) -- (V1);

		\draw (0,2) -- (0,-2);	
	\end{scope}
	\end{tikzpicture}
	\caption{Some of the diagrams contributed by the backreaction. All diagrams of interest have two external lines, at which we place gravitational fields.  Here, we depict diagrams where there are no operators placed on the brane: these would correspond to wavy lines connected to the brane.  Summing over all diagrams such as those depicted will give rise to the propagator of a field in the backreacted geometry.}  
\end{figure}
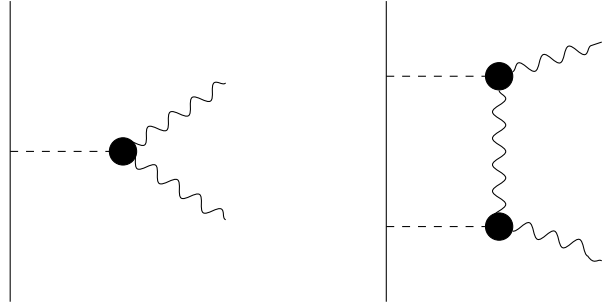

To compute the deformed Koszul dual algebra, we use the same strategy as we did before.  Order by order, we compute anomalies to diagrams including the new type of edge, on which we place the field sourced by the brane.  If such a diagram has an anomaly, we change the OPEs of the currents which couple to the bulk fields so that the anomaly is cancelled.

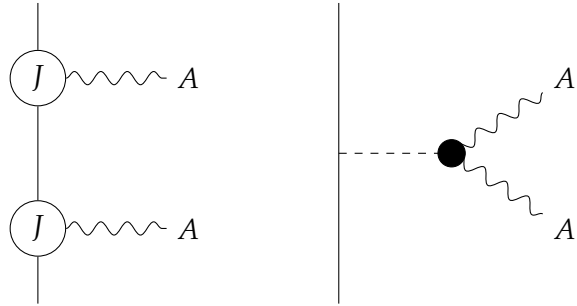
\begin{figure}
	\begin{tikzpicture}
	\begin{scope}
		\node[circle, draw] (J1) at (0,1) {$J$};
		\node[circle, draw] (J2) at (0,-1) {$J$};
		\node (A1) at (2,1)  {$A$};
		\node (A2) at (2,-1)  {$A$};
		\draw[decorate, decoration={snake}] (J1) --(A1);
		\draw[decorate, decoration={snake}] (J2) --(A2);
		\draw (0,2) -- (J1) --(J2) -- (0,-2); 
	\end{scope}	
	\begin{scope}[shift={(4,0)}];
		\node (A1) at (3,1)  {$A$};
		\node (A2) at (3,-1)  {$A$};
		\node[circle,draw,fill=black, minimum size = 0.2pt]  (V) at (1.5,0) {};  
		\draw[dashed] (0,0) -- (1.5,0);
		\draw[decorate, decoration={snake}] (1.5,0) --(A1);
		\draw[decorate, decoration={snake}] (1.5,0) --(A2);
		\draw (0,2)-- (0,-2);
	\end{scope}
	\end{tikzpicture}
	\caption{Cancellation of the gauge anomaly of these two diagrams, where the dashed line indicates the Beltrami differential field, gives a central extension to the current $J[0,0]$. \label{fig:central} } 
\end{figure}
The first two relevant diagrams are depicted in figure \ref{fig:central}.  The second diagram indicates the Beltrami differential field $\mu_{BR}$ interacting with two bulk holomorphic Chern-Simons gauge fields. 

The coupling is, as we have seen above,
\begin{equation} 
	\tfrac{1}{2} \int_{\C^3}  A^a \mu_{BR}  A^a \d z \d w_1 \d w_2 =  -\tfrac{1}{2}	N \int_{\C^3} \frac{\eps^{ij} \wbar_i \d \wbar_j   }{ (w_1 \br{w}_1 + w_2 \br{w}_2)^2  }  A^a \partial_z A^a.   
\end{equation}
The gauge variation of this is
\begin{equation} 
	 \int_{z} A^a_{\zbar} \partial_z \c^a. 
\end{equation}
Note that this expression does not couple to whatever fields we have placed on the defect wrapping the $z$-plane, so we can think of this expression as being accompanied by the identity operator in the chiral algebra of the defect.

\begin{figure}
	\begin{tikzpicture}
	\begin{scope}
		\node[circle, draw] (J1) at (0,1) {$J$};
		\node[circle, draw] (J2) at (0,-1) {$J$};
		\node (A1) at (2,1)  {$A$};
		\node (A2) at (2,-1)  {$A$};
		\draw[decorate, decoration={snake}] (J1) --(A1);
		\draw[decorate, decoration={snake}] (J2) --(A2);
		\draw (0,2) -- (J1) --(J2) -- (0,-2); 
	\end{scope}

	\begin{scope}[shift={(5,0)}]		
		\node (A1) at (3,1.5) {};
		\node (A2) at (3,-1.5){};

		\node[circle,draw,fill=black]  (V1) at (1.5,1) {}; 
		\node[circle,draw,fill=black]  (V2) at (1.5,-1) {};  
		\draw[decorate, decoration={snake}]  (V1) --  (A1);
		\draw[decorate, decoration={snake}] (V1) -- (V2) -- (A2);
		\draw[dashed](0,-1) -- (V2);
		\draw[dashed] (0,1) -- (V1);

		\draw (0,2) -- (0,-2);	
	\end{scope}

	\end{tikzpicture}
	\caption{Cancellation of the gauge anomaly of these two diagrams gives the two-point function $\ip{J^a[1,0] J^a[0,1]}$. \label{fig:central2} } 
\end{figure}
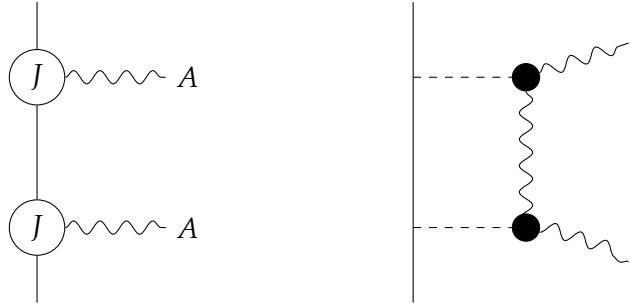

We would like this expression to cancel with the gauge variation of the coupling to the defect, given as before by
\begin{equation} 
	\exp \left( \sum  J^a[k,l] \frac{1}{k! l!}  \partial_{w_1}^{l} \partial_{w_2}^{l} A_{\zbar}^a    \right) 
\end{equation}
Since the anomaly associated to the Beltrami differential has no $w$-derivatives, only the coupling to the operators $J[0,0]$ will be relevant. Further, only the term involving two such operators can appear, as we need to cancel an anomaly involving one bulk ghost and one bulk gauge field.  Therefore, the OPEs of $J^a[0,0]$ must be such that the gauge variation of 
\begin{equation} 
	\tfrac{1}{2}\int_{z_1,z_2} J^a[0,0] (z_1) J^b[0,0](z_2) A_{\zbar}^a(z_1)A_{\zbar}^b(z_2) 
\end{equation}
is
\begin{equation} 
	\int_{z} N \op{Id} A_{\zbar}^a(z) \partial_z \c^a(z). 
\end{equation}
By integrating by parts and using the identity
\begin{equation} 
	\dbar \left( \frac{1}{(z_1 - z_2)^2}  \right)  = \partial_{z_2} \delta_{z_1 = z_2} 
\end{equation}
(where factors of $\pi$ have been absorbed into various normalizations) we find that the currents $J^a[0,0]$ must have the term 
\begin{equation} 
	J^a[0,0] (0) J^a[0,0](z) \sim N \op{Id} / z^2 + \dots 
\end{equation}
in their OPE. Incorporating what we found earlier, the full OPE of these currents is given by that of the Kac-Moody algebra,
\begin{equation} 
	J^a[0,0] (0) J^b[0,0](z) \sim f^{ab}_c \frac{1}{z} J^c[0,0] + \delta^{ab} N \op{Id} \frac{1}{z^2} 
\end{equation}
at level $N$.   One can compute other two-point functions by similar, but more complicated, diagrams.  For example, the two-point function of the operators $J^a[1,0]$ and $J^a[0,1]$ will be constrained by the diagram in figure \ref{fig:central2}.  

One can convince oneself that, if we sum over all numbers of dashed lines in these diagrams, the diagrams we are drawing can be re-interpreted as Witten diagrams in the backreacted geometry (in this case, $SL_2(\C)$).  The point is that when we insert an arbitrary number of the Beltrami differential fields, the flat-space propagator gets replaced by that on the backreacted geometry.  

The structure constants of the deformed Koszul dual chiral algebra are scattering amplitudes in  the backreacted geometry.  If we take two generators $J^a[k,l]$, $J^b[r,s]$ of the chiral algebra, their OPEs will be expressed as a sum of products of the generators and their derivatives.  This corresponds to the scattering of two single-particle states in the backreacted geometry, being expressed as a sum of multi-particle states.  

\section{(Deformed) Koszul duality and $\Sym^N T^4$ holography}\label{sec:deformed}

We have seen that the effective $6$-dimensional theory obtained by compactifying twisted type IIB supergravity on $T^4$ is a version of Kodaira-Spencer theory, where all fields are, in addition to being functions on the space-time $\C^3$, functions of four odd variables $\eta_a$.   The backreaction of the $D1$-$D5$ branes wrapping $T^4$ give rise to the Beltrami differential
\begin{equation} 
	\mu_{BR} = F^{ab} \eta_a \eta_b \frac{\eps^{ij} \wbar_i \d \wbar_j   }{ (w_1 \br{w}_1 + w_2 \br{w}_2)^2  } \partial_z 
\end{equation}
where $F \in H^2(T^4)$ is the flux sourced by the branes, working in a duality frame where the $D1$ and $D5$ branes are both $D3$'s wrapping two-cycles in $T^4$. We have added the subscript to emphasize that the Beltrami differential encodes the backreaction of the branes on the gravitational theory.

	We now have all the ingredients at hand to study the universal chiral algebra that can be placed at $w_i = 0$ in the presence of the field $\mu_{BR}$.  Indeed, we have already done some representative computations in the example of holomorphic Chern-Simons theory above.  Before we get to the details of the computation for our Kodaira-Spencer theory, let us explain a very important simplification.

	To compute the deformed Koszul dual, we study anomalies for Feynman diagrams as above, with some number of dashed lines at which we place the Beltrami differential field.  In the case of holomorphic Chern-Simons, we saw that we could include an arbitrary number of dashed lines.  

	In the Kodaira-Spencer theory we consider, however, it turns out that these diagrams can have at most two dashed lines! This greatly reduces the complexity of the calculations. To see this, we note that the propagator for our Kodaira-Spencer theory depends on the fermionic fields only via the fermionic $\delta$-function
	\begin{equation} 
		\delta_{\eta_a = \eta'_a} = \prod_a (\eta_a - \eta'_a). 
	\end{equation}
	The amplitude for each Feynman diagram is a product of an integral over the fermionic variables with an ordinary bosonic integral.  Since the fermionic propagator is a $\delta$-function, the fermionic integral for a connected diagram reduces to an integral over a single set of fermionic variables.  If there are three or more dashed lines, the integrand includes  $(F^{ab} \eta_a \eta_b)^3$, which is zero. 
	
A similar argument shows that any diagram that has a loop of bulk propagators is zero. If we have such a loop, then the fermionic integral is obtained by contracting all vertices to a single vertex, which has a self-loop.  The self-loop contributes the fermionic $\delta$-function evaluated at $\eta = \eta'$, which is zero.  

\begin{figure}

\begin{tikzpicture}[scale={0.75}]

	\begin{scope}[shift={(-5,0)}]
		\node[circle, draw] (J) at (0,0) {$J$};
		\node (A1) at (3,1)  {};
		\node (A2) at (3,-1)  {};
		\node[circle,draw,fill=black, minimum size = 0.2pt]  (V) at (1.5,0) {};  
		\draw[decorate, decoration={snake}] (J) -- (1.5,0);
		\draw[decorate, decoration={snake}] (1.5,0) --(A1);
		\draw[decorate, decoration={snake}] (1.5,0) --(A2);
		\draw (0,2)-- (J) -- (0,-2);
		\node at (0.75,-3) {\small $(a)$};
	\end{scope}

	\begin{scope}[shift={(0,0)}];
		\node (A1) at (3,1)  {};
		\node (A2) at (3,-1)  {};
		\node[circle,draw,fill=black, minimum size = 0.2pt]  (V) at (1.5,0) {};  
		\draw[dashed] (0,0) -- (1.5,0);
		\draw[decorate, decoration={snake}] (1.5,0) --(A1);
		\draw[decorate, decoration={snake}] (1.5,0) --(A2);
		\draw (0,2)-- (0,-2);
		\node at (0.75,-3) {\small $(b)$};
	\end{scope}

	\begin{scope}[shift={(5,0)}]
		\node (J1) at (0,1) {};
		\node[circle, draw] (J2) at (0,-1) {$J$};
		\node (A1) at (3,1.5)  {};
		\node (A2) at (3,-1.5)  {};

		\node[circle,draw,fill=black ]  (V1) at (1.5,1) {}; 
		\node[circle,draw,fill=black]  (V2) at (1.5,-1) {};  
		\draw[decorate, decoration={snake}]  (V1) --  (A1);
		\draw[decorate, decoration={snake}] (J2) -- (V2) -- (A2);
		\draw[decorate, decoration={snake}] (V1) -- (V2);
		\draw[dashed] (J1) to (V1);
		\draw (0,2) --  (J2) -- (0,-2);
		\node at (0.75,-3) {\small $(c)$};
	\end{scope}

	\begin{scope}[shift={(10,0)}]
		\node (J1) at (0,1) {};
		\node (J2) at (0,-1) {};
		\node (A1) at (3,1.5)  {};
		\node (A2) at (3,-1.5)  {};

		\node[circle,draw,fill=black ]  (V1) at (1.5,1) {}; 
		\node[circle,draw,fill=black]  (V2) at (1.5,-1) {};  
		\draw[decorate, decoration={snake}]  (V1) --  (A1);
		\draw[decorate, decoration={snake}] (V2) -- (A2);
		\draw[dashed] (J1)--(V1);
		\draw[dashed] (J2)--(V2);
		\draw[decorate, decoration={snake}] (V1) -- (V2); 
		\draw (0,2)  -- (0,-2);
		\node at (0.75,-3) {\small $(d)$};
	\end{scope}

	\end{tikzpicture}
	\caption{All diagrams that contribute in the planar limit.  Diagrams $(a)$ and $(c)$ will contribute to the planar $3$-point function, i.e.\ the terms in the OPE which takes generators of the universal chiral algebra and gives another generator. Diagrams $(b)$ and $(d)$ will contribute to the two-point function, i.e.\ the terms in the OPE which take two generators of the universal chiral algebra and yield the identity operator. \label{fig:alldiagramsplanar}}
\end{figure}
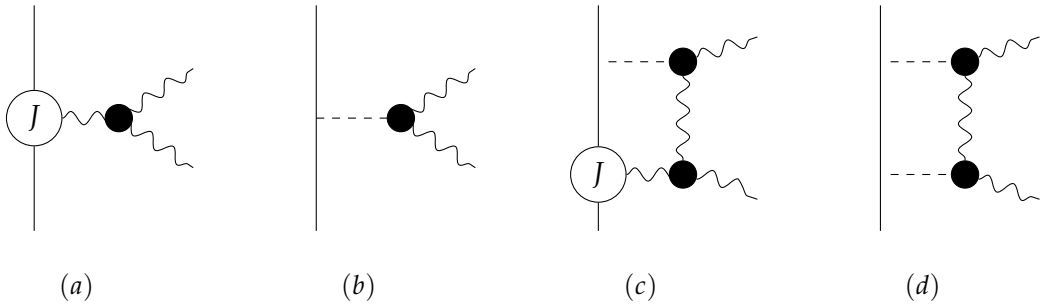

	We conclude that all diagrams that can contribute to the Koszul duality computation are those enumerated in figure \ref{fig:alldiagrams}.  The diagrams that can occur in the planar limit are drawn in figure \ref{fig:alldiagramsplanar}. When we refer to the planar limit in our theory in what follows, we simply mean the limit where $N \rightarrow \infty$ or, equivalently, the topological string coupling $\lambda \rightarrow 0$. As we explained earlier,  a rescaling of the fermionic variables brings the theory  with parameters $(\lambda, N)$ is equivalent to the theory with parameters $(\lambda N^{-1/2}, 1)$.
	
\begin{figure}
	\begin{tikzpicture}
	\begin{scope}
		\node [circle,draw,fill=black]  (A1) at (1.5,3) {};
		\node [circle,draw,fill=black]  (A2) at (1.5,2){};
		\node [circle,draw,fill=black]  (A3) at (1.5,1) {};
		\node [circle,draw,fill=black]  (A4) at (1.5,0) {};

		\node [circle,draw,fill=black]  (A6) at (1.5,-3) {};
		\draw[decorate, decoration={snake}]  (0,1) --  (A3);
		\draw[decorate, decoration={snake}] (0,0) -- (A4);
		\draw[decorate, decoration={snake}] (0,-3) -- (A6);
		\draw[decorate, decoration={snake}] (3,3.5) -- (A1)-- (A6)--(3,-3.5); 
			
		\draw[dashed](0,3) -- (A1);
		\draw[dashed] (0,2) -- (A2);

		\draw (0,4) -- (0,-4);
		\draw[white, fill=white] (1.5,-1.5)  circle (0.7);  
		\node [rotate=90] (A5) at (1.5,-1.5) {$\dots$  };
		\draw [decorate, decoration={brace,amplitude=10pt}] (-0.5,-3.5)--(-0.5,1.4) node[midway, xshift=-0.8cm]{$k$};
		\draw [decorate, decoration={brace,amplitude=10pt}] (-0.5,1.6)--(-0.5,3.5) node[midway, xshift=-0.8cm]{$\le 2$};
	\end{scope}
	\end{tikzpicture}
	\caption{The most general diagrams that can contribute to the Koszul duality calculations. The defect is attached to the bulk vertices by $\le 2$ dashed lines, corresponding to the backreaction, and an arbitrary number $k$ of wavy lines, which couple to non-trivial operators on the defect. We have only depicted the diagrams where the dashed lines are at the top of the ladder: they can be in any position.  In the planar limit, only diagrams with $k \le 1$ contribute. \label{fig:alldiagrams}} 
\end{figure}
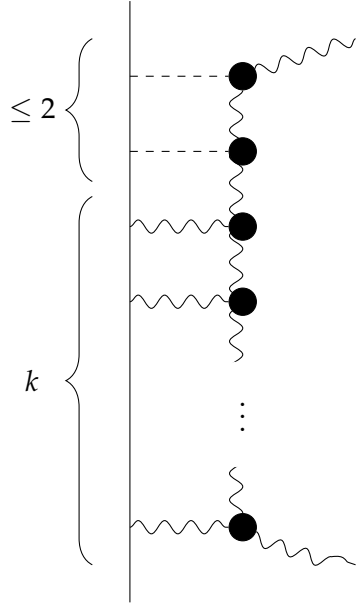

\subsection{The generators of the Koszul dual chiral algebra}

Now let us start analyzing the universal chiral algebra that can couple to our gravitational theory. Our first task will be to enumerate the set of generators of this chiral algebra, which will correspond to single-trace operators in the holographically dual CFT.  The collection of generators does not depend on the backreaction, and so can be determined on flat space. (As in the example of holomorphic Chern-Simons theory, the backreaction only affects the OPEs).  Recall that the fields of the gravitational theory are all superfields which are polynomials in the fermionic variables $\eta_a$. The ghost number $0$ fields are each $(0,1)$-forms on $\C^3$.  There are two fermionic fields (which by slight abuse of notation we will still denote by the same symbol as their $(2,*)$-form counterparts) $\alpha,\gamma \in \Omega^{0,1}(\C^3)\otimes \C[\eta_a]$, and three bosonic fields $\mu_z,\mu_i$ ($i = 1,2$)  which are also in $\Omega^{0,1}(\C^3) \otimes \C[\eta_a]$, and which satisfy (recall \ref{eq:constraint})
\begin{equation} 
	\partial_z \mu_z + \partial_{w_i} \mu_i  = 0.  
\end{equation}
This constraint will lead to an extra complication when compared with the analysis of holomorphic Chern-Simons theory.

Just like the fields of the bulk theory, all the operators of the universal defect theory will be polynomials in the fermionic variables $\eta_a$. 

The basic operators we find, before imposing the constraints on the bulk fields, are the following.
\begin{enumerate} 
	\item  Fermionic Virasoro primaries $G_\alpha[r,s] (\eta_a )$, and $G_\gamma[r,s](\eta_a )$, for each pair $r,s$ of non-negative integers, which couple to the fields $\alpha$, $\gamma$ by the expressions
\begin{equation}
	\begin{split}
		&	\frac{1}{r! s!}\int_{\C^{1 \mid 4}} G_\alpha[r,s](\eta_a) \partial_{w_1}^r \partial_{w_2}^s \alpha (\eta_a)   \d z  \d^4 \eta \\
		& \frac{1}{r! s!}\int_{\C^{1 \mid 4}} G_\gamma[r,s](\eta_a) \partial_{w_1}^r \partial_{w_2}^s \gamma (\eta_a) \d z \d^4 \eta
	\end{split}
\end{equation}
		These operators are of spin $1 + r/2 + s/2$. Notice that now $r+s$ plays the role of $k$ from section \ref{sec:states}, where we enumerated the gravitational states sourcing local boundary modifications, in terms of indexing the spin and conformal weight quantum numbers. That is, the operators $G_\alpha[r,s]$ form part of a representation of $SU(2)_R$ of spin $(r+s)/2$.  Each operator is a function of the fermionic variables $\eta_a$, and so can be expanded into $2^4$ component operators. 

	\item Similarly, we have bosonic Virasoro primaries $\til{T}[r,s](\eta_a)$, of spin $2 + r/2 + s/2$, which couple to the fields $\mu_z$ by 
		\begin{equation} 
			\frac{1}{r! s!}\int_{\C^{1 \mid 4}} \til{T}[r,s] (\eta_a) \partial_{w_1}^r \partial_{w_1}^s \mu_z(\eta_a) 
		\end{equation}
		
	\item Next, we have bosonic operators $\til{J}^i[r,s](\eta_a)$, which couple to the fields $\mu_i$ by
		\begin{equation} 
			\frac{1}{r! s!}\int_{\C^{1 \mid 4}}  \til{J}^i[r,s] (\eta_a) \partial_{w_1}^r \partial_{w_1}^s \mu_i(\eta_a).  
		\end{equation} 
\end{enumerate}
However, not all such operators can actually be sourced by bulk fields, because the bulk fields satisfy the constaint $\partial^i \mu_i + \partial_z \mu_z = 0$.   To implement this constraint at the level of the universal chiral algebra, it is convenient to make this a cohomological constraint, which we sketch below.

We can enumerate the operators that can be sourced by bulk fields by writing down bulk field configurations which are localized at $z = 0$ and which satisfy the equations of motion as well as this constraint. This constraint does not affect the operators $G_{\alpha}$, $G_{\gamma}$, so we will not discuss them.

The fields satisfying this constraint and localized at $z = 0$ are given by
\begin{equation} 
	\begin{split}
		\mathcal{E}_{r, s}&:= (r+1)(s+1) w_1^r w_2^s \delta_{z = 0} \partial_z - \frac{(s+1)}{2} w_1^{r+1} w_2^s \delta^{(1)} _{z = 0} \partial_{w_1} -   \frac{(r+1)}{2} w_1^{r} w_2^{s+1} \delta^{(1)}_{z = 0} \partial_{w_1}  \\
		& r  w_1^{r-1} w_2^s \delta_{z = 0} \partial_{w_2} - s w_1^r w_2^{s-1} \delta_{z = 0} \partial_{w_1}.
	\end{split}
\end{equation}
These fields are analogous to those constructed in section \ref{sec:states}, but now in the absence of the deformation, and in our new chosen flat coordinate system. These fields source the operators
\begin{equation} 
	\begin{split}
		T[r,s] &:=  \til{T}[r,s] - \frac{1}{2(r+1)} \partial_z \til{J}^1[r+1,s] - \frac{1}{2(s+1)} \partial_z \til{J}^2[r,s+1] \\
		J[r,s] &:= r \til{J}^2[r-1,s] - s \til{J}^1[r,s-1] 
	\end{split}
\end{equation}
which, because of the constraint, are the only operators (apart from the $G_\alpha$, $G_\gamma$) which generate the universal chiral algebra.

The operators $T[r,s]$ are of conformal dimension $2 + (r+s)/2$ and live in a representation of spin $(r+s)/2$ of $SU(2)_R$.  The operators $J[r,s]$ are of conformal dimension $(r+s)/2$ and of and live in a representation of spin $(r+s)/2$ of $SU(2)_R$.  The operators $J[r,s]$ only exist for $r+s \ge 1$.

In \cite{CL12}, it was shown that one can impose the constraint on the fields $\mu$ homologically, by introducing extra fields and extra terms to the BRST operator.  If we use this homological approach, then the universal chiral algebra becomes a differential graded vertex algebra, generated by the operators $\til{J}^i[r,s]$ and $\til{T}[r,s]$ together with a new operator $V[r,s]$ of cohomological degree $1$.  The differential is 
\begin{equation} 
	\begin{split}
		Q \til{J}^1[r,s] &= r V[r-1,s] \\
		Q \til{J}^2[r,s] &= s V[r,s-1] \\
		Q \til{T}[r,s] &= \partial_z V[r,s]. 
	\end{split}
\end{equation}
The $Q$-closed operators are then given by $J[r,s]$ and $T[r,s]$.

		It will often be more convenient to expand these operators in terms of Fourier dual fermionic variables $\what{\eta}^a$ (cf. equation \ref{eq:FFT}), by setting
		\begin{equation} 
			G_\alpha[r,s](\what{\eta}^a) = \int_{\C^{0\mid 4}} e^{\what{\eta}^a \eta_a} G_\alpha[r, s](\eta_a) \d^4 \eta 
		\end{equation}
		and similarly for all the other fields $G_\gamma[r,s] (\what{\eta}^a)$, $\til{J}_i[r,s](\what{\eta}_a)$, $\til{T}[r,s](\what{\eta}^a)$. Note that this fermionic Fourier transform simply identifies the coefficient of $\eta_{a_1} \dots \eta_{a_s}$ with the coefficient of $\what{\eta}^{a_{s+1}} \dots \what{\eta}^{a_{4}} \eps_{a_1 \dots a_4}$.

	Let us expand each operator as a polynomial in $\what{\eta}^a$:
		\begin{equation}
			\begin{split}
				G_\alpha[r,s] (\what{\eta}^a) &= G_\alpha^0[r,s] + G_{\alpha,a}^1[r,s] \what{\eta}^a + \dots \\
				G_\gamma[r,s] (\what{\eta}^a) &= G_\gamma^0[r,s] + G_{\gamma,a}^1[r,s] \what{\eta}^a + \dots \\
				J[r,s] (\what{\eta^a})&= J^0[r,s] + J^1_a[r,s] \what{\eta}^a +\dots \\ 
				T[r,s](\what{\eta^a})&= T^0[r,s] + T^1_a[r,s] \what{\eta}^a + \dots 
			\end{split}
		\end{equation}
In this expansion, the operators $G_\alpha^i[r,s]$, $J^i[r,s]$ etc.\  transforms in the $i$'th exterior power of the spin representation of the $\op{Spin}(5)$ global symmetry which rotates the $\what{\eta}^a$.

The leading terms in this expansion, for particular small values of $r,s$, will give the currents for the $\mf{psu}(1,1 \mid 2)$ global subalgebra of the $\mc{N}=4$ superconformal algebra.  The fermionic elements of the superconformal algebra are given by the two  	$SU(2)_R$ doublets  
		\begin{equation}
			\begin{split}
				& G_\alpha^0 [1,0] \ \ G_\alpha^0 [0,1] \\
				& G_\gamma^0 [1,0] \ \ G_\gamma^0 [0,1] 
			\end{split}
		\end{equation}
The bosonic operator $T^0[0,0]$ is the stress-energy tensor.  The operators $J^0[r,s]$ for $r+s = 2$ are the components of the $SU(2)_R$ current.

\subsection{The relation to the supergravity analysis and to the dual CFT}

Let us pause to discuss the relation of these states with what we found when we analyzed states at $\infty$ in the backreacted geometry, and to the states in the holomorphic twist of the dual CFT $Sym^{N \rightarrow \infty}(T^4)$. As in section \ref{sec:states}, one can preliminarily observe that these states admit a natural organization into short representations of $\mf{psu}(1, 1|2)$.   For every positive integer $n$, there is a short representation with highest weight, under $L_0$ and $J_0$, $(n/2,n/2)$. Each such short representation appears with multiplicity the cohomology groups $H^\ast(T^4)$.   The short representation with this highest weight consists of the operators 
\begin{equation} 
	\begin{split}
		J[r,s] & \ \ \ \ r + s = n \\
		G_\alpha[r,s],\ G_\beta[r,s] & \ \ \ \  r + s = n-1 \\
		T[r,s] & \ \ \ \  r+s = n-2.
	\end{split}
\end{equation}
The highest weight vector is $J[n,0]$, and the remaining operators are the superconformal descendents. 

The states we find in this Koszul duality analysis match exactly what we found in the more standard analysis using states at $\infty$ in the backreacted geometry.  Indeed, the components of $J[m,0]$ for $m \ge 1$, as we expand in the fermionic variables, give rise to the highest weight vector of the short representation $H^\ast(T^4)\otimes (\mbf{\tfrac{m}{2}})_S$.  The generators of the Koszul dual algebra are 
\begin{equation} 
	\bigoplus_{m \ge 1} H^\ast(T^4)\otimes (\mbf{\tfrac{m}{2}})_S. 
\end{equation}
In the case of $K3$, we would replace the cohomology of $T^4$ with that of $K3$.  

These states are closely related to the $\mf{psu}(1,1|2)$ descendents of the holomorphic chiral primary operators (1/2-BPS states) of the Fock space   $\oplus_N Sym^N(T^4)$ \cite{DMVV}.     See e.g.\ \cite{DMW02} for a review of how to construct these states in the CFT from twist fields and $T^4$ cohomology classes and see \cite{CP2} for a description of their twisted counterparts.   The single-particle states in the Fock space, which are descendents of $1/2$-BPS states, are
\begin{equation} 
	\bigoplus_{p,q} \bigoplus_{m \ge p} H^{p,q}(T^4) \otimes (\mbf{\tfrac{m}{2}})_S. 
\end{equation}
(As always, we are only considering the chiral sector: right-moving states are assumed to be highest weight states of a short representation).

This is almost the same as what we find by our Koszul duality analysis.  In our analysis, we are missing the states in $H^{0,q}(T^4) \otimes (\mbf{0})_S$.  These states are also removed in de Boer's analysis, to take the large $N$ limit.  Our construction also yields extra states, in $H^{2,q}(T^4) \otimes (\mbf{\tfrac{1}{2}})_S$, as discussed in section \ref{sec:states}.   

Let us now formulate a conjecture concerning the deformed Koszul dual chiral algebra and the $\Sym^N T^4$ chiral algebra. Rather than asking to produce an algebra which is \emph{isomorphic} to the supergravity chiral algebra, we will instead conjecture that there is a \emph{homomorphism} from the supergravity chiral algebra to the $\Sym^N T^4$ chiral algebra, for all values of $N$. As $N$ becomes large, we could hope that this homomorphism becomes closer to being an isomorphism.   

Let us denote the universal chiral algebra by $\mc{A}_{F}$, where $F$ is the flux, which is an element of $H^2(T^4)$.  The integral over $T^4$ of $F^2$ is $N_1 N_5$ (equivalently, this is the integral of $F^{ab}\eta_a \eta_b$ over the fermionic variables $\eta_a$).  

There are two important features to note about the definition of $\mc{A}_{F}$. First, we are  \emph{not} taking the planar limit.  After all, if we want to relate $\mc{A}_{F}$ to any particular finite $N$ chiral algebra, we must include non-planar corrections.  Defining the algebra beyond the planar limit is in principle not a problem, given the relatively small number of diagrams that contribute to its definition. 

A second important point is that we are setting the string coupling constant to $1$. Indeed, the string coupling constant can be absorbed into a rescaling of $N$: this theory has only one parameter. 
\begin{conjecture}
	There exists a homomorphism from the universal chiral algebra $\mc{A}_{F}$ of our Kodaira-Spencer theory on our prescribed deformation of $\C^{3|4}$ to the chiral algebra of the $\Sym^{N}(T^4)$ CFT, where $N$ and $F$ are related by $N= \int_{\C^{0\mid 4}} F^2 \d^4 \eta $.   
\end{conjecture}
Optimistically, one can hope that this homomorphism is surjective.

We can view this conjectural homomorphism as analogous to the truncation of $\mc{W}_{\infty}[\lambda]$ to $\mc{W}_N$ when the parameter of $\mc{W}_{\infty}[\lambda]$ is specialized to an integer, $\lambda = -N \in \Z$ (and in the $\lambda = -N \rightarrow \infty$ limit one recovers the linear $\mc{W}_{\infty}$ algebra). 

\subsection{OPEs in the universal chiral algebra}\label{sec:OPEs}
In this section, we will compute, for illustration, some of the OPEs in the universal chiral algebra in the limit when we turn off the backreaction.  This means we will obtain non-centrally extended algebras in what follows.  The full algebras can then be obtained by incorporating the backreaction in the manner explained above. The analysis is in many ways similar to the toy model of holomorphic Chern-Simons theory presented above.  

We will calculate the OPEs with the ``off-shell'' operators $\til{J}^i[r,s]$, and then restrict to the $Q$-closed operators $J[r,s]$. We will defer any computation of the OPEs of the operators $T[r,s]$ to future work.  

We can summarize the result of the computation as follows. Let $w_{\infty}$ be the Lie algebra of holomorphic Hamiltonian vector fields on $\C^2$.  Let $\mc{L}w_{\infty}$ be its loop Lie algebra: an element of this is a map $S^1 \to w_{\infty}$, or a holomorphic map $\C^\times \to w_{\infty}$.  We can also form the super-loop Lie algebra where we replace $S^1$ by $S^{1 \mid 4}$, or $\C^\times$ by $\C^\times \times \C^{0 \mid 4}$.  We refer to this as $\mc{L}^{1 \mid 4} w_{\infty}$.
\begin{theorem} 
	In the absence of the backreaction, and in the planar limit, the Lie algebra of modes of the operators $J[r,s]$ is the super-loop Lie algebra $\mc{L}^{1 \mid 4} w_{\infty}$.  
\end{theorem}
(We also calculate the leading-order contribution of the backreaction, which centrally extends this algebra in a certain way.)  

Now let us turn to the calculations which lead to this result.  Let us use the notation $D_{r,s}$ for the differential operator
\begin{equation} 
	D_{r,s} = \frac{1}{r!} \frac{1}{s!} \partial_{w_1}^r \partial_{w_2}^s. 
\end{equation}
Let us first calculate the OPEs between the fields $\til{J}^i[r,s]$.  Consider the gauge variation of 
\begin{equation} 
	\int_{(z,\eta_a) \C^{1 \mid 4}} \til{J}^1[r,s](z,\eta_a)  D_{r,s} \mu_1(z,w_i = 0,\eta_a) . 
\end{equation}
The gauge variation of $\mu_1$ is
\begin{equation} 
	\delta \mu_1 = \dbar \mf{c}_1 + \mu_i \partial_{w_i} \mf{c}_1 + \mu_z \partial_{z} \mf{c}_1 - \mf{c}_i \partial_{w_i} \mu_1 - \mf{c}_z \partial_z \mu_1. 
\end{equation}
Inserting this gauge variation into the coupling to $\til{J}^i[r,s]$, we see that the first term, $\dbar \mf{c}_1$, vanishes by integration by parts.  Cancellation of the remaining terms will give us constraints on the OPE coefficients, just as it did in the case of holomorphic Chern-Simons theory. 
The remaining terms are 
\begin{equation} 
	\int_{(z,\eta_a) \in \C^{1 \mid 4}} \til{J}^1[r,s] (z,\eta_a)  D_{r,s}\left( \mu_i \partial_{w_i} \mf{c}_1 + \mu_z \partial_{z} \mf{c}_1 - \mf{c}_i \partial_{w_i} \mu_1 - \mf{c}_z \partial_z \mu_1 \right)(z,w_i = 0, \eta_a). 
\end{equation}
Let us focus on the term in this expression which involves the fields $\mu_1$ and $\mf{c}_1$. This is 
\begin{equation} 
	 \int_{(z,\eta_a) \in \C^{1 \mid 4}} \til{J}^1[r,s] (z,\eta_a)  D_{r,s}\left( \mu_1 \partial_{w_1} \mf{c}_1   - \mf{c}_1 \partial_{w_1} \mu_1  \right)(z,w_i = 0, \eta_a). 
\end{equation}
Because this expression involves both $\mf{c}_1$ and $\mu_1$, which are fields (and a corresponding ghost) that couple to $\til{J}^1$, we find that it can only be cancelled by a gauge variation of an integral involving two copies of the operators $\til{J}^1$, at separate points $z,z'$:  
\begin{equation} 
	\tfrac{1}{2} \int_{z,z', \eta_a,\eta_a'} \til{J}^1[m,n] (z,\eta_a) D_{m,n} \mu_1(z, w_i = 0, \eta_a)  \til{J}^1[r,s] (z',\eta'_a) D_{r,s} \mu_1(z', w'_i = 0, \eta'_a) . 
\end{equation}
Applying the gauge variation of $\mu_1$ to this expression, and retaining only the terms involving $\dbar \mf{c}_1$, gives us
\begin{equation} 
	\int_{z,z', \eta_a,\eta_a'} \til{J}^1[m,n] (z,\eta_a) D_{m,n} \mu_1(z, w_i = 0, \eta_a)  \til{J}^1[r,s] (z',\eta'_a) D_{r,s} \dbar \mf{c}_1 (z', w'_i = 0, \eta'_a) . 
\end{equation}
Here the $\dbar$ operator is that in the $z$ direction, because restricting to $w_i= 0$ sets any $\d \wbar_i$ to zero. We can integrate by parts to move the location of the $\dbar$ operator. Every field $\mu_i$ contains a $\d \zbar$, as otherwise it would restrict to zero at $w_i = 0$, so that $\partial_{\zbar} \mu_i = 0$. 

This discussion shows that in order for the anomaly to cancel we need
\begin{multline} 
	\int_{z,z', \eta_a,\eta_a'} \dbar_{\zbar} \left( \til{J}^1[m,n] (z,\eta_a)  \til{J}^1[r,s] (z',\eta'_a) \right)  D_{m,n} \mu_1(z, w_i = 0, \eta_a)  D_{r,s}  \mf{c}_1 (z', w'_i = 0, \eta'_a)  \\
	= \int_{(z'',\eta''_a) \in \C^{1 \mid 4}} \til{J}^1[k,l] (z'',\eta''_a)  D_{k,l}\left( \mu_1 \partial_{w_1} \mf{c}_1   - \mf{c}_1 \partial_{w_1} \mu_1  \right)(z'',w_i = 0, \eta''_a).   
\end{multline}
In these expressions, we sum over the indices $r,s,k,l,m,n$.  This equation must hold for all values of the field $\mu_1$, $\mf{c}_1$. To constrain the OPEs, we can test the equation by setting $\mu_1 = G(z,\zbar,\eta_a) \d \zbar w_1^m w_2^n$, $\mf{c}_1 = H(z,\zbar,\eta_a) w_1^r w_2^s$, for $G,H$ arbitrary functions of the variables $z,\zbar,\eta_a$. 

Inserting these values for the fields into the anomaly-cancellation condition gives
\begin{multline} 
	\int_{z,z', \eta_a,\eta_a'} \dbar_{\zbar} \left( \til{J}^1[m,n] (z,\eta_a)  \til{J}^1[r,s] (z',\eta'_a) \right)  G(z,\zbar,\eta_a) H(z',\zbar',\eta_a') \\ 
	= \int_{z'',\eta''_a} (r-m)  \til{J}^1[m+r-1, n+s] (z'',\eta''_a)  G(z'',\zbar'',\eta''_a) H(z'',\zbar'', \eta_a'').
\end{multline}
Since this must hold for all values of the functions $G,H$ we get an identity of the integrands:
\begin{equation} 
	\dbar_{\zbar} \left( \til{J}^1[m, n] (z,\eta_a)  \til{J}^1[r,s] (z',\eta'_a) \right)  = \delta_{z = z',\br{z} = \br{z}'} \delta_{\eta_a = \eta'_a} (r-m) \til{J}^1[m + r -1, n+s] .  
\end{equation}
(Recall that the fermionic $\delta$-function $\delta_{\eta_a = \eta'_a}$ has the simple expression $\prod_a (\eta_a - \eta'_a)$).  

This in turn leads to the OPE:
\begin{equation} 
		\til{J}^1[m,n](0,\eta_a)  \til{J}^1[r,s](z,\eta'_a)  
	\sim \frac{1}{z} (r-m)  \til{J}^1 [m+r-1,n+s] (0,\eta_a) \delta_{\eta_a = \eta'_a}. 
\end{equation}

Finally, applying the fermionic Fourier transform to exchange the variables $\eta_a$ with $\what{\eta}^a$, we find	 
\begin{equation} 
	\til{J}^1[m,n](0,\what{\eta}^a)  \til{J}^1[r,s](z,\what{\eta'}^a)  
	\sim \frac{1}{z} (r-m)  \til{J}^1 [m+r-1,n+s] (0,\what{\eta}^a + \what{\eta'}^a ). 
\end{equation}
Similarly, we have 
\begin{equation} 
		\til{J}^2[m,n](0,\what{\eta}^a)  \til{J}^2[r,s](z,\what{\eta'}^a)  
	\sim \frac{1}{z} (s-n)  \til{J}^2 [m+r,n+s-1] (0,\what{\eta}^a + \what{\eta'}^a ). 
\end{equation}

To understand the OPE between $\til{J}^1$ and $\til{J}^2$, we proceed similarly.  The key point is that the coefficient of $\til{J}^1$ in the OPE between $\til{J}^1$ and $\til{J}^2$ must cancel the components of the BRST variation of $\sum \til{J}^1[k,l] D_{k,l} \mu_1$ which involve $\mf{c}_1 \mu_2$ and $\mu_1 \mf{c}_2$, and similarly for the coefficient of $\til{J}^2$ in this OPE.  Arguing as above leads to the results:
\begin{equation} 
	\til{J}^1[r,s] (0,\what{\eta}^a) \til{J}^2[k,l](z,\what{\eta'}^a) = - \frac{1}{z} s  \til{J}^1[r+k, l+s - 1] (0,\what{\eta}^a + \what{\eta'}^a)    + \frac{1}{z} k \til{J}^2[k+r-1,l+s] (0,\what{\eta}^a + \what{\eta'}^a)  
\end{equation}
and
\begin{equation} 
	\til{J}^2[r,s] (0,\what{\eta}^a) \til{J}^1[k,l](z,\what{\eta'}^a) = - \frac{1}{z} r  \til{J}^2[r+k-1, l+s ] (0,\what{\eta}^a + \what{\eta'}^a)    + \frac{1}{z} l \til{J}^1[k+r,l+s-1] (0,\what{\eta}^a + \what{\eta'}^a).   
\end{equation}

Let us use the these calculations to calculate the OPEs of the on-shell operators 
\begin{equation} 
	 	J[r,s] = r \til{J}^2[r-1,s] - s \til{J}^1[r,s-1]. 
\end{equation}
We find
\begin{multline}
	J[r,s] (0,\what{\eta}^a) J[k,l](z,\what{\eta'}^a) =   \frac{1}{z} (l-s) k r  \til{J}^2 [k+r-2,l+s-1]   \\
	+ \frac{1}{z} ls(k-r) \til{J}^1[k+r-1, l+s-2] \\
	+  \frac{1}{z} r (r-1) l  \til{J}^2[r+k-2, l+s -1 ]    - \frac{1}{z} l (l-1) r \til{J}^1[k+r-1,l+s-2]  \\
	+ \frac{1}{z} k s(s-1)  \til{J}^1[r+k-1, l+s - 2]    - \frac{1}{z}k s  (k-1) \til{J}^2[k+r-2,l+s-1] 
\end{multline}
(On the right hand side, all operators are evaluated at $z = 0$ and with the fermionic variables $\what{\eta}^a+\what{\eta'}^a$.  We have dropped this dependence for clarity.)

Collecting the terms, we find the OPE is
\begin{align*} 
	&\frac{1}{z} \left( (l-s)kr + r (r-1) l - ks (k-1)     \right)  \til{J}^2 [k+r - 2, l +s - 1]  \\
	&+ \frac{1}{z} \left( ls (k-r) - l (l-1) r + k s (s-1)  \right) \til{J}^1 [ k + r -1, l + s - 2]. 
\end{align*}

Since 
\begin{equation} 
	 J[k+r-1,l+s-1] = (k+r-1)\til{J}^2 [k+r-2,l+s-1] - (l+s-1) \til{J}^1 [ k+r - 1, l +s - 2] 
\end{equation}
we find that the OPE is 
\begin{equation} 
	J[r,s](0,\what{\eta}^a)J[k,l](z,\what{\eta'}^a) = \frac{1}{z} (rl-ks)  J[r+k-1,l+s-1](z, \what{\eta}^a + \what{\eta'}^a).     
\end{equation}

Note that the operators with $r + s = 2$ which are independent of $\what{\eta}^a$ satisfy the OPE of an $\mf{su}(2)$ Kac-Moody algebra at level zero.  

This OPE leads to a particularly nice mode algebra. If we let
\begin{equation} 
	J[r,s]_n (\what{\eta}^a) = \oint z^{-n- 1 + (r+s)/2} J[r,s](z,\what{\eta}^a) \d z 
\end{equation}
be the $n$th mode, then the OPE we have described gives rise to the following commutation relation:
\begin{equation} 
	[ J[r,s]_n ( \what{\eta}^a ) , J[k,l]_m (\what{\eta'}^a)  ]  =(rl-ks)  J[r+k-1,l+s-1]_{n+m}     (\what{\eta}^a + \what{\eta'}^a) \label{eqn:mode} 
\end{equation}

Note that the Lie algebra of Hamiltonian functions on the plane has the Poisson bracket
\begin{equation} 
	\{w_1^r w_2^s, w_1^k w_2^l \}  = (lr-ks) w_1^{r+k -1} w_2^{s+l-1}  . 
\end{equation}
This has a clear similarity to our mode algebra.  If we set the fermionic variables $\what{\eta}^a$ to zero, we see that the Lie algebra of zero modes is   precisely the Lie algebra of volume-preserving symmetries on the plane. 

Further, the Lie algebra of all modes, where we set $\what{\eta}^a = 0$, is the loop algebra of the Lie algebra of volume-preserving symmetries of the plane.  An element of this algebra is a vector field on $S^1 \times \R^2$, which is tangent to $\R^2$ and preserves the volume along the fibration $\R^2 \times S^1 \to S^1$. (In our context it is perhaps more natural to describe the algebra in terms of holomorphic vector fields on $\C^2 \times \C^\times$; the result is the same).  

The Lie algebra including fermionic variables $\what{\eta}^a$ has a similar description.  Consider the Lie algebra of functions on $\C^{2 \mid 4}$ with Lie bracket given by the Poisson bracket $\partial_{w_1} \wedge \partial_{w_2}$.  This Lie algebra is spanned by expressions like $w_1^r w_2^s F(\eta_a)$, with commutation relation
\begin{equation} 
	[w_1^r w_2^s F(\eta_a), w_1^k w_2^l G(\eta_a) ] = (rl-ks)w_1^{r+k-1}w_2^{s+l-1} F(\eta_a) G( \eta_a). 
\end{equation}
We can build elements of this algebra which depend on an auxiliary variable $\what{\eta}^a$ by
\begin{equation} 
	\rho_{r,s}(\what{\eta}^a) = e^{\what{\eta}^a \eta_a} w_1^r w_2^s.
\end{equation}
Note that each term in the Taylor expansion of $\rho_{r,s}(\what{\eta}^a)$ in the $\what{\eta}^a$ variables is a function on $\C^{2 \mid 4}$.   These operators have the commutation relations
\begin{equation} 
	[\rho_{r,s}(\what{\eta}^a), \rho_{k,l}(\what{\eta'}^a)] = (rl-ks) \rho_{r+k-1,s+l-1} (\what{\eta}^a + \what{\eta'}^a ).  
\end{equation}
Because this is the same commutation relation as we found in equation \eqref{eqn:mode},  

From this we find that there is a Lie algebra homomorphism from $\C[w_i,\eta_a]$, with its Poisson bracket, to the Lie algebra of single-particle modes of our theory. 

We can easily uplift this to a similar statement for the entire Lie algebra of single-particle modes.  We find that there is a homorphism from the infinite-dimensional Lie algebra of loops into $\C[w_i,\eta_a]$ into the entire mode algebra.  This loop algebra is simply $\C[w_i,z,z^{-1}, \eta_a]$ with the Lie bracket coming from the Poisson tensor $\partial_{w_1}\wedge \partial_{w_2}$.  If we let
\begin{equation} 
	\rho_{r,s,n} (\what{\eta}^a) =  e^{\what{\eta}^a \eta_a} w_1^r w_2^s
\end{equation}
then it is easy to check that these satisfy the commutation relations
\begin{equation} 
	[\rho_{r,s,n} (\what{\eta}^a), \rho_{k,l,m}(\what{\eta'}^a) ] = (rl-ks) \rho_{r+k-1,s+l-1,m+n} (\what{\eta}^a + \what{\eta'}^a).  
\end{equation}
We find therefore find that the modes of $J[r,s](\what{\eta}^a)$ are the Lie algebra of polynomial functions on $\C^{2 \mid 4} \times \C^\times$, with the Poisson bracket $\partial_{w_1} \wedge \partial_{w_2}$. 

This is very reasonable: this Lie algebra is the algebra of  symmetries of $\C^{2 \mid 4} \times \C^\times$, viewed as a bundle over $\C^{0 \mid 4} \times \C^\times$ with fibres the holomorphic symplectic manifold $\C^2$.  In particular, these are certain gauge symmetries of our bulk gravitational theory on $\C^{2} \times \C^\times$.  

We have not included calculations of all the OPEs: we intend to do this in a sequel.  We believe that when we do so, we will find that the mode algebra consists of \emph{all} gauge symmetries on $\C^2 \times \C^\times$. In particular, this will include the Lie algebra of divergence-free vector fields on $\C^2 \times \C^\times$, which are also functions of the fermionic variables of $\eta_a$.  

\subsection{Central extensions}
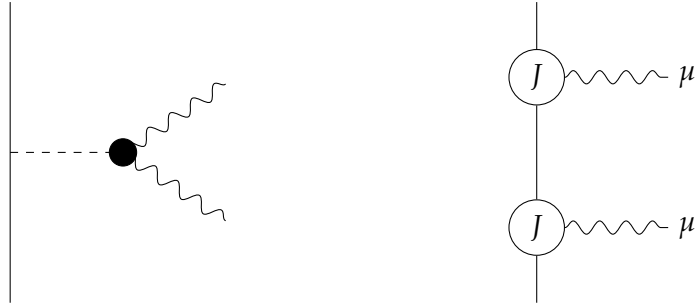
\begin{figure}
	\begin{tikzpicture}
	\begin{scope}[shift={(-7,0)}]
		
		\node (A1) at (3,1) {};
		\node (A2) at (3,-1) {};
		\node[circle,draw,fill=black, minimum size = 0.2pt]  (V) at (1.5,0) {};  
		\draw[dashed] (0,0) -- (1.5,0); 
		\draw[decorate, decoration={snake}] (1.5,0) --  (A1);
		\draw[decorate, decoration={snake}] (1.5,0) -- (A2);
		\draw (0,2) -- (0,-2);
	\end{scope}
	
	\begin{scope}[shift={(0,0)}]
		\node[circle, draw] (J1) at (0,1) {$J$};
		\node[circle, draw] (J2) at (0,-1) {$J$};
		\node (A1) at (2,1)  {$\mu$};
		\node (A2) at (2,-1)  {$\mu$};
		\draw[decorate, decoration={snake}] (J1) --(A1);
		\draw[decorate, decoration={snake}] (J2) --(A2);
		\draw (0,2) -- (J1) --(J2) -- (0,-2); 
	\end{scope}

	\end{tikzpicture}
	\caption{The new vertex in the presence of the backreaction (left) and one of the diagrams contributing to the anomaly (right) \label{fig:br}}
\end{figure}

Let us include the very simplest contribution of the backreaction to the OPEs of the currents $J[r,s](\what{\eta}^a)$.  This comes from the anomaly for the diagram in figure \ref{fig:br}.

This diagram contributes
\begin{equation} 
	\int_{\C^3 \mid 4}  \mu_{BR}(\eta_a)\mu_1 (\eta_a)  \mu_2 (\eta_a). 
\end{equation}
Since
\begin{equation} 
	\dbar \mu_{BR}= \delta_{w_i = 0} F^{ab} \eta_a \eta_b \partial_z 
\end{equation}
the gauge variation of the diagram gives us
\begin{equation} 
	\int_{\C^{1 \mid 4}} F^{ab} \eta_a \eta_b \mu_i (\eta_a) \c_j(\eta_b) \eps^{ij}.  
\end{equation}
Since $\til{J}^i[0,0]$ couples to $\mu_i$, this anomaly can be cancelled by the gauge variation of 
\begin{equation} 
	\int_{z,z',\eta_a,\eta'_a} \til{J}^1[0,0] (z,\eta_a) \mu_1(z,\eta_a) \til{J}^2[0,0](z',\eta'_a) \mu_2 (z',\eta'_a) \label{eqn:extensioncoupling} 
\end{equation}
provided that the operators $\til{J}^i[0,0]$ satisfy an appropriate OPE. The required OPE is
\begin{equation} 
	\til{J}^i[0,0] (0,\eta) \til{J}^j[0,0](z,\eta') \simeq \eps_{ij} \delta_{\eta = \eta'} F^{ab} \eta_a \eta_b  \frac{1}{z}, 
\end{equation}
it is easy to verify that reinserting this into \eqref{eqn:extensioncoupling} cancels the anomaly. 

Applying the fermionic Fourier transform, this becomes
\begin{equation} 
	\til{J}^i[0,0] (0, \what{\eta}^a) \til{J}^j[0,0](z,\what{\eta'}^a) \simeq \frac{1}{z} \what{F} (\what{\eta}^a + \what{\eta'}^a) . 
\end{equation}
Here $\what{F}$ is the Fourier transform of $F^{ab} \eta_{a} \eta_b$, which is of the form $\what{F}_{ab} \what{\eta}^a \what{\eta}^b$.

If we expand out each $\til{J}^i[0,0]$ as a function of the variables $\what{\eta}^a$, we get component operators
\begin{equation} 
	\til{J}^i[0,0] (\what{\eta}^a) = \til{J}^i_0[0,0] + \what{\eta}^a \til{J}^i_a[0,0] + \what{\eta}^a \what{\eta}^b \til{J}^i_{ab}[0,0] + \dots. 
\end{equation}
In components, the OPE we have just derived tells us that
\begin{equation}
	\begin{split}
		\til{J}^i_a[0,0](0) \til{J}^j_b[0,0](z) & \simeq \eps^{ij} \frac{1}{z} \what{F}_{ab}\\ 
		\til{J}^i_0[0,0](0) \til{J}^j_{ab} [0,0](z) & \simeq \eps^{ij} \frac{1}{z} \what{F}_{ab}. 
	\end{split}
\end{equation}

The operators $\til{J}^i[0,0]$ are actually on-shell operators, because of the relation
\begin{equation}
	\begin{split}
		J[1,0] (\what{\eta}^a) &=  \til{J}^2[0,0]  (\what{\eta}^a) \\
		J[0,1] (\what{\eta}^a)  &=  -\til{J}^1[0,0] (\what{\eta}^a) 
	\end{split}
\end{equation}
This means that the equations we have derived apply to the components of the operators $J[1,0]$, $J[0,1]$. In particular, the components $J_a[1,0]$ and $J_b[0,1]$ acquire the term in the OPE
\begin{equation} 
	J_a[1,0] J_b[0,1] \simeq \frac{1}{z} \what{F}_{ab},,  
\end{equation}
and so forth. 

We have seen above that the algebra of modes of the operators $J[r,s]$ is the Lie algebra of polynomial functions on $\C^{2 \mid 4} \times \C^\times$, under the Poisson bracket $\partial_{w_1} \wedge \partial_{w_2}$.  The OPE we have just computed gives us a central term in the commutator of the mode algebra, whereby if $G(w_i,z,\eta_a)$, $H(w_i, z,\eta_a)$ are functions on $\C^{2 \mid 4} \times \C^\times$, the commutator is
\begin{equation} 
	[G,H] = \eps_{ij} \partial_{w_i} G \partial_{w_j} H + c \oint_{z, \eta_a, w_i = 0} \eps_{ij} \partial_{w_i} G  \partial_{w_j} H   F^{ab} \eta_a \eta_b  
\end{equation}	 
where $c$ is the central element.  (There is one other diagram which we have not yet computed which involves only one copy of the flux $F$: this should add a further correction to this expression).

\appendix

\section{Koszul duality from TQFT}\label{app:koszulTFT}

Here we briefly recapitulate some basic facts about Koszul duality and its natural appearance in topological field theory.

Suppose that $\mc{A}$ is a differential graded associative algebra and that $h: \mc{A} \to \C$ is homomorphism of dg algebras:
\begin{equation} 
h(ab) = h(a)h(b), \ \ \ \ h(\d a) = 0 . 
\end{equation} 
 The homomorphism $h$ is called the \emph{augmentation}. The augmentation makes the one-dimensional vector space $\C$ into a module for $\mc{A}$, the module structure given by multiplication with $h(a)$.  We refer to this module as $\C_{h}$. In the context of topological line defects described in section \ref{sec:koszul}, if the theory on the line one obtains upon reducing the theory along the transverse directions is the trivial theory, then we have the augmentation map $\mc{A} \rightarrow \C$, viewing $\C$ as a rank-1 $\mc{A}$-module. This reduction requires a choice of vacuum, viewed as a suitable boundary condition at infinity, in the noncompact transverse directions \footnote{These considerations should enable one to connect the boundary condition and universal defect pictures of Koszul duality that we have presented in this note, though we will not attempt to do so here.}.

The mathematical definition of the Koszul dual algebra $\mc{A}^!$ is
\begin{equation} \label{eq:koszuldef}
\mc{A}^! = \op{Ext}^\ast_{\mc{A}}(\C_{h},\C_{h}). 
\end{equation}
That is, it is the self-Ext's (roughly speaking, symmetries) of the module $\C_{h}$.  

One way to interpret this more physically is the following. Using TFT axiomatics \cite{L09, L11}(see also \cite{BBBDN18} for recent related discussions), we can build a two-dimensional topological field theory whose category of left (or right) boundary conditions is the category $\mc{A}$-$\op{mod}$ of left (or right)  $\mc{A}$-modules. Given any module $M$, the algebra of operators on the corresponding boundary condition is $\op{Ext}^\ast_\mc{A}(M,M)$, the self-Ext's of $M$. 

The algebra $\mc{A}$ itself is a right $\mc{A}$-module under right multiplication: therefore it defines a right boundary condition, $B_R$. The algebra of self-Ext's of $\mc{A}$, as a right $\mc{A}$-module, is simply $\mc{A}$ itself, acting by left multiplication.  Therefore $\mc{A}$ is the algebra of local operators on the boundary $B_R$. 

The module $\C_{h}$ defines a left boundary condition, $B_L$. The operators on the boundary with this boundary condition are, by definition \eqref{eq:koszuldef}, the Koszul dual algebra $\mc{A}^!$. 

The states of the two-dimensional TFT on a strip, with $B_L$ on one side and $B_R$ on the other, are the one-dimensional vector space $\C_{h}$. This strip configuration is one convenient physical interpretation of the augmentation.

In general, following this picture, we can propose a physical origin of Koszul duality: if we have a two-dimensional TFT with left and right boundary conditions $B_L$, $B_R$, such that the states on a strip with these boundary conditions is one dimensional, then the algebras of boundary operators for the two boundary conditions are Koszul dual.

\end{document}